\documentclass[aps,pra,reprint,showpacs,groupedaddress]{revtex4-1}
\pdfoutput=1
\usepackage{graphicx,color} 
\usepackage{amsmath, amssymb}
\usepackage{longtable}
\usepackage{natbib}

\begin{document}
\title{
Exchange interaction between $J$-multiplets
}
\author{Naoya Iwahara}
\author{Liviu F. Chibotaru}
\affiliation{Theory of Nanomaterials Group, 
Katholieke Universiteit Leuven, 
Celestijnenlaan 200F, B-3001 Leuven, Belgium}
\date{\today}

\begin{abstract}
Analytical expressions for the exchange interaction between $J$-multiplets of 
interacting metallic centers are derived on the basis of 
a complete electronic model which includes the intrasite relativistic effects.
%a microscopic electronic model.
A common belief that this interaction can be approximated by an isotropic form $\propto {\mathbf{J}}_1\cdot{\mathbf{J}}_2$ (or $\propto {\mathbf{J}}_1\cdot{\mathbf{S}}_2$ in the case of interaction with an isotropic spin) is found to be ungrounded. 
It is also shown that the often used ``$1/U$ approximation'' for the description of the kinetic contribution of the exchange interaction is not valid in the case of $J$-multiplets. 
The developed theory can be used for microscopic description of exchange interaction in materials containing lanthanides, actinides and some transition metal ions. 
%
%The crystal-field levels in lanthanides and other metal complexes with unquenched orbital momentum
%originate from the ground atomic $J$-multiplet.
%It was long believed that the exchange interaction between $J$-multiplets is basically
%described by Heisenberg form. %, $\hat{\mathbf{J}}_1 \cdot \hat{\mathbf{J}}_2$.
%In this work, the direct and the superexchange models are applied
%for analytical derivation of exchange interaction between
%arbitrary $J$-multiplets. % $\hat{\mathbf{J}}_1$ and $\hat{\mathbf{J}}_2$. 
%The structure and the energy spectrum of the obtained superexchange Hamiltonian
%are significantly different from those of a Heisenberg Hamiltonian.
%Besides, it is also found that
%the $1/U$ approximation is not applicable for the description of exchange spectrum,
%since it gives qualitatively different predictions compared to the present treatment,
%where $U$ is a promotion energy.
%Similar results are obtained for
%the exchange interaction between
%$J$-multiplet 
%($\hat{\mathbf{J}}_1$) 
%and isotropic magnetic center. % ($\hat{\mathbf{S}}_2$).
\end{abstract}

\pacs{
75.30.Et % kinetic exchange
71.70.Ej % spin-orbit coupling in condensed matter
75.50.Xx % molecular magnets (magnetic materials)
}

\maketitle 

\section{Introduction}
\label{Sec:Introduction}
% Introduction
Strong magnetic anisotropy induced by spin-orbit coupling on the metal sites is a key ingredient for a number of 
intriguing properties of magnetic materials, such as single-molecule magnet behavior 
\cite{Gatteschi2006, LayfieldMurugesu2015}, 
magnetic multipole ordering \cite{Santini2009},
and various exotic electronic phases \cite{Witczak-Krempa2014, Gingras2014}.
If the spin-orbit coupling exceeds the crystal-field splitting of the ground term $LS$ on the metal site, the latter acquires unquenched orbital momentum $\hat{\mathbf{L}}$ and the low-lying spectrum is well described as crystal-field split eigenstates of the total angular momentum $\hat{\mathbf{J}}=\hat{\mathbf{L}}+\hat{\mathbf{S}}$, where $\hat{\mathbf{S}}$ is the spin of the metallic term. 
This situation takes place in lanthanides \cite{Wybourne1965}, actinides \cite{Santini2009} 
and some transition metal ions in a cubic symmetry environment \cite{Griffith1971, Abragam1970}. 

The exchange interaction between such split-$J$ crystal-field levels (or groups of levels) 
is significantly more complicated than the exchange interaction between pure spin terms ($L=0$) 
described by the Heisenberg Hamiltonian $\mathcal{J} \hat{\mathbf{S}}_1 \cdot \hat{\mathbf{S}}_2$.
For the weak spin-orbit coupling, 
the discrepancy of the exchange Hamiltonian from the isotropic form 
was first pointed out by Stevens \cite{Stevens1953}.
Later, the anisotropic exchange interaction was extensively developed by Moriya \cite{Moriya1960} 
based on the Anderson's microscopic approach \cite{Anderson1959, Anderson1963}.
In the case of the strong spin-orbit coupling, 
the exchange interaction including the higher order terms of $\hat{\mathbf{J}}$
was also phenomenologically treated since long time ago \cite{Levy1964, Birgeneau1969}.
The microscopic description was addressed for the first time 
by Elliott and Thorpe \cite{Elliott1968} for uranium oxides, 
and by Hartmann-Boutron \cite{Hartmann-Boutron1968} for transition metal compounds
on the basis of simplified analysis based on so-called $1/\bar{U}$ approximation
($\bar{U}$ is the average electron promotion energy between the sites).
%On the basis of simplified analysis based on so-called $1/U$ approximation 
%\cite{Elliott1968, Hartmann-Boutron1968} 
%it has been pointed out that the $J$-$J$ exchange Hamiltonian is not of the simple bilinear form 
%of the pseudospin or total angular momentum 
Recently, within the same approximation, the microscopic derivation of the 
exchange Hamiltonian between $J$-multiplets was completed by Santini {\it et al}. \cite{Santini2009}. 
%\cite{Elliott1968, Mironov2003UO2}. 

Despite this early evidence of complexity of exchange interaction between metal ions with unquenched orbital moments, 
it was repeatedly conjectured that %long time ago that 
%the low-energy crystal-field levels of the magnetic center 
%are characterized by the ground atomic $J$-multiplet,
%Thus, the magnetic interactions between the metal sites with unquenched orbital momentum 
%cannot be described by spin but by total angular momentum, $\hat{\mathbf{J}}$.
the exchange interaction between fully degenerate $J$-shells, involving $(2J_1 +1)$ and $(2J_2 +1)$ angular momentum eigenstates on the first and the second magnetic centers, respectively, is described by an isotropic exchange Hamiltonian written in terms of $\hat{\mathbf{J}}_i$ momenta:
\begin{eqnarray}
 \hat{H}_\text{Heis} &=& \mathcal{J} \hat{\mathbf{J}}_1 \cdot \hat{\mathbf{J}}_2.
\label{Eq:JJ}
\end{eqnarray}
Contrary to Heisenberg Hamiltonian for isotropic spins, 
there is no {\it a priori} justification for the Hamiltonian (\ref{Eq:JJ}). 
Nevertheless, this form is often used for the description of interaction between lanthanides or actinides (or a similar form, $\propto \hat{\mathbf{J}}_1\cdot \hat{\mathbf{S}}_2$, in the case of their interaction with an isotropic spin), especially, in the last years 
\cite{Molavian2007, Talbayev2008, Curnoe2008, Arnold2010, 
Magnani2010, Lukens2012, Carretta2013, Przychodzen2007, Yamaguchi2008, 
Klokishner2009, Klokishner2012, Reu2013, Dreiser2012, Kofu2013}.
One of the reasons that the simple bilinear form has been often used is that 
the large numbers of the phenomenological exchange parameters cannot be easily determined.

It is not clear, however, how important are ``non-Heisenberg'' terms in the actual $J$-$J$ coupling, 
nor is the $1/\bar{U}$ approximation {\it a priori} justified for metal ions with unquenched orbital momenta. 
Both these questions can only be answered after a 
%decent 
more complete
derivation of exchange interaction between $J$ multiplets on the basis of a reliable microscopic model. 
Besides, a microscopic description of $J$-$J$ ($J$-$S$) exchange interaction is desirable due to a very large number of phenomenological parameters, % parametrizeing such interaction, 
in contrast to weakly-anisotropic spin systems containing only a few of them \cite{Bencini1990, Chibotaru2015}.
%which is a reason why Eq. (\ref{Eq:JJ}) is used so often.
Given that many microscopic electronic parameters describing individual magnetic centers and their interaction can be accurately derived via density functional theory \cite{dft} 
or {\it ab initio} calculations \cite{Ungur2015}, a microscopically derived Hamiltonian for 
$J$ multiplets can become a powerful tool for the investigation of exchange interaction in materials 
containing lanthanides, actinides and transition metal ions with unquenched orbital momentum. 
To this end the %$J$-$J$ ($J$-$S$) exchange 
electronic Hamiltonians only need to be downfolded on the reduced manifold of low-lying states at the corresponding metal ions.    
%For example, the superexchange pseudospin Hamiltonian between the degenerate ground crystal-field states 
%was found to have higher order terms as well as the bilinear term \cite{Mironov2003UO2}.
%Similarly, the superexchange Hamiltonian between $J$-multiplets is not Heisenberg form 
%\cite{Santini2009}.
%However, because of the lack of the detailed experimental data \cite{Elliott1968} 
%or the complexity of the exchange coupling \cite{Santini2009, Gingras2014}, 
%the investigation based on the adequate microscopic exchange mechanism has been avoided.
%Consequently, the development of the methodology has not been accomplished
%and the investigation of concrete systems. %based on the adequate Hamiltonian has been avoided. 

%On the other hand, recently, 
%the requirement of the use of the adequate magnetic interaction is gradually increasing.
%For example, the lanthanide complexes containing radical as a magnetic center
%exhibit significantly strong intramolecular antiferromagnetic interaction 
%between the metal center and the radical \cite{Demir2014}. 
%, implying the importance to take into account of the entire $J$-multiplets.
%Fortunately, these complexes turn out to have simple core structures, 
%which enables us to concentrate on the development and the application of 
%the general form of the microscopic exchange interaction. 
% and understanding of the nature of the exchange states.

In this work, we 
%apply the direct and the superexchange mechanisms \cite{Anderson1959, Anderson1963}
derive analytically the exchange Hamiltonians for interacting $J$-multiplets 
and for interacting $J$-multiplet and isotropic spin, starting from %a complete electronic model. 
a  microscopic electronic Hamiltonian including the relativistic interactions on the metal sites.
The obtained exchange parameters are expressed via electronic matrix elements 
which can be derived from electronic structure calculations. 
The structure of the exchange Hamiltonian is discussed and 
the result is applied for the analysis of some systems with different geometries.
Comparison with the predictions given by the Hamiltonian (\ref{Eq:JJ}) and the simplified treatment 
on the basis of $1/\bar{U}$ approximation shows that both of them are not suitable approaches 
to describe the exchange interaction of ions with unquenched orbital momentum.
Finally, the relative contributions to the kinetic exchange interaction from intermediate states 
is analyzed.

\section{Microscopic description of intersite interaction}
\label{Sec:Microscopic}
We derive the expression for the interaction between metal ions with unquenched orbital moments. 
The derivation is based on a complete electronic Hamiltonian, including all intrasite relativistic effects, 
and employs adequate approximations. 
The multipolar intersite interactions of electromagnetic type, 
such as the electric quadrupolar and magnetic dipolar interactions, 
have been described elsewhere \cite{Santini2009, Birgeneau1969, Birgeneau1967, Zvezdin1985} 
and are not considered here. 
Their effect can be taken into account as additive contribution to the exchange parameters.
%The exchange interaction between localized magnetic sites is described here. 
%We first explain the electronic structure of single ion,
%and then derive the exchange Hamiltonian.
%For simplicity, we consider the exchange interaction between two sites since 
%the generalization to many site system is straightforward.

\subsection{Electronic multiplets on sites}
\label{Sec:ion}
%Consider the electronic structure of a free metal ion 
%with partially filled $nl$ shell \cite{LandauQM},
%where $n$ is the main quantum number and $l$ is the atomic orbital angular momentum.
% where the $d$ or $f$ orbitals are partially filled.
The nonrelativistic electronic state of an ion with partially filled $nl$ shell ($n$ is the main quantum number and $l$ is the one-electron orbital angular momentum) 
%in the absence of spin-orbit coupling 
corresponds to an $LS$-term characterized by the total orbital $\hat{\mathbf{L}}$ 
and spin $\hat{\mathbf{S}}$ angular momenta \cite{LandauQM}. The eigenfunctions 
$\{|\alpha_{LS} L M_L S M_S\rangle\}$ are described by orbital and spin quantum numbers, $L$ and $S$, and the projections of $\hat{\mathbf{L}}$ and $\hat{\mathbf{S}}$ on a given axis $z$,
$M_L$ and $M_S$, respectively; 
$\alpha_{LS}$ indicates the other quantum numbers.
The $(2L+1)(2S+1)$-fold degenerate term is further split by the spin-orbit interaction into $J$-multiplets which are eigenstates of the total angular momentum 
$\hat{\mathbf{J}}=\hat{\mathbf{L}}+\hat{\mathbf{S}}$. The corresponding eigenfunctions $\{|\alpha_J JM\rangle\}$ are characterized by quantum numbers of the total angular momentum
$J$ and its projection $M=-J,-J+1,\cdots J$ ($\alpha_J$ stands for other quantum numbers).

In general, the spin-orbit interaction mixes multiplets with the same $J$ belonging to different $LS$ terms (the so-called $J$-$J$ mixing \cite{Zvezdin1985}).
In the case when this mixing can be neglected,
each $J$-multiplet is attributed to one $LS$-term and the corresponding wave functions become of the form ($\alpha_J = (\alpha_{LS}, L,S)$):
\begin{eqnarray}
 |\alpha_J JM\rangle &=& \sum_{M_L M_S}|\alpha_{LS} LM_LSM_S\rangle C_{LM_L SM_L}^{JM},
\label{Eq:JM}
\end{eqnarray}
where %$C_{j_1m_1 j_2m_2}^{jm}$ 
$C_{LM_L SM_L}^{JM}$ is the Clebsch-Gordan coefficient (\ref{Eq:CG}) \cite{Varshalovich1988}.
This is a good approximation, in particular, for the ground $J$-multiplet of trivalent ions from the 
late lanthanides series, Ln$^{3+}$.

When the metal ions are embedded in complexes or crystals, 
their electronic structure is modified due to covalent and electrostatic interaction of the 
magnetic $nl$ orbitals with the environment.
In the case of $f$-metals the magnetic orbitals are usually strongly localized
and the effect of the surrounding is relatively weak.
For example, in the case of lanthanide,
the intraionic bielectronic interaction leading to atomic terms separation is ca 5-7 eV 
and the spin-orbit splitting is ca 1 eV for lanthanide ions, thus 
exceeding several times the 
the crystal-field splitting, which is usually of the order of 0.1 eV \cite{vanderMarel1988}.
In this situation the low-energy electronic states are well approximated as crystal-field split atomic $J$-multiplets.
%though the degeneracy is slightly lifted by the crystal-field.
%slightly split by the crystal-field.
Due to the weak hybridization of the $4f$ orbitals with the surrounding,
the Wannier functions of the corresponding magnetic orbitals %of lanthanide is well described 
practically coincide with the atomic $4f$ orbitals.
Similar holds true for actinide ions although 
$5f$ orbitals are more delocalized than $4f$ orbitals.

On the other hand, the effect of the hybridization of $d$ orbitals 
with the ligand orbitals is usually much stronger than in lanthanides and actinides resulting in a crystal-field splitting which often overcomes the atomic $LS$-term splitting. 
Therefore, 
%in general, the $(2L+1)$-fold degeneracy due to the spherical symmetry is lifted and 
the orbital angular momentum is generally not a good quantum number for 
nondegenerate ground state of 
embedded transition metal ions. 
Moreover this orbital angular momentum is quenched as a rule, $\langle \hat{\mathbf{L}} \rangle = 0$, in most compounds.
%The exception for unquenche orbital angular momentum, $\hat{\mathbf{L}} \ne 0$,
%The orbital angular momentum $\hat{\mathbf{L}}$ can be partially preserved 
%if the environment has high-symmetry.
%As is well known, in the case of 
The exception is the cubic environment, in which
the $d$ orbitals split into doubly degenerate $e$ and triply degenerate $t_2$ levels. 
When the $t_2$ orbital are partially filled, the electronic state 
is characterized by the nonzero fictitious orbital angular momentum $\tilde{L}=1$, which
%And the electronic state with unquenched orbital angular momentum couples 
couples to the total spin of the site via spin-orbit coupling and gives molecular multiplets characterized by fictitious total angular momentum
$\tilde{\mathbf{J}} = \tilde{\mathbf{L}} + \hat{\mathbf{S}}$
(see Ref. \onlinecite{Abragam1970} for details).
%The other example would be trigonal environment:
%the $d$ orbital does split into two doublets and one nondegenerate orbitals. 
%If the gap between the nondegenerate and one of the doublets is small, 
%the fictious angular momentum is still a good quantum number and the ground state is again 
%described by the spin-orbit state.

\subsection{Intersite interaction}
\label{Sec:interaction}
%\section{Microscopic derivation of an effective interaction Hamiltonian between magnetic centers}
%The model Hamiltonian $\hat{H}$ for the localized system with two sites consists of 
%the intrasite Hamiltonian $\hat{H}^i_0$, 
%the intersite bielectronic interaction $\hat{H}_{\rm bi}'$, 
%and the electron transfer Hamiltonian $\hat{H}_{\rm t}$.
The electronic Hamiltonian $\hat{H}$ for electrons localized at two sites can be divided into 
the intrasite contributions $\hat{H}_0^i$ $(i=1,2)$, the 
intersite bielectronic $\hat{H}_\text{bi}'$ and electron transfer $\hat{H}_\text{t}$ parts:
\begin{eqnarray}
 \hat{H} &=& \sum_{i=1,2} \hat{H}_0^i + \hat{H}'_\text{bi} + \hat{H}_\text{t}.
\label{Eq:H}
\end{eqnarray}
%$\hat{H}^i_0$ includes all the important interactions such as 
%the bielectronic, the spin-orbit and other relativistic effects, crystal-field.
The intrasite Hamiltonian for site $i$, $\hat{H}_0^i$, 
contains all effects discussed in Sec. \ref{Sec:ion}
such as the non-relativistic atomic terms, the spin-orbit term and other relativistic corrections, 
and the crystal-field.
The eigenstate of $\hat{H}_0^i$ is determined by 
the number of electrons $N_i$ in magnetic orbitals and crystal-field level $p$, $|iN_i,p\rangle$.
$\hat{H}_{\rm bi}'$ consists of intersite Coulomb interaction $\hat{H}_{\rm Coul}$ and direct exchange (multipole) part $\hat{H}_{\rm DE}$:
%the orbital energy term $\hat{H}_{\rm orb}^i$,
%the intrasite $\hat{H}_{\rm bi}^i$ and intersite $\hat{H}_{\rm bi}'$ bielectronic interactions, 
%the spin-orbit coupling $\hat{H}_{\rm so}^i$, 
%the crystal field $\hat{H}_{\rm cf}^i$, 
%and the electron transfer Hamiltonian $\hat{H}_{\rm t}$. 
%In the case of the lanthanide system, 
%where the magnetic orbital is localized in comparison with actinide and transition metal compounds,
%the strengths of the bielectronic ($\sim$ 5-7 eV) and the spin-orbit splitting ($\sim 1$ eV) are 
%at least several times larger than the crystal field splitting ($\sim 0.1$ eV) 
%and electron transfer ($\sim$ 0.1 eV) \cite{Abragam1970, vanderMarel1988, Mironov2003}. 
%In comparison with the energy scale of $\hat{H}_0^i$, 
%the electron transfer Hamiltonian $\hat{H}_{\rm t}$ is weak. 
%Thus, $\hat{H}_{\rm t}$ is treated as the perturbation to the unperturbed Hamiltonian, 
%$\hat{H}_0 =\sum_{i=1}^2 \hat{H}_0^i + \hat{H}'_{\rm bi}$.
%the orbital energy term $\hat{H}_{\rm orb}^i$,
%the intrasite $\hat{H}_{\rm bi}^i$ and intersite $\hat{H}_{\rm bi}'$ bielectronic interactions, 
%the spin-orbit coupling $\hat{H}_{\rm so}^i$, 
%the crystal field $\hat{H}_{\rm cf}^i$, 
%and the electron transfer Hamiltonian $\hat{H}_{\rm t}$. 
%The intersite bielectronic interaction $\hat{H}'_{\rm bi}$ 
%has the Coulomb $\hat{H}_{\rm Coul}$ and the direct exchange $\hat{H}_{\rm DE}$ (\ref{Eq:HDE}) terms
%\cite{Anderson1959, Anderson1963}:
\begin{eqnarray}
 \hat{H}'_{\rm bi} &=& \hat{H}_{\rm Coul} + \hat{H}_{\rm DE}, 
\\
 \hat{H}_{\rm Coul} &=&
  \sum_{mn\sigma\sigma'} 
 U'
 \hat{n}_{1m\sigma}
 \hat{n}_{2n\sigma'},
\label{Eq:HCoul}
\\
 \hat{H}_{\rm DE} &=&
 -\sum_{mnm'n'\sigma\sigma'} V_{mm'n'n}
 \hat{c}_{1m\sigma}^\dagger \hat{c}_{1n\sigma'}
 \hat{c}_{2m'\sigma'}^\dagger \hat{c}_{2n'\sigma},
\label{Eq:HDE}
\end{eqnarray}
where $m,n$ indicate the projection of the orbital angular momentum $l_i$,
$\sigma$ is the projection of the electron spin momentum,
$\hat{c}_{im\sigma}^\dagger$ ($\hat{c}_{im\sigma}$)
is the electron creation (annihilation) operator in spin-orbital $(m,\sigma)$ of site $i$ $(=1,2)$,
$\hat{n}_{im\sigma} = \hat{c}^\dagger_{im\sigma} \hat{c}_{im\sigma}$, 
$U'$ is the intersite electron repulsion,
$V_{mm'n'n}$ is the intersite exchange integral,
\begin{eqnarray}
 V_{mm'n'n} &=& \int d\mathbf{r}_1 d\mathbf{r}_2 
                \psi_{1m}^*(\mathbf{r}_1) \psi_{2m'}^*(\mathbf{r}_2) 
\nonumber\\
            &\times&
                v(|\mathbf{r}_1-\mathbf{r}_2|)
                \psi_{2n}(\mathbf{r}_1) \psi_{1n'}(\mathbf{r}_2),
\label{Eq:VDE}
\end{eqnarray}
$\psi_{im}$ is the Wannier function at site $i$ and component $m$, and 
$v(|\mathbf{r}_1-\mathbf{r}_2|)$ is the two body interaction.
Note that $\hat{H}'_{\rm bi}$ does not change the number of the electrons on each sites.
The transfer Hamiltonian is written as 
\begin{eqnarray}
 \hat{H}_{\rm t} &=& \sum_{i\ne j}\sum_{mm'\sigma} t^{ij}_{mm'} \hat{c}_{im\sigma}^\dagger \hat{c}_{jm'\sigma},
\label{Eq:Ht}
\end{eqnarray}
where $t^{ij}_{mm'}$ is the electron transfer parameter between orbitals $m$ of site $i$ and $m'$ of $j$.

As discussed by Anderson \cite{Anderson1959, Anderson1963},
the direct exchange parameter (\ref{Eq:VDE}) and the transfer parameters 
are finite due to the delocalization of the magnetic orbital $\psi_{im}$ on 
the ligand between metal sites (or the other magnetic site).
%The electron transfer parameter is proportional to the scale of the atomic orbital level,
%whereas the direct exchange parameter is to the intrasite exchange interaction,
%which is about one order of magnitude smaller than the intrasite Coulomb interaction. 
The electron transfer parameter %($\approx$ 0.1 eV) between metal sites 
between metal sites is at least several times smaller than the intrasite %Hamiltonian. % \cite{Mironov2003}.
electron repulsion.
Thus, the electronic Hamiltonian $\hat{H}$ (\ref{Eq:H}) can be divided into zeroth order Hamiltonian 
\begin{eqnarray}
\hat{H}_0 =\sum_{i=1}^2 \hat{H}_0^i + \hat{H}_{\rm Coul}
\label{Eq:H0}
\end{eqnarray}
and small terms $\hat{H}_{\rm DE}$ and $\hat{H}_{\rm t}$. 
For localized magnetic electrons, the latter can be treated in the first and the second order of perturbation theory, respectively \cite{Anderson1959, Anderson1963}. 
%Thus, $\hat{H}_{\rm t}$ is treated as the perturbation to the unperturbed Hamiltonian, 
%$\hat{H}_0 + \hat{H}_{\rm DE}=\sum_{i=1}^2 \hat{H}_0^i + \hat{H}'_{\rm bi}$.
This is done here via a unitary transformation,
\begin{eqnarray}
 \hat{H}_{\rm eff} = e^{-\hat{S}}\hat{H}e^{\hat{S}},
\label{Eq:Heff}
\end{eqnarray}
removing $\hat{H}_{\rm t}$ from the initial Hamiltonian.  
%$e^{-\hat{S}}\hat{H}e^{\hat{S}}$, 
Neglecting the terms higher than second order after the transfer parameters, % and the crystal field splitting,
we obtain the effective Hamiltonian acting on the ground $J$-multiplets on sites, 
%$ \hat{H}_{\rm eff} = \hat{H}_0 + \bar{H}$.
\begin{eqnarray}
 \hat{H}_{\rm eff} = \hat{H}_0 + \bar{H}.
\end{eqnarray}
%The effective Hamiltonian $\hat{H}_{\rm eff}$ for the low-energy states 
%is derived performing the unitary transformation of the model Hamiltonian, 

In the unperturbed Hamiltonian (\ref{Eq:H0}), there are no terms which vary the numbers of the electrons.
Then the electronic states can be written as follows:
\begin{eqnarray}
 |N_1, N_2, r\rangle &=& \sum_{p,q}|1N_1,p\rangle |2N_2,q\rangle C^{N_1N_2}_{pq,r},
\label{Eq:r}
\end{eqnarray}
where $r$ indicates the eigenstate of the system and $C^{N_1N_2}_{pq,r}$ is a coefficient.
%Among them, $\hat{H}_0^i=\hat{H}_{\rm orb}^i + \hat{H}_{\rm bi}^i + \hat{H}_{\rm so}^i + \hat{H}_{\rm cf}^i$ 
%is written only by the operators of site $i$.
%On the other hand, since the main part of $\hat{H}'_{\rm bi}$ is the Coulomb repulsion, 
%it is expressed by the product of the number operators on both sites.
%\begin{eqnarray}
% \hat{H}_0^i |iN_i, p\rangle &=& E_p^i|iN_i, p\rangle,
%\end{eqnarray}
%Here, the second term corresponds to the direct exchange part, $\hat{H}_{\rm DE}$,
%of the exchange Hamiltonian.
%and hence, $\hat{H}'_{\rm bi}$ is expressed by the product of the number operators on sites.
%as a product of the eigenstates of one-site Hamiltonians, $\hat{H}_0^i$.
In the derivation of the effective Hamiltonian, we consider the truncated vector space $\mathcal{B}$
of the electron configurations which include up to one electron transfer 
with respect to the numbers of the electrons in the ground electron configurations.
Hereafter, $N_i$ is used as the number of electrons on site $i$ in the ground electron configurations.
%$\mathcal{B} = \{|N_1,N_2,r\rangle, |N_1-1,N_2+1,s\rangle, |N_1+1, N_2-1, s'\rangle \}$.
For simplicity, the numbers of electrons in the ground configurations will not be written explicitly, 
and the configurations $s$ with $N_1 - 1$ and $N_2 + 1$ 
($N_1 + 1$ and $N_2 - 1$) electrons on sites 1 and 2 
are expressed by the type of the virtual electron transfer, $1\rightarrow 2$ ($2\rightarrow 1$).
Therefore, $\mathcal{B}$ is defined as follows:
\begin{eqnarray}
 \mathcal{B} &=& \{|r\rangle, |1\rightarrow 2,s\rangle, |2\rightarrow 1, s'\rangle \}.
\label{Eq:B}
\end{eqnarray}
%where $|r\rangle = |N_1,N_2,r\rangle$,
%$i\rightarrow j$ $(i,j=1,2)$ indicates the electron transfer with respect to the ground configuration,
%and the numbers of electrons in the ground configuration are omit.
The eigenenergies of the states $|r\rangle$, $|i\rightarrow j,s\rangle$ %in $\mathcal{B}$ belong to 
are denoted as $E_r^0$ and $E_s^{i\rightarrow j}$, respectively.
%Within the space, the electron transfer Hamiltonian is 
%\begin{eqnarray}
% \hat{H}_{\rm t} &=& \sum_{i\ne j} 
% \left(\hat{P}^{i\rightarrow j} \hat{H}_{\rm t} \hat{P}^0
% + \hat{P}^0 \hat{H}_{\rm t} \hat{P}^{i\rightarrow j}\right).
%\end{eqnarray}

The exponent of the unitary operator $e^{\hat{S}}$ is given as
\begin{eqnarray}
 \hat{S} &=& \sum_{i\ne j}\sum_{r} \sum_{s} 
 \left(
 \frac{\hat{P}_{s}^{i\rightarrow j} \hat{H}_{\rm t} \hat{P}_{r}^0}
 {E_{r}^{0} - E_{s}^{i\rightarrow j}}
 - 
 \frac{\hat{P}_{r}^0 \hat{H}_{\rm t} \hat{P}_{s}^{i\rightarrow j}}
 {E_{r}^{0} - E_{s}^{i\rightarrow j}}
 \right),
\label{Eq:S}
\end{eqnarray}
where $\hat{P}_{r}^0$ and $\hat{P}_{s}^{i\rightarrow j}$ are the projection operators,
\begin{eqnarray}
 \hat{P}_r^0 &=& |r\rangle \langle r|, 
\\
 \hat{P}_s^{i\rightarrow j} &=& |i\rightarrow j, s\rangle \langle i\rightarrow j, s|.
\end{eqnarray}
The exponent $\hat{S}$ is chosen to fulfill the condition
\begin{eqnarray}
 [\hat{S}, \hat{H}_0 + \hat{H}_{\rm DE}] &=& \hat{H}_{\rm t}
\label{Eq:commutatorSH0}
\end{eqnarray}
within the space $\mathcal{B}$.
The effective Hamiltonian (\ref{Eq:Heff}) within $\mathcal{B}_0=\{|r\rangle\}$ is obtained as 
\begin{eqnarray}
 \hat{H}_{\rm eff}
 &=& 
 \hat{H}_0 + \hat{H}_{\rm DE} - \frac{1}{2}[\hat{S}, \hat{H}_{\rm t}],
\label{Eq:SHS}
\end{eqnarray}
%Here, the terms up to second order after $\hat{H}_{\rm t}$ is left and the higher terms are neglected.
%where, the terms up to second order after $\hat{H}_{\rm t}$ is left and the higher terms are neglected.
up to second order after $\hat{H}_{\rm t}$. % is left and the higher terms are neglected.
The second and the third terms in Eq. (\ref{Eq:SHS}) correspond to $\bar{H}$ defined above, 
\begin{eqnarray}
 \bar{H} &=& 
 \hat{H}_{\rm DE} + \hat{H}_{\rm KE}, 
\label{Eq:Hbar}
\\
 \hat{H}_{\rm KE} &=& 
 \frac{1}{2}
 \sum_{i\ne j}\sum_{r} \sum_{s} 
% \left(
 \frac{\hat{H}_{\rm t} \hat{P}_{s}^{i\rightarrow j} \hat{H}_{\rm t} \hat{P}_{r}^0}
 {E_{r}^{0} - E_{s}^{i\rightarrow j}}
 + 
 \text{H.c.}
% \frac{\hat{P}_{r}^0 \hat{H}_\text{t} \hat{P}_{s}^{i\rightarrow j}\hat{H}_\text{t}}
% {E_{r}^{0} - E_{s}^{i\rightarrow j}}
% \right).
%\nonumber\\
\end{eqnarray}
Note that the terms 
$\hat{P}_{s}^{i\rightarrow j}\hat{H}_\text{t}\hat{P}_r^0 \hat{H}_\text{t}$
and 
$\hat{H}_\text{t}\hat{P}_r^0 \hat{H}_\text{t}\hat{P}_{s}^{i\rightarrow j}$
do not enter here because they map the states outside the domain $\mathcal{B}_0$.

Neglecting the effects of the crystal-field splitting %and direct exchange 
in the denominator of $\hat{H}_\text{KE}$, 
which is a reasonable approximation for our systems,
the eigenstates $r,s$ reduce to the sets of the $J$-multiplet states:
\begin{eqnarray}
 r &\rightarrow& (\alpha_{1} J_1M_1,\alpha_{2} J_2M_2), 
\\
 s &\rightarrow& (\alpha_J JM, \alpha_J' J'M'),
\end{eqnarray}
where, $J_i$ is the total angular momentum with the ground electron configuration, 
and $J, J'$ %and $\alpha_J, \alpha_J'$ 
are the total angular momenta %and the other quantum numbers 
for intermediate states arising from the transfer of one electron between the sites.
The kinetic exchange Hamiltonian becomes 
%\begin{widetext}
\begin{eqnarray}
 \hat{H}_\text{KE} &=& 
 \frac{1}{2}
 \sum_{i\ne j}\sum_{\alpha_{i} J_i, \alpha_{j} J_j} \sum_{\alpha_J J, \alpha_J' J'} 
%\nonumber\\
% &\times&
% \left(
 \frac{\hat{H}_\text{t} \hat{P}_{\alpha_J J,\alpha_J' J'}^{i\rightarrow j} 
      \hat{H}_{\rm t} \hat{P}_{\alpha_{i} J_i,\alpha_{j} J_j}^0}
 {E_{\alpha_{i} J_i,\alpha_{j} J_j}^{0} - E_{\alpha_J J,\alpha_J' J'}^{i\rightarrow j}}
\nonumber\\
 &+& \text{H.c.},
% \frac{\hat{P}_{J_i,J_j}^0 \hat{H}_{\rm t} \hat{P}_{\mu J,\nu J'}^{i\rightarrow j}\hat{H}_{\rm t}}
% {E_{J_i,J_j}^{0} - E_{\mu J, \nu J'}^{i\rightarrow j}}
% \right).
\end{eqnarray}
where the projection operators are 
\begin{eqnarray}
 \hat{P}_{\alpha_i J_i,\alpha_j J_j}^0 &=& 
 \hat{P}^{N_i}_{i\alpha_i J_i} \hat{P}^{N_j}_{j\alpha_j J_j},
% \sum_{M_1,M_2} |J_1M_1\rangle \langle J_1M_1|
% \sum_{M_2} |J_2M_2\rangle \langle J_2M_2|,
%\\
\\
 \hat{P}_{\alpha_J J,\alpha_J' J'}^{i\rightarrow j}  &=&
 \hat{P}^{N_i-1}_{i\alpha_J J} \hat{P}^{N_j+1}_{j\alpha_J' J'},
\end{eqnarray}
and $\hat{P}^N_{i\alpha_J J}$ is the projection operator on site $i$, 
\begin{eqnarray}
\hat{P}^N_{i\alpha_J J} = \sum_{M=-J}^J |iN\alpha_J JM\rangle \langle iN\alpha_J JM|.
\label{Eq:P0}
\end{eqnarray}
%$\alpha_J$ of the ground $J$-multiplet is omitted for the sake of simplicity.
%\end{widetext}
%In the derivation, the effects of the crystal-field splitting
%and the $j$-$j$ mixing in the intermediate states $\alpha_J J$ are neglected
%because they are weaker than the multiplet splitting. 
In the space of the ground $J$-multiplets on sites, 
$\mathcal{B}_J = \{|J_1M_1, J_2M_2\rangle: -J_i\le M_i\le J_i\}$,
the kinetic exchange Hamiltonian reduces to 
\begin{eqnarray}
% \bar{H} &=& 
 \hat{H}_{\rm KE} &=& 
 \sum_{i\ne j}\sum_{\alpha_J J, \alpha_J' J'} 
 \frac{\hat{H}_{\rm t} \hat{P}_{\alpha_J J,\alpha_J' J'}^{i\rightarrow j} \hat{H}_{\rm t}} %\hat{P}_{J_i,J_j}^0}
 {E_{J_i,J_j}^{0} - E_{\alpha_J J,\alpha_J' J'}^{i\rightarrow j}}.
\label{Eq:barH}
\end{eqnarray}
Here, $\alpha_i$ of the ground $J$-multiplet is not written for the sake of simplicity and 
$\hat{P}_{\alpha_iJ_i,\alpha_jJ_j}^0$ is omitted because it is the unit operator within $\mathcal{B}_J$.

Substituting Eq. (\ref{Eq:Ht}) into Eq. (\ref{Eq:barH}), we obtain %Eq. (2) in the main text.
%\begin{widetext}
\begin{eqnarray}
% \bar{H} &=& 
 \hat{H}_{\rm KE} &=& 
 \sum_{i \ne j}
 \sum_{\alpha_J J} \sum_{\alpha_J' J'} 
 \sum_{mn\sigma}\sum_{m'n'\sigma'} 
% \frac{
 -t_{mm'}^{ij} t_{n'n}^{ji}
\nonumber\\
 &\times&
 \frac{%-t_{mm'}^{ij} t_{n'n}^{ji}
 \left(\hat{c}_{im\sigma}^\dagger \hat{P}^{N_i-1}_{i\alpha_J J} \hat{c}_{in\sigma'}\right)
 \left(\hat{c}_{jm'\sigma} \hat{P}^{N_j+1}_{j\alpha_J' J'} \hat{c}_{jn'\sigma'}^\dagger \right)
 }{U_{ij} + \Delta E_{i\alpha_J J}^{N_i-1} + \Delta E_{j\alpha_J' J'}^{N_j+1}},
%\nonumber\\
% &\times&
% \left(\hat{c}_{im\sigma}^\dagger \hat{P}^{N_i-1}_{i\mu J} \hat{c}_{in\sigma'}\right)
% \left(\hat{c}_{jm'\sigma} \hat{P}^{N_j+1}_{j\nu J'} \hat{c}_{jn'\sigma'}^\dagger \right),
\label{Eq:HKE}
\end{eqnarray}
%\end{widetext}
where 
%$t_{mn}^{ij}$ is the transfer parameter, 
%$N_i$ is the number of electrons occupying the magnetic orbitals of site $i$, 
$U_{ij}$ is the smallest promotion energy for the electron transfer from site $i$ to site $j$, and
$\Delta E_{i\alpha_J J}^{N}$ is the excitation energy from the ground intermediate state with $N$ electrons.
In the derivation of Eq. (\ref{Eq:HKE}), we have not used the approximate form (\ref{Eq:JM}) for the
$J$-multiplets in both the ground and the virtual states. 
The quantum number $J$ for the virtual states fulfills the condition $|J_i-l_i-1/2| \le J \le J_i+l_i+1/2$.
Although the crystal-field splitting is neglected, 
the multiplet structures of sites are completely retained in Eq. (\ref{Eq:HKE}).
Hence, important effects such as the Goodenough's mechanism \cite{Goodenough1963} 
are included in the kinetic exchange Hamiltonian.

\section{Exchange Hamiltonian in $J$-representation}
\label{Sec:exchangeJJ}
The Hamiltonian $\bar{H}$ (\ref{Eq:Hbar}) is transformed into tensor form 
with the use of irreducible (double) tensor technique \cite{Judd1967}, % and of
method of equivalent operator \cite{Abragam1970}, and the form of $J$-multiplet state (\ref{Eq:JM}):
\begin{equation}
 \bar{H} = 
 \sum_{kqk'q'} 
  \mathcal{J}_{kqk'q'} 
 \frac{O_k^{q}(\hat{\mathbf{J}}_1) O_{k'}^{q'}(\hat{\mathbf{J}}_2)}{O_k^0(J_1)O_{k'}^0(J_2)}.
\label{Eq:Hbar_tensor}
\end{equation}
Here, $O_k^q(\hat{\mathbf{J}}_1)$ and $O_{k'}^{q'}(\hat{\mathbf{J}}_2)$ are 
Stevens operators \cite{Stevens1952} (see also Appendix \ref{Sec:equiv})
whose ranks $k$ and $k'$ have to obey the relation $k+k'=$ even due to the invariance of 
the Hamiltonian with respect to time inversion \cite{Abragam1970} 
(see Appendix \ref{Appendix:J}),
$q$ and $q'$ are component, 
and $O_k^0({J}_1)$ and $O_{k'}^{0}({J}_2)$ are the scalars obtained by replacing 
$\hat{\mathbf{J}}_i^2$ and $\hat{J}_{iz}$ in
$O_k^0(\hat{\mathbf{J}}_i)$ %and $O_{k'}^0(\hat{\mathbf{J}}_2)$
with eigenvalues $J_i(J_i+1)$ and $J_i$, respectively ($i=1,2$).
The exchange coupling constant $\mathcal{J}_{kqk'q'}$ is a sum of the 
direct exchange $\mathcal{J}^{\rm DE}_{kqk'q'}$ and the kinetic $\mathcal{J}^{\rm KE}_{kqk'q'}$ contributions 
\footnote{For interacting lanthanide ions, 
one should add also the contribution from magnetic dipolar interaction, 
which is of known first-rank form:
$-g_{\rm L}^2 \mu_{\rm B}^2/R_{12}^3 
[\hat{\mathbf{J}}_1 \cdot \hat{\mathbf{J}}_2 
-3 (\hat{\mathbf{J}}_1 \cdot \mathbf{n}_{12}) (\hat{\mathbf{J}}_2 \cdot \mathbf{n}_{12})]$, 
where $g_{\rm L}$ is the Land\'{e} g factor, $\mu_\text{B}$ is Bohr magneton, 
$R_{12}$ is the distance between sites, and 
$\mathbf{n}_{12}$ is the unit direction vector from site 1 to site 2}:
%\cite{comment_1}.
%The direct exchange coupling constant $\mathcal{J}^{\rm DE}_{kqk'q'}$ is written as 
\begin{eqnarray}
 \mathcal{J}_{kqk'q'} &=& \mathcal{J}^{\rm DE}_{kqk'q'} + \mathcal{J}^{\rm KE}_{kqk'q'}.
\label{Eq:J}
\end{eqnarray}

The advantages of using the exchange Hamiltonian in the tensorial form is that 
with Eq. (\ref{Eq:Hbar_tensor}) it is easier 
(i) to obtain physical insight on the exchange interaction and 
(ii) to combine it with other terms 
such as crystal-field and Zeeman interaction included in $\hat{H}_0$.
The latter can be treated at {\it ab initio} level \cite{Chibotaru2015, Ungur2015}.

\subsection{Direct exchange interaction}
\label{Sec:DE}
The outline of the derivation for the direct exchange part in $\bar{H}$ is given here,
whereas details of the calculations are given in Appendix \ref{Appendix:DE}.
%More precisely, the reduction of the reducible operators,
%the method of equivalent operator (see for example Ref. \onlinecite{Abragam1970}).
%and the form of $J$-multiplet state. % (Eq. (4) in the main text). 
%This is done by reducing the tensor operators, and the method of equivalent operator
%with the use of the form of $J$-multiplet state (\ref{Eq:JM}).
The direct product of the double tensors $\hat{c}_{im\sigma}^\dagger \hat{c}_{im\sigma'}$
appearing in $\hat{H}_{\rm DE}$ (\ref{Eq:HDE}) is reduced as follows:
\begin{eqnarray}
 \hat{c}_{im\sigma}^\dagger \hat{c}_{in\sigma'}
% &=& 
% \hat{c}_{im\sigma}^\dagger (-1)^{l_i+n+\frac{1}{2}+\sigma'}\bar{c}_{i-n-\sigma'}
%\nonumber\\ 
 &=&
 (-1)^{l_i+n+\frac{1}{2}+\sigma'}
 \sum_{a\alpha b \beta} 
 \left\{\hat{c}_i^\dagger \otimes \bar{c}_i\right\}^{a\alpha}_{b\beta}
\nonumber\\
 &\times&
 C_{l_iml_i-n}^{a\alpha} C_{\frac{1}{2}\sigma \frac{1}{2}-\sigma'}^{b\beta},
\label{Eq:cc}
\end{eqnarray}
where $\bar{c}_{in\sigma'} = (-1)^{l_i+n+\frac{1}{2}+\sigma'} \hat{c}_{i-n-\sigma'}$ 
(\ref{Eq:cbar}) is a double tensor (see Appendix \ref{Appendix:ITO}), the curly bracket 
$\{\hat{c}_i^\dagger \otimes \bar{c}_i\}^{a\alpha}_{b\beta}$ 
indicates the irreducible operator of ranks $(a,b)$ and components
$(\alpha, \beta)$ constructed from the product of two tensors,
where the superscripts and subscripts are the orbital and spin parts, respectively.
%$\{\hat{c}_i^\dagger \otimes \bar{c}_i\}^{a\alpha}_{b\beta}$ 
The irreducible tensor operator is replaced by the total angular momentum operator $\hat{\mathbf{J}}_i$
using the method of equivalent operator for double tensor (\ref{Eq:T}):
%and formula (\ref{Eq:sum4CG}) 
\begin{eqnarray}
 \left\{\hat{c}_i^\dagger \otimes \bar{c}_i\right\}^{a\alpha}_{b\beta}
 &=& 
 \sum_{kq} C_{a\alpha b\beta}^{kq} \mathcal{D}^i_{abk} \frac{O_k^q(\hat{\mathbf{J}}_i)}{O_k^0(J_i)}.
\label{Eq:cxc}
\end{eqnarray}
Here, $\mathcal{D}_{abk}^i$ is a tensor with three indices $a,b,k$ (\ref{Eq:D}). 
Note that when the method of equivalent operator (\ref{Eq:T}) is used, 
the form of $|JM\rangle$ (\ref{Eq:JM}) is assumed.
%and the ranges of the ranks $a,b,k$ %in Eqs. (\ref{Eq:cc}), (\ref{Eq:cxc}) satisfy 
%$0 \le a \le 2\min[l_i, L_i], 0 \le b \le 2\min[1/2, S_i]$ and for a given set of $a,b$, 
%$k$ holds $|a-b| \le k \le \min[a+b, 2J_i]$.
%In the ground $LS$ term of the magnetic ion, 
%$L_i \ge l_i$ and $S_i \ge 1/2$ except for the case of the half-filling 
%($L_i=0$ and $J_i=S_i$).
%Thus, the ranks $a,b,k$ hold
%are given as 
%\begin{eqnarray}
% 0 \le a \le 2\min[l_i, L_i], \quad
% 0 \le b \le 1, 
%\nonumber\\
% 0 \le k \le \min[2l_i+1, 2L_i+1, 2J_i].
%\label{Eq:abk_DE}
%\end{eqnarray}

Substituting Eqs. (\ref{Eq:cc}), (\ref{Eq:cxc}) into $\hat{H}_\text{DE}$ (\ref{Eq:HDE}), 
we obtain the tensor form of the direct exchange Hamiltonian.
The exchange parameter $\mathcal{J}^\text{DE}$ is obtained as
\begin{eqnarray}
\mathcal{J}^{\rm DE}_{kqk'q'} 
 &=&
% \sum_{a\alpha a'\alpha' b\beta} 
% -V_{a\alpha a'\alpha'} 
 -\sum_{aa'b} \mathcal{V}^{aa'b}_{kqk'q'}
 \mathcal{D}_{abk}^1 \mathcal{D}_{a'bk'}^2,
 \label{Eq:JDE}
\end{eqnarray}
where
\begin{eqnarray}
 \mathcal{V}^{aa'b}_{kqk'q'} &=& 
 \sum_{mn} \sum_{m'n'} \sum_{\alpha \alpha' \beta}
 (-1)^{l_1+l_2+n+n'-\beta} V_{mm'n'n} 
\nonumber\\
 &\times&
 C_{l_1ml_1-n}^{a\alpha} 
 C_{a\alpha b\beta}^{kq} 
 C_{l_2m'l_2-n'}^{a'\alpha'}
 C_{a'\alpha' b-\beta}^{k'q'}.
 \label{Eq:V}
\end{eqnarray}
%
%Here, $L_i, S_i, J_i$ are the orbital, spin, and total angular momenta for the ground state of site $i$.
%and $\{\hat{A} \otimes \hat{B}\}^{a\alpha}_{b\beta}$ 
%is the irreducible double tensor of rank $(a,b)$ \cite{Judd1967} constructed from the direct product 
%of double tensors $\hat{A}$ and $\hat{B}$,
%where the super- and subscripts are orbital and spin indices, respectively (the range of variation of all indices is specified in Appendix).

\subsection{Kinetic exchange interaction}
\label{Sec:KE}
The derivation of the kinetic exchange parameter $\mathcal{J}^\text{KE}$
is similar to that of $\mathcal{J}^\text{DE}$.
As in the previous case, only the outline of the derivation is given here and the
details can be found in Appendix \ref{Appendix:KE}.
In comparison with the direct exchange, the derivation of $\mathcal{J}^\text{KE}$ is cumbersome
%The difference comes from 
because of the projection operator $\hat{P}_{i\alpha_J J}^N$ appearing in Eq. (\ref{Eq:HKE}).
The latter is reducible within the product group $SO(3) \otimes SU(2)$ 
where the creation $\hat{c}^\dagger$ and annihilation $\bar{c}$ 
(\ref{Eq:cbar}) operators are irreducible.
Therefore, we first transform $\hat{P}_{i\alpha_J J}^N$ into the sum of the 
irreducible double tensors $\hat{P}_{i\alpha_J a\alpha a'\alpha'}^N$ (\ref{Eq:P1}), and then the product 
such as $\hat{c}_{im\sigma}^\dagger \hat{P}^{N_i-1}_{i\alpha_J a\alpha a'\alpha'} \hat{c}_{in\sigma'}$ 
is reduced.
The irreducible tensor operator 
$\{\hat{c}_{i}^\dagger \otimes \{\hat{P}^{N_i-1}_{i\alpha_J aa} \otimes \bar{c}_{i}\}^b_d\}^{c\gamma}_{e\epsilon}$
is replaced by the total angular momentum $\hat{\mathbf{J}}_i$ 
using the method of the equivalent operator (\ref{Eq:T}),
and finally we obtain the kinetic exchange parameter:
%The superexchange coupling constant $\mathcal{J}^{\rm SE}_{kqk'q'}$ is expressed as 
\begin{eqnarray}
\mathcal{J}^{\rm KE}_{kqk'q'} 
 &=&
 \sum_{fxx'} \sum_{\alpha_J J} \sum_{\alpha_J' J'} 
 \frac{\{t \times t\}^{fxx'}_{kq k'q'}
 \mathcal{F}^{1}_{\alpha_J Jfxk}\mathcal{G}^{2}_{\alpha_J' J'fx'k'}}
  {U_{12} + \Delta E_{1\alpha_J J}^{N_1-1} + \Delta E_{2\alpha_J' J'}^{N_2+1}}
\nonumber\\
 &+&
 \sum_{fxx'} \sum_{\alpha_J J} \sum_{\alpha_J' J'} 
 \frac{\{t \times t\}^{fxx'}_{kq k'q'}
 \mathcal{G}^{1}_{\alpha_J Jfxk}\mathcal{F}^{2}_{\alpha_J' J'fx'k'}}
       {U_{21} + \Delta E_{1\alpha_J J}^{N_1+1} + \Delta E_{2\alpha_J' J'}^{N_2-1}},
\nonumber\\
 \label{Eq:JKE}
\end{eqnarray}
where %$\{t \times t\}$ part is 
\begin{eqnarray}
 \{t \times t\}^{fxx'}_{kq k'q'} &=&
 (-1)^{l_1-l_2-f+q'} 
 \sum_{mn}\sum_{m'n'} 
 \sum_{\xi \xi' \phi} 
t_{mm'}^{12} t_{n'n}^{21}
\nonumber\\
 &\times&
 C_{l_1n kq}^{x\xi} C_{f\phi l_1m}^{x\xi}
 C_{l_2-n' k'q'}^{x'\xi'} C_{f-\phi l_2-m'}^{x'\xi'}.
 \label{Eq:tt}
\end{eqnarray}
The range of variation of indices of the tensors described above, 
as well as in the subscripts of $\mathcal{F}$ and $\mathcal{G}$ is specified in Appendix \ref{Appendix:KE}.
%In the equations above, 
%$L_i, S_i, J_i$ are the ground orbital, spin, and total angular momenta of site $i$,
%$L, S, J$ are those for the intermediate states, 
%$\{\hat{A} \otimes \hat{B}\}^{a\alpha}_{b\beta}$ 
%is the irreducible double tensor of rank $(a,b)$ constructed from the direct products 
%of double tensors $\hat{A}$ and $\hat{B}$,
%where the super- and subscripts are the orbital and spin parts, respectively.
%In the superexchange mechanism, the intermediate multiplets $\mu J$ 
%which contribute to $\mathcal{J}^{\rm SE}$ are restricted to those with 
%$|L_i-l_i| \le L \le L_i+l_i$, $|S_i-1/2| \le S \le S_i+1/2$, 
%and $|J_i-l_i-1/2| \le J \le J_i+l_i+1/2$.
As in the case of the derivation of $\mathcal{J}^\text{DE}$, Eq. (\ref{Eq:JM}) is used. 
%In Eq. (\ref{Eq:JKE}), $L, S, J$ refer to %without subscript $i$ denote 
%the intermediate states.
%Under the electron transfer, the multiplets which contribute to $\mathcal{J}^{\rm KE}$ 
Because of this assumption, 
the quantum numbers of the %multiplets 
intermediate multiplets contributing to $\mathcal{J}^{\rm KE}$ 
obey the relations: $|L_i-l_i| \le L \le L_i+l_i$, $|S_i-1/2| \le S \le S_i+1/2$, 
and $|J_i-l_i-1/2| \le J \le J_i+l_i+1/2$,
where $L,S,J$ without subscript $i$ refer to the intermediate states
%\cite{comment_2}.
\footnote{The correctness of Eq. (\ref{Eq:JKE}) was checked numerically 
by comparing the resulting exchange spectrum with the one predicted by Eq. (\ref{Eq:HKE})}.
%\footnote{
%The correctness of Eqs. (\ref{Eq:JKE} - \ref{Eq:F}) was checked numerically by comparing the resulting exchange spectrum
%with the one predicted by Eq. (\ref{Eq:HKE}).
%}.

\subsection{Structure of the exchange Hamiltonian}
\label{Sec:structure}
The domains of variation of $k$ and $q$ characterize the structure of the Hamiltonian (\ref{Eq:Hbar_tensor}).
The upper bound for the rank $k$ and $k'$ in Eq. (\ref{Eq:Hbar_tensor}) is only determined 
by the electronic state of site 1 and 2, respectively
\footnote{
The rank of the direct exchange part is 
also influenced by the orbital angular momentum $L_i$ (\ref{Eq:k_DE}).
}:
\begin{eqnarray}
k_{\text{max}} = \min[2l_i +1, 2J_i].
\label{Eq:k}
\end{eqnarray}
%
%This constraints are obtained from the $6j$ and $9j$ symbols in Eqs. (\ref{Eq:D}), (\ref{Eq:F}).
%For example, in the case of the $f$-electron systems, $f^N$, the maximum of the rank $k_{\rm max}$ is 7 
Thus the maximum rank for $f$-electron system, $f^N$, 
is 7 for $N=$ 2-4 and 7-13, $k_\text{max}=$ 5 for $N=$ 1,5 and $k_\text{max}=$ 0 for $N=6$
\footnote{
After we submitted this work, new manuscript appeared on preprint server \cite{Rau2015},
where the authors reached similar conclusions %on the rank for $f$ elements
}.

On the other hand, the range of $q$ %for each allowed rank $k$ 
is determined by the nonzero parameters describing the intersite interactions, $V_{mm'n'n}$ and $t_{mm'}^{12} t_{n'n}^{21}$.
%From the conservation of the components of the Clebsch-Gordan coefficients in 
%From the structure of 
%Eqs. (\ref{Eq:V}), (\ref{Eq:tt}), $q = m-n+\eta$, 
%where $\eta=\beta, \phi$ for the direct and the kinetic exchanges, respectively.
If $\Delta_\text{max}$ is the maximal difference of the indices corresponding to one site in the above parameters ($m-n$ for site 1 and $m'-n'$ for site 2) then the upper bound for $q$ ($q'$) is
\begin{eqnarray}
q_{\text{max}} = \min[\Delta_\text{max} + 1, k_\text{max}].
\label{Eq:q}
\end{eqnarray}
%
%Here, we used $|\eta| \le 1$ and that the parameters with $m = -m_{\rm max}$ 
Note that terms with $-q_{\text{max}}$ will also be present in the Hamiltonian (\ref{Eq:Hbar_tensor}) due to the time reversal symmetry, implying the following range: $ -q_{\text{max}} \le q \le q_{\text{max}}$. 
%On the same reason we have $|V_{mm'n'n}| = |V_{-m-m'-n'-n}|$ and $|t_{mm'}^{12}| = |t_{-m-m'}^{12}|$.
%For the kinetic exchange, the maximal $q$ corresponds to the expression (\ref{Eq:q}),
%whereas, for the direct exchange, the maximal $|q|$ may be smaller than the estimate.
%In general, the set of $(m,n)$ which gives maximal $|m-n|$ with nonzero $V_{mm'n'n}$ 
%does not necessarily correspond to $\pm (m_{\rm max},-m_{\rm max})$.

The effective Hamiltonian (\ref{Eq:Hbar_tensor}) is further divided into the exchange $\hat{H}_{\rm ex}$ 
and the zero-field splitting parts. %$\hat{H}_{\rm zfs}$ parts.
The latter is defined as comprising terms with either $k=0$ or $k'=0$.

\subsection{Decomposition of $\bar{H}$}
%Now we know the form of the general exchange Hamiltonian (\ref{Eq:Hbar_tensor}) 
%and the domain of rank $k$ (\ref{Eq:k}), 
%it is possible 
Knowledge of the domain of rank $k$ (\ref{Eq:k}) in the general exchange Hamiltonian (\ref{Eq:Hbar_tensor}) 
allows us to calculate $\mathcal{J}_{kqk'q'}$ by using the orthogonality of the Stevens operators:
\begin{eqnarray}
 \mathcal{J}_{kqk'q'} &=& 
 (-1)^{q+q'} 
% \frac{\Pi_{kkk'k'}}{\Pi_{J_iJ_iJ_jJ_j}}
%\nonumber\\
% &\times&
 \left(
 \frac{\Pi_{kk'}}{\Pi_{J_1J_2}}
  C_{J_1J_1k0}^{J_1J_1} C_{J_2J_2k'0}^{J_2J_2}\right)^2
\nonumber\\
&\times&
 \text{Tr}\left[\hat{Q}_{k-qk'-q'} \bar{H}\right],
\label{Eq:Jprojection}
\end{eqnarray}
where $\Pi_j=\sqrt{2j+1}$ and $\Pi_{jj'}=\Pi_j\Pi_{j'}$,
the trace (Tr) is taken over the ground $J$-multiplets, and 
%$\hat{Q}^{kq}_{k'q'} = O_k^q(\hat{\mathbf{J}}_i) O_{k'}^{q'}(\hat{\mathbf{J}}_j)/
%[O_k^0(J_i)O_{k'}^0(J_j)]$.
\begin{eqnarray}
\hat{Q}_{kqk'q'} = \frac{O_k^q(\hat{\mathbf{J}}_1) O_{k'}^{q'}(\hat{\mathbf{J}}_2)}
{O_k^0(J_1)O_{k'}^0(J_2)}.
\end{eqnarray}
The form (\ref{Eq:Jprojection}) for the exchange parameters offer some advantages for practical calculations. 
$\bar{H}$ enters Eq. (\ref{Eq:Jprojection}) in the form of the numerical matrices 
in the basis of the products of multiplet wave functions on sites, $|J_1M_1,J_2M_2\rangle$.
%such as Eqs. (\ref{Eq:HDE}), (\ref{Eq:HKE}) will be used.
The %kinetic 
exchange parameters obtained by Eqs. (\ref{Eq:JDE}), (\ref{Eq:JKE}) and those calculated 
by the projection (\ref{Eq:Jprojection}) have been compared with each other for some test examples.
%Eq. (\ref{Eq:Jprojection}) is useful for practical purpose.

\section{Exchange interaction between $J$-multiplet and isotropic spin}
\label{Sec:exchangeJS}
When the orbital angular momentum is zero in the ground state term of one of the sites,
the low-energy states of this site are characterized by the corresponding spin $\hat{\mathbf{S}}$.
This situation is encountered in mixed lanthanide-transition metal 
and lanthanide-radical complexes \cite{Gatteschi2006, Chibotaru2015, Demir2015}.
The exchange Hamiltonian between a $J$-multiplet and an isotropic spin is obtained 
in a similar way as Eq. (\ref{Eq:Hbar_tensor}):
\begin{eqnarray}
 \bar{H} &=& 
 \sum_{kq} \mathcal{J}_{kq00} 
 \frac{O_k^q(\hat{\mathbf{J}}_1) \hat{I}_2}{O_k^0(J_1)}
+
 \sum_{kq q'} \mathcal{J}_{kq1q'} 
 \frac{O_k^q(\hat{\mathbf{J}}_1) \hat{S}_{2q'}}{O_k^0(J_1)S_2}.
\nonumber\\
\label{Eq:HJS}
\end{eqnarray}
The expressions for exchange coupling constants are similar to 
Eqs. (\ref{Eq:JDE}), (\ref{Eq:JKE}) and are listed in Appendix \ref{Appendix:JS}.
Because of the lack of orbital degrees of freedom on site $2$ ($l_2=0$), 
the rank $k'$ of the spin operator does not exceed 1.
Due to the time-reversal symmetry, 
$k$ is even and odd for the first and the second terms in Eq. (\ref{Eq:HJS}), respectively.
As in the previous case, the former ($k'=0$) is zero-field splitting term 
and the latter ($k' = 1$) is the exchange interaction.
%Since the first term is not an interaction between sites, this is regarded as the zero-field splitting,
%and the second term as the exchange part.

\begin{table*}[tb]
\caption{Excitation energies of Dy$^{2+}$ and Dy$^{4+}$ (meV). 
The number in the parenthesis indicates quantum number $\alpha_{LS}$ in the main text, 
where the enumeration follows Ref. \onlinecite{Nielson1963}.}
\label{Table:DeltaE}
\begin{ruledtabular}
\begin{tabular}{ccccccccc}
 \multicolumn{3}{c}{Dy$^{2+}$} & \multicolumn{3}{c}{Dy$^{4+}$} & \multicolumn{3}{c}{Dy$^{4+}$} \\
 $LS$-term & $J$ & $\Delta E_{\alpha_J J}$ & $LS$-term & $J$ & $\Delta E_{\alpha_J J}$ & $LS$-term & $J$ & $\Delta E_{\alpha_J J}$ \\
\hline
${}^5I$ & 8 &     0.000 & ${}^7F$   & 6 &     0.000 & ${}^5H$(2) & 4 &  8738.547 \\
        & 7 &   458.212 &           & 5 &   309.504 & ${}^5I$(1) & 8 & 10579.798 \\
        & 6 &   859.148 &           & 4 &   508.583 &            & 7 & 10778.918 \\
        & 5 &  1202.808 &${}^5D(1)$ & 4 & 13934.562 &            & 6 & 10917.180 \\
        & 4 &  1489.190 &${}^5D(2)$ & 4 &  7624.320 &            & 5 & 11003.924 \\
${}^5D$ & 4 &  5667.150 &${}^5D(3)$ & 4 &  3298.723 &            & 4 & 11049.978 \\
${}^5F$ & 5 &  2369.357 &${}^5F(1)$ & 5 &  5838.268 & ${}^5I$(2) & 8 &  5827.337 \\
        & 4 &  2655.740 &           & 4 &  5936.668 &            & 7 &  6035.414 \\
${}^5G$ & 6 &  3412.346 &${}^5F(2)$ & 5 & 11206.930 &            & 6 &  6173.415 \\
        & 5 &  3756.005 &           & 4 & 11118.121 &            & 5 &  6242.469 \\
        & 4 &  4042.388 &${}^5G(1)$ & 6 &  6925.862 &            & 4 &  6261.863 \\
        &   &           &           & 5 &  7221.721 & ${}^5K$    & 9 &  6516.513 \\
        &   &           &           & 4 &  7423.960 &            & 8 &  6766.583 \\
        &   &           &${}^5G(2)$ & 6 &  4540.837 &            & 7 &  6935.080 \\
        &   &           &           & 5 &  4539.168 &            & 6 &  7051.111 \\
        &   &           &           & 4 &  4613.355 &            & 5 &  7130.959 \\
        &   &           &${}^5G(3)$ & 6 & 13458.495 & ${}^5L$    &10 &  4277.003 \\
        &   &           &           & 5 & 13323.410 &            & 9 &  4402.240 \\
        &   &           &           & 4 & 13191.767 &            & 8 &  4537.021 \\ 
        &   &           &${}^5H(2)$ & 7 &  8096.483 &            & 7 &  4674.613 \\ 
        &   &           &           & 6 &  8392.826 &            & 6 &  4809.212 \\
        &   &           &           & 5 &  8603.858 &            &   &  \\
\end{tabular}
\end{ruledtabular}
\end{table*}

\begin{figure}[bt]
\begin{center}
\begin{tabular}{ll}
(a) & (b) \\
\multicolumn{1}{c}{
\includegraphics[bb=0 0 264 138, width=3.7cm]{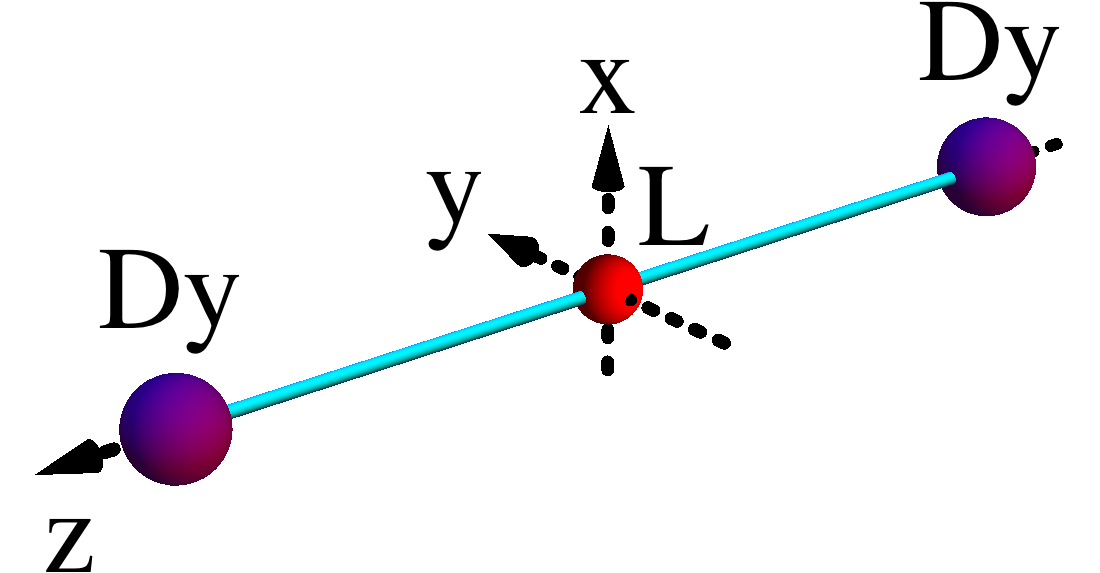}
}
&
\multicolumn{1}{c}{
\includegraphics[bb=0 0 264 138, width=3.7cm]{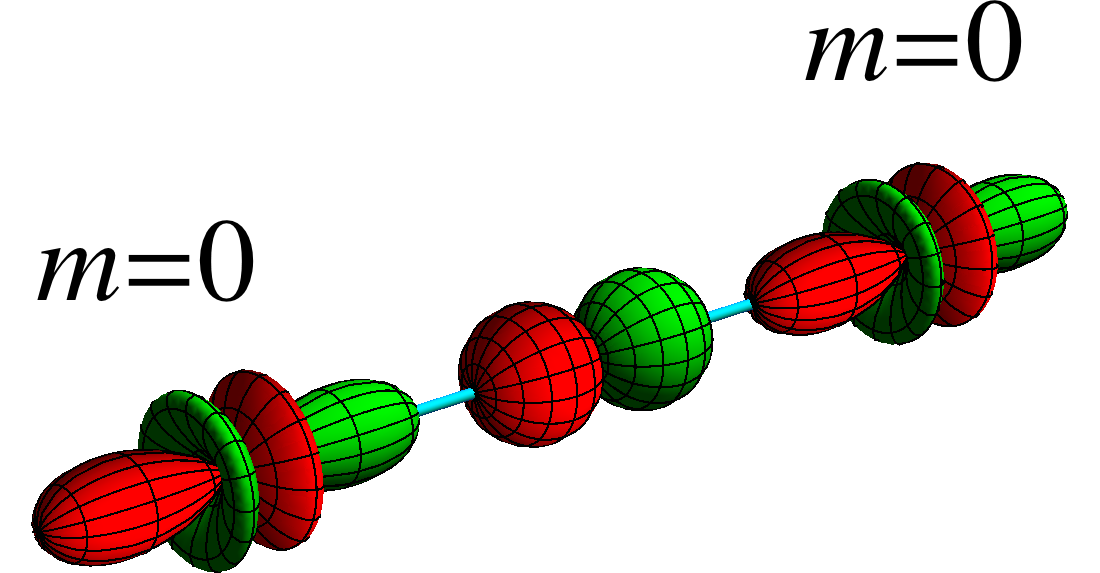}
}
\\
(c) 
\\
\multicolumn{2}{c}{
\includegraphics[bb= 0 0 1040 940, width=8.6cm]{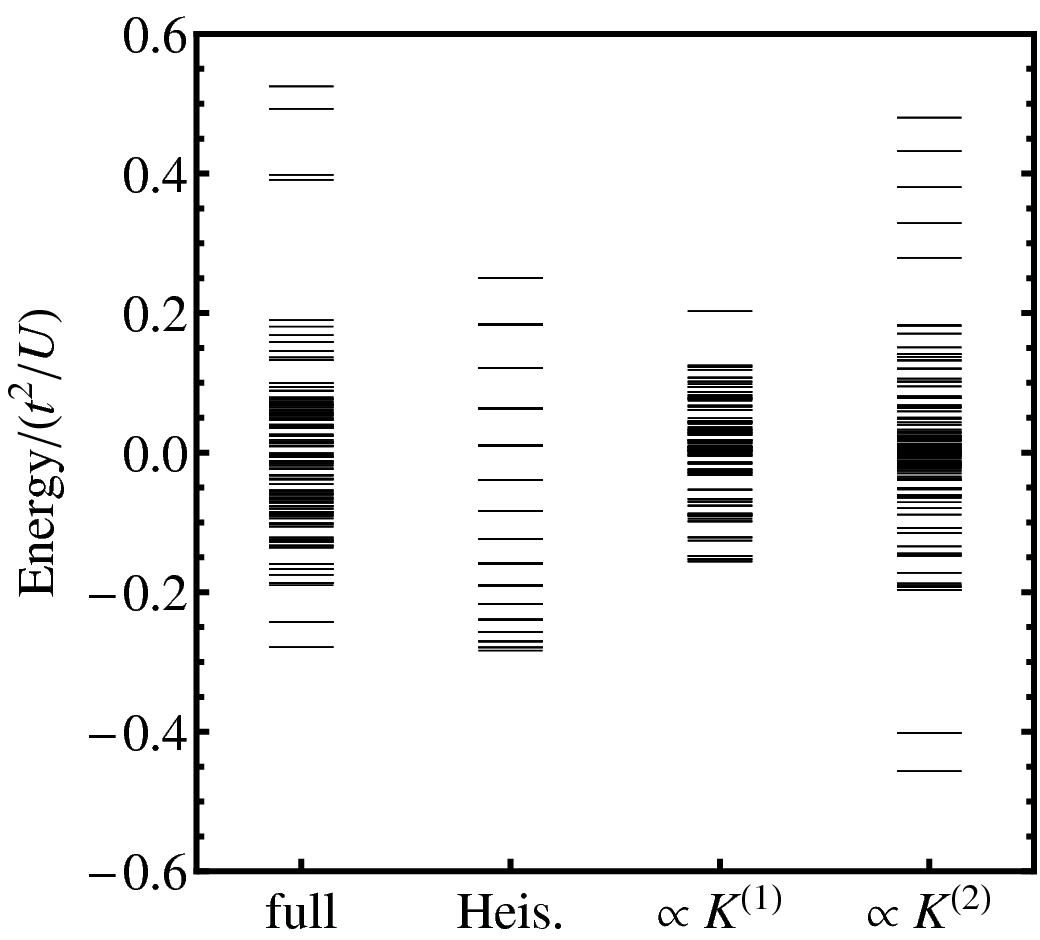}
}
\end{tabular}
\end{center}
\caption{(Color online) (a) Linearly bridged Dy dimer with one ligand atom (L). 
(b) Kinetic exchange interaction between $4f_{(5z^2-3r^2)z}$ orbitals. (c) Calculated exchange spectrum with full $\hat{H}_{\rm ex}$  (\ref{Eq:Hlinear}) and with its different contributions ($U\equiv U_{12}=U_{21}= 5$ eV). 
The excitation energies of the intermediate states entering Eq. (\ref{Eq:JKE}) have been calculated {\it ab initio} (Sec. \ref{Sec:excitation}).}
\label{Fig1}
\end{figure}

\begin{table}[tb]
\caption{$\mathcal{K}_{kk'}^{(1)}$ and $\mathcal{K}_{kk'}^{(2)}$.}
\label{Table:K}
\begin{ruledtabular}
\begin{tabular}{cccc}
$\mathcal{K}_{00}^{(1)}$ & $ 4.4458\times 10^{-3}$ & $\mathcal{K}_{11}^{(2)}$ & $ 2.2245\times 10^{-3}$ \\
$\mathcal{K}_{02}^{(1)}$ & $-3.7903\times 10^{-3}$ & $\mathcal{K}_{13}^{(2)}$ & $-3.2216\times 10^{-4}$ \\
$\mathcal{K}_{04}^{(1)}$ & $ 1.7181\times 10^{-4}$ & $\mathcal{K}_{15}^{(2)}$ & $ 1.5888\times 10^{-5}$ \\
$\mathcal{K}_{06}^{(1)}$ & $-1.7242\times 10^{-6}$ & $\mathcal{K}_{17}^{(2)}$ & $-1.8049\times 10^{-7}$ \\
$\mathcal{K}_{11}^{(1)}$ & $ 1.0908\times 10^{-2}$ & $\mathcal{K}_{22}^{(2)}$ & $ 4.8355\times 10^{-5}$ \\
$\mathcal{K}_{13}^{(1)}$ & $-1.1514\times 10^{-3}$ & $\mathcal{K}_{24}^{(2)}$ & $-4.6269\times 10^{-6}$ \\
$\mathcal{K}_{15}^{(1)}$ & $ 3.5361\times 10^{-5}$ & $\mathcal{K}_{26}^{(2)}$ & $ 7.4785\times 10^{-8}$ \\
$\mathcal{K}_{17}^{(1)}$ & $-2.9084\times 10^{-7}$ & $\mathcal{K}_{33}^{(2)}$ & $ 4.3815\times 10^{-5}$ \\
$\mathcal{K}_{22}^{(1)}$ & $ 2.9710\times 10^{-3}$ & $\mathcal{K}_{35}^{(2)}$ & $-2.1423\times 10^{-6}$ \\
$\mathcal{K}_{24}^{(1)}$ & $-1.3026\times 10^{-4}$ & $\mathcal{K}_{37}^{(2)}$ & $ 2.3716\times 10^{-8}$ \\
$\mathcal{K}_{26}^{(1)}$ & $ 1.2614\times 10^{-6}$ & $\mathcal{K}_{44}^{(2)}$ & $ 4.4502\times 10^{-7}$ \\
$\mathcal{K}_{33}^{(1)}$ & $ 1.1745\times 10^{-4}$ & $\mathcal{K}_{46}^{(2)}$ & $-7.2365\times 10^{-9}$ \\
$\mathcal{K}_{35}^{(1)}$ & $-3.3571\times 10^{-6}$ & $\mathcal{K}_{55}^{(2)}$ & $ 1.0531\times 10^{-7}$ \\
$\mathcal{K}_{37}^{(1)}$ & $ 2.6060\times 10^{-8}$ & $\mathcal{K}_{57}^{(2)}$ & $-1.1732\times 10^{-9}$ \\
$\mathcal{K}_{44}^{(1)}$ & $ 5.4544\times 10^{-6}$ & $\mathcal{K}_{66}^{(2)}$ & $ 1.1839\times 10^{-10}$\\
$\mathcal{K}_{46}^{(1)}$ & $-4.9672\times 10^{-8}$ & $\mathcal{K}_{77}^{(2)}$ & $ 1.3155\times 10^{-11}$\\
$\mathcal{K}_{55}^{(1)}$ & $ 7.4109\times 10^{-8}$ &  \\
$\mathcal{K}_{57}^{(1)}$ & $-3.9138\times 10^{-10}$&  \\
$\mathcal{K}_{66}^{(1)}$ & $ 4.1115\times 10^{-10}$&  \\
%$\mathcal{K}_{77}^{(1)}$ & $ 0.00000$              &  \\
\end{tabular}
\end{ruledtabular}
\end{table}

\begin{figure*}[tb]
\begin{center}
\begin{tabular}{llll}
(a) & (b) & (c) & (d)\\
\multicolumn{1}{c}{
\includegraphics[bb=0 0 288 194, width=4.0cm]{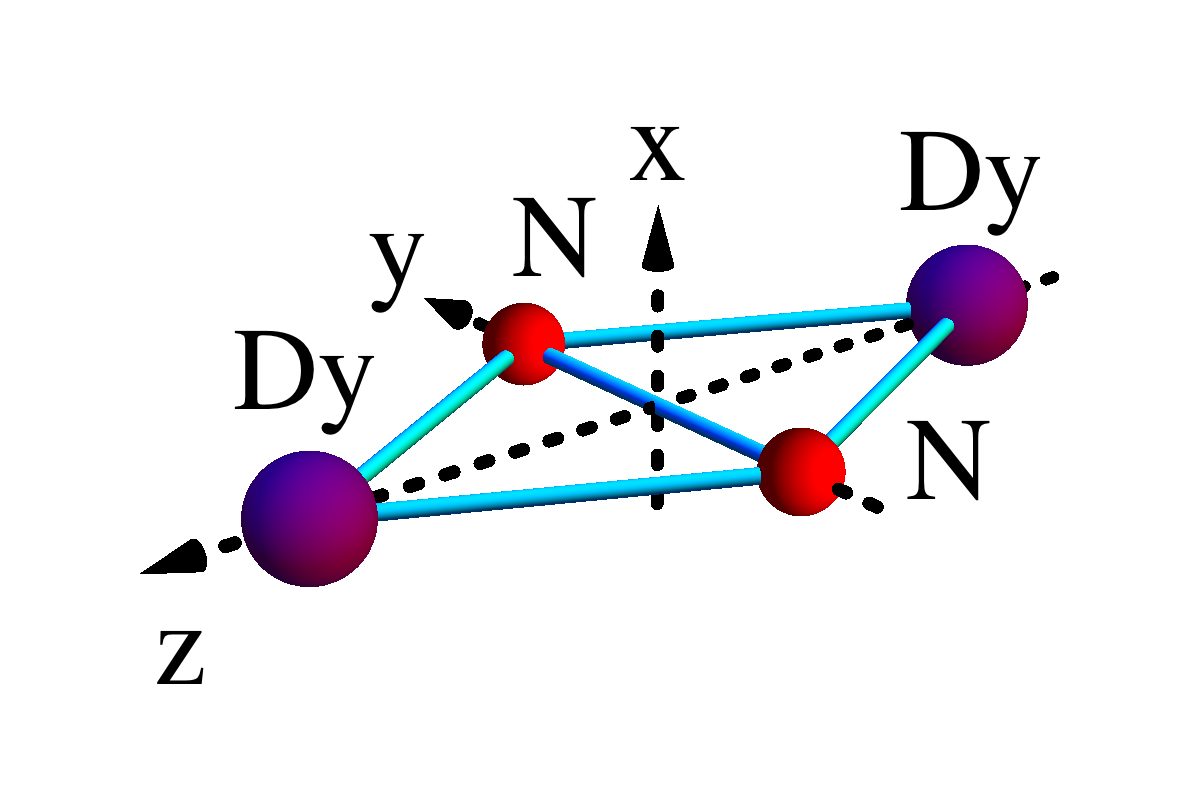}
}
&
\multicolumn{1}{c}{
\includegraphics[bb=0 0 288 194, width=4.0cm]{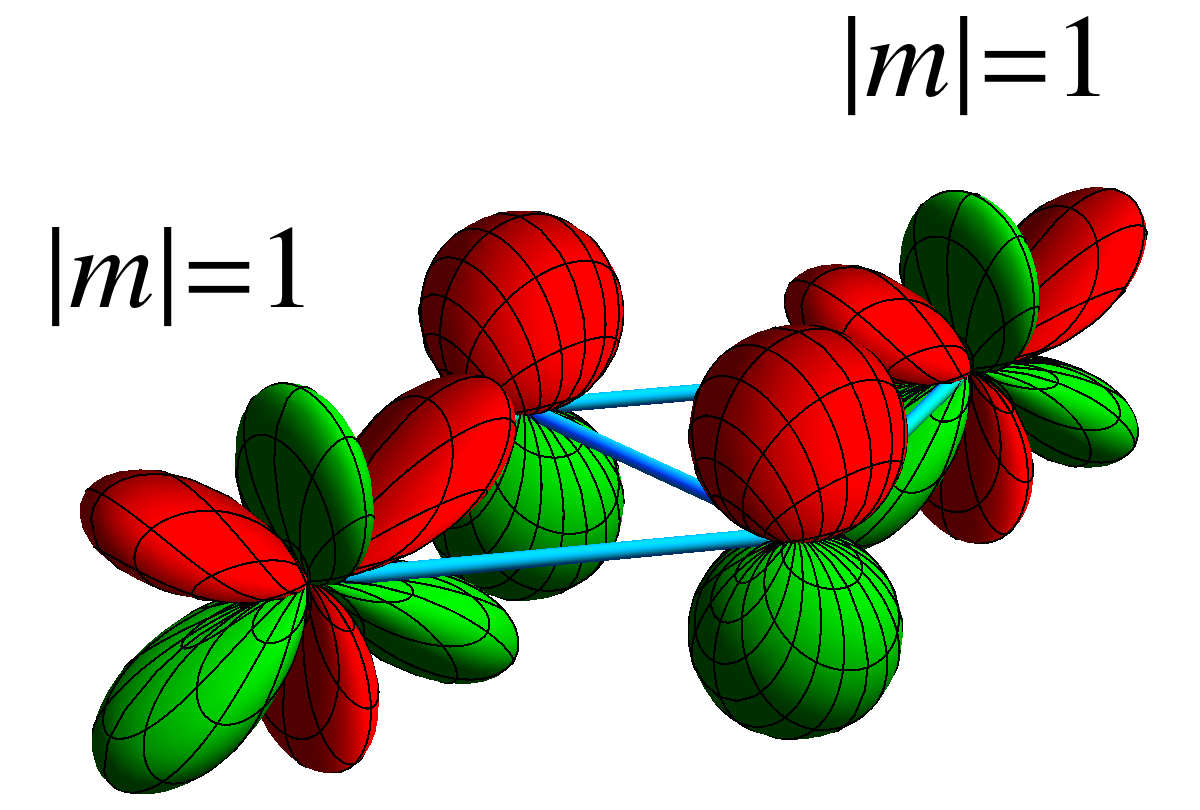}
}
&
\multicolumn{1}{c}{
\includegraphics[bb=0 0 288 194, width=4.0cm]{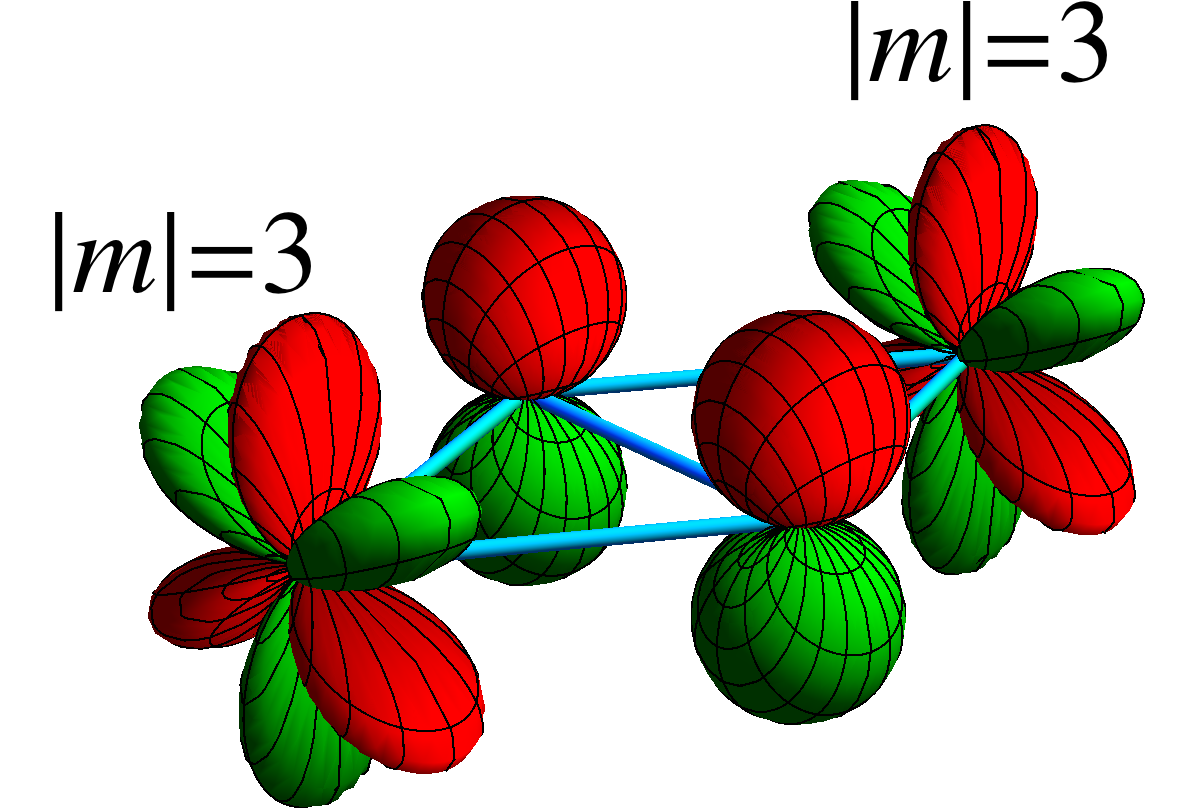}
}
&
\multicolumn{1}{c}{
\includegraphics[bb=0 0 288 194, width=4.0cm]{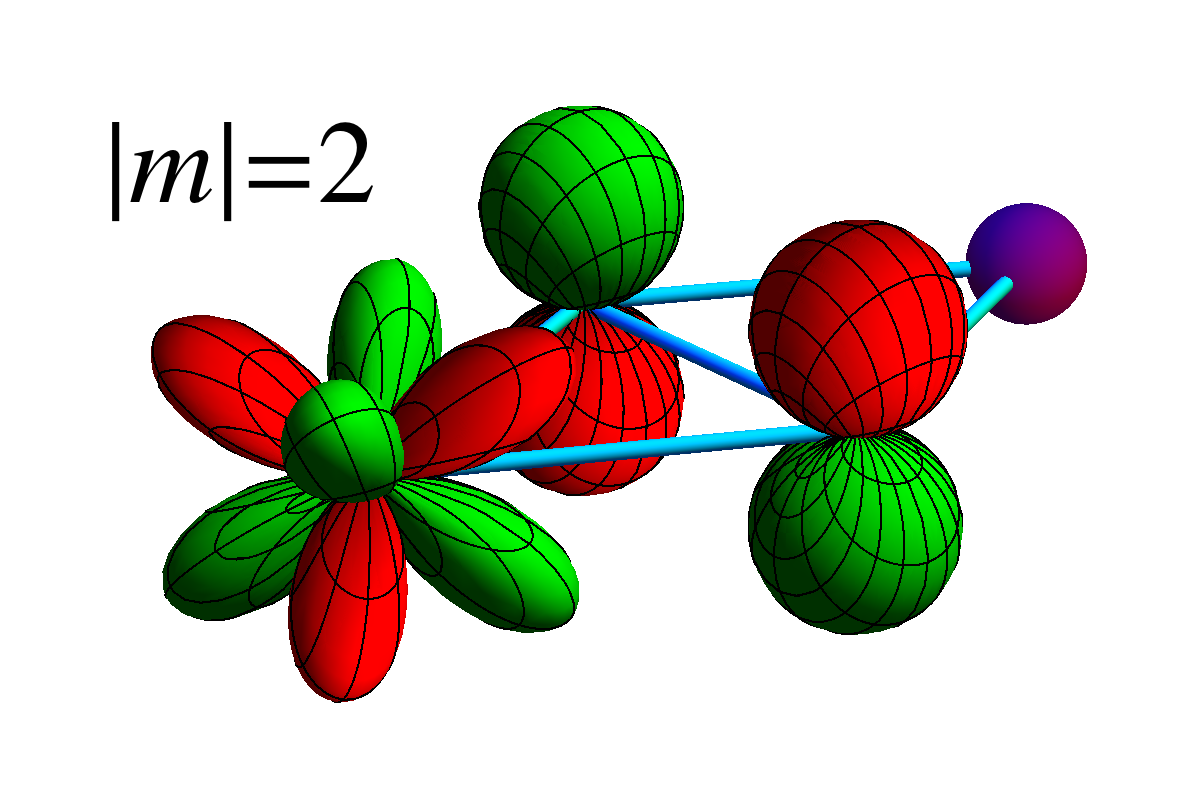}
}
\\
(e) && (f) \\
\multicolumn{2}{c}{\includegraphics[bb=0 0 1040 940, width=8.6cm]{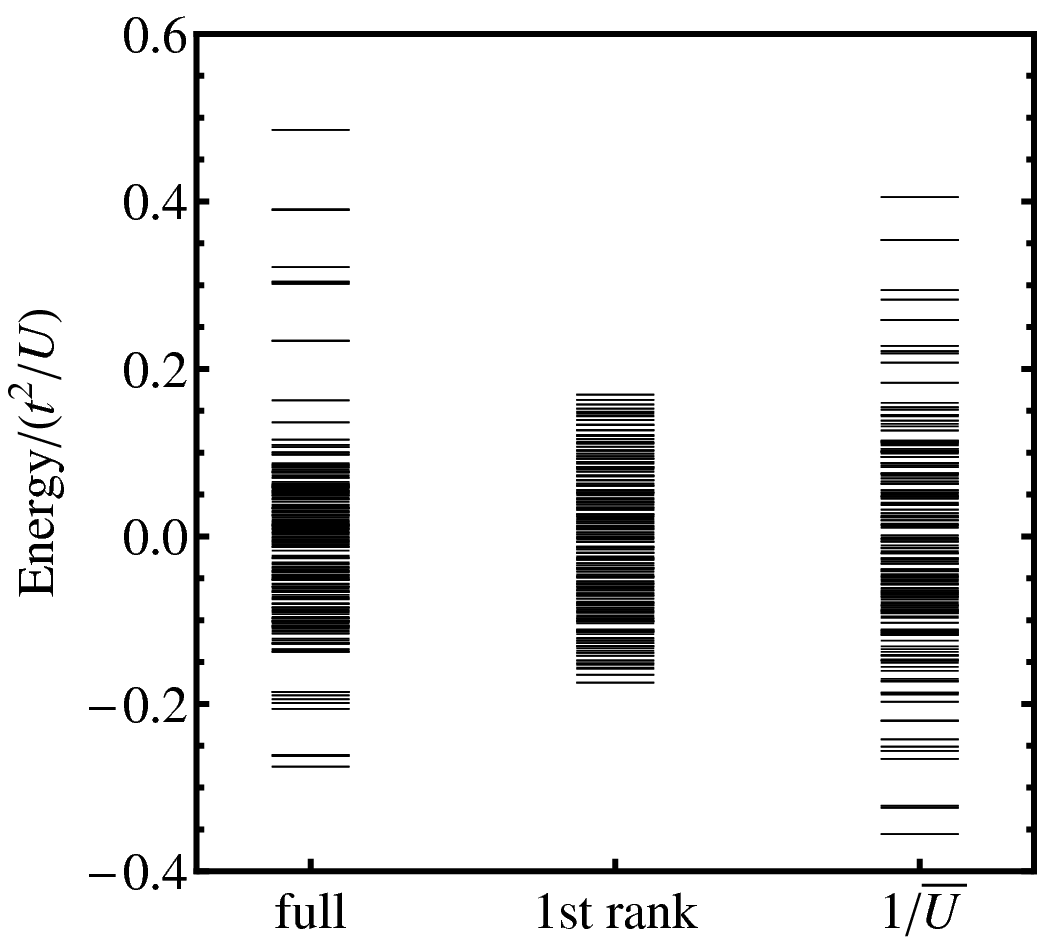}}
%\multicolumn{2}{c}{\includegraphics[width=8.6cm]{./Fig2e.eps}}
&
\multicolumn{2}{c}{\includegraphics[bb=0 0 1040 940, width=8.6cm]{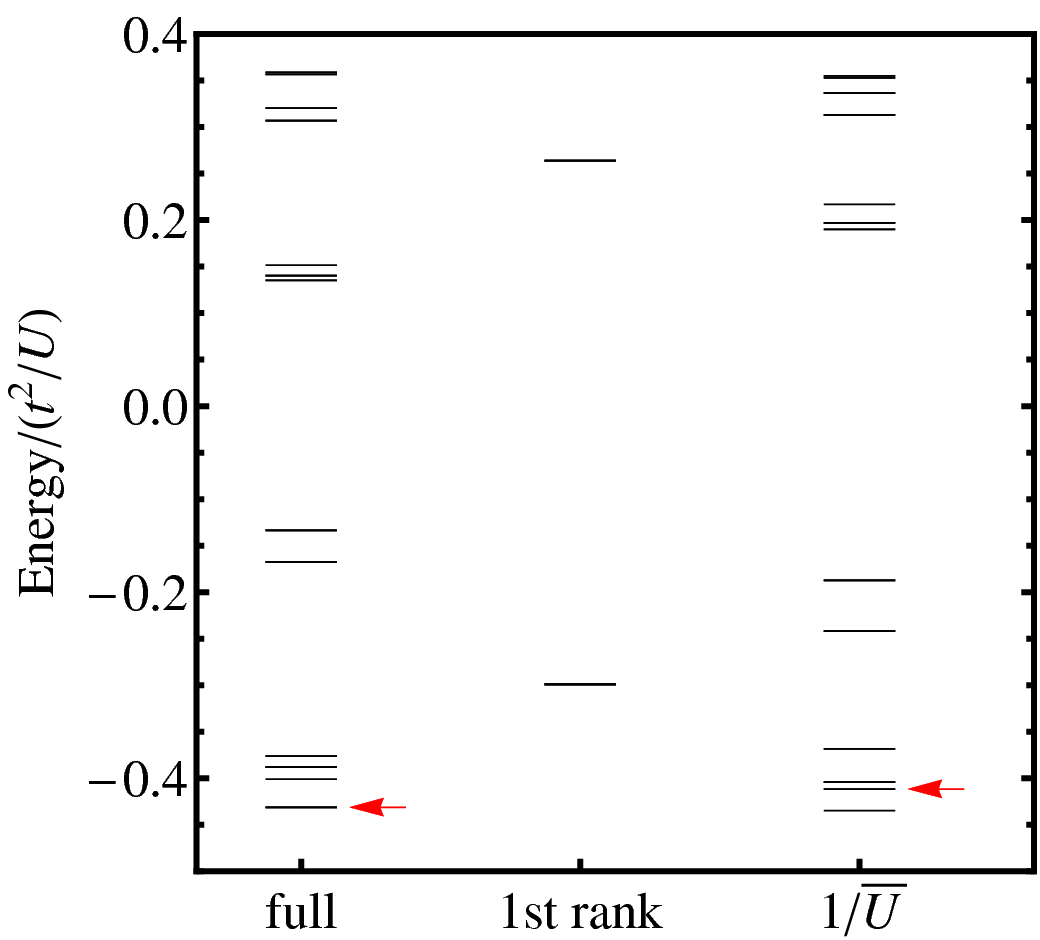}}
%\multicolumn{2}{c}{\includegraphics[width=8.6cm]{./Fig2f.eps}}
\end{tabular}
\end{center}
\caption{
(Color online)
(a) Dy dimer bridged via N$_2^{n-}$ anion ($D_{2h}$ core symmetry).
Large (purple) and small (red) balls are Dy and N, respectively.
%The core is assumed to have $D_{2h}$ symmetry. 
(b), (c) Kinetic exchange interaction between $4f_{(5z^2-r^2)x}$ orbitals and $4f_{x^3-3xy^2}$ orbitals, respectively, via the HOMO of N$_2^{2-}$.
%is large, while the overlap between the $4f_{x^3-3xy^2}$ orbital and the HOMO is small. 
%The electron transfer between $4f_{(5z^2-r^2)x}$ orbitals via the HOMO of N$_2^{2-}$ is dominant.
(d) Kinetic exchange interaction between the $4f_{xyz}$ orbital of Dy and the LUMO of N$_2^{2-}$.
(e) Calculated exchange spectrum with full $\hat{H}_{\rm ex}$ and its first-rank contribution ($U\equiv U_{12}=U_{21}= 5$ eV), and in the $1/\bar{U}$ approximation for kinetic exchange pattern (b).
%The exchange levels of between $4f_{(5z^2-r^2)x}$ orbitals $(|m|=1)$ with $U_{12}=U_{21}=$ 5 eV. 
(f) Calculated exchange spectrum with full $\hat{H}_{\rm ex}$ and its first-rank contribution ($U_{21}=$ 3 eV), and in the $1/\bar{U}$ approximation for kinetic exchange pattern (d).
%The exchange levels of between $4f_{xyz}$ orbital $(|m|=2)$ and the LUMO with $U_{21}=$ 3 eV. 
%The excitation energies in the intermediate states are calculated {\it ab initio} \cite{SM}.
The calculations in (e) and (f) involved exchange parameters (\ref{Eq:JKE}) with excitation energies of the intermediate states on Dy evaluated {\it ab initio}. $\bar{U}$ in the calculation within $1/\bar{U}$ approximation was chosen to reproduce the width of the spectrum for full $\hat{H}_{\rm ex}$.
}
\label{Fig2}
\end{figure*}

\section{Examples}
\label{Sec:example}
We further consider some typical examples 
of $J$-$J$ and $J$-$S$ exchange interaction. %states for simple systems are shown here. 
Since the kinetic exchange interaction is usually much stronger than the direct exchange interaction
\cite{Anderson1959, Anderson1963}, we only take into account the former. % kinetic exchange interaction.
In order to include the multiplet structure of intermediate states in Eq. (\ref{Eq:JKE}), 
first we calculate {\it ab initio} the excitation energies of the virtual electron-transfer states. 
The calculated exchange levels are compared with those arising from the bilinear form (\ref{Eq:JJ})
and corresponding to the $1/\bar{U}$ approximation. 

\subsection{Excitation energies of Dy$^{2+}$ and Dy$^{4+}$ ions}
\label{Sec:excitation}
The excitation energies $\Delta E_{\alpha_J J}$ appearing in the denominator of the 
kinetic exchange Hamiltonian (\ref{Eq:HKE}) are calculated {\it ab initio}.
Since the effect of the crystal-field splitting in the intermediate states is negligible 
(Sec. \ref{Sec:ion}),
%$\Delta E_\text{cf}$ is negligible (its magnitude is the third order of smallness, 
%$t^2/U\cdot \Delta E_\text{cf}/U$ and $\Delta E_\text{cf}/U \alt$ 0.01),
we used the energy levels of the free Ln$^{2+}$ and Ln$^{4+}$ ions.
In this work, we only calculated the energies for Ln $=$ Dy that we will use in the following sections.
Apart from the crystal-field splitting, there is totally symmetric electrostatic potential 
which only depends on the number of electrons and shifts uniformly the $J$-multiplet energies.
This effect is absorbed in the minimum promotion energy $U_{ij}$ in Eq. (\ref{Eq:JKE}).
The energies are calculated using the complete active space self-consistent field (CASSCF) 
and the restricted active space SCF state interaction (RASSI) methods 
with ANO-RCC QZP basis set \cite{Molcas}.
With the CASSCF method, the $LS$-term energies are obtained, while with the RASSI method, 
the spin-orbit ($J$-multiplet) energy levels are calculated.
For the CASSCF calculations, all $4f$ orbitals are included into the active space. 
%The $J$-multiplet energies are obtained by the RASSI calculations including 
The terms included in the RASSI mixing are 
${}^6P$, ${}^6F$, ${}^6H$ for Dy$^{2+}$ and 
${}^7F$, ${}^5S$, ${}^5P$, three ${}^5D$, two ${}^5F$, three ${}^5G$, two ${}^5H$, two ${}^5I$, 
${}^5K$, ${}^5L$ for Dy$^{4+}$.
%Each $J$-multiplet state mainly comes from one set of $LS$-term, 
%which supports the assumption Eq. (2) in the main text.
The excitation energies are tabulated in Table \ref{Table:DeltaE}.

There are several $LS$-terms which appear more than once, i.e., 
${}^5D$, ${}^5F$, ${}^5G$, ${}^5H$, ${}^5I$ terms of Dy$^{4+}$ ion.
These terms obtained by the CASSCF calculations are assigned to the symmetrized $LS$-states 
within the shell model, $|l^N \alpha_{LS}LSJM\rangle$, 
comparing the patterns of the {\it ab initio} and the model spin-orbit splittings of each $LS$-term.
The symmetrized $LS$-states are constructed by using the coefficient of fractional parentage \cite{Nielson1963}.
In the basis of symmetrized states, the matrix element of the spin-orbit Hamiltonian
$\hat{H}_{\rm so} = \zeta \sum_{i=1}^N \hat{\mathbf{l}}_i \cdot \hat{\mathbf{s}}_i$
$(\zeta > 0)$ is given by
%\begin{eqnarray}
% \langle l^N \alpha LSJM_J|\hat{H}_{\rm so}|l^N \alpha' L'S'J'M_J'\rangle &=&
% \delta_{LL'} \delta_{SS'} \delta_{JJ'} \delta_{M_J M_J'} 
% \zeta \frac{-\Pi_{lJ}\sqrt{l(l+1)}}{2\sqrt{3}} 
%\nonumber\\
% &\times&
% \langle l^N \alpha LS\Vert V^{11} \Vert l^N \alpha' LS\rangle
% \begin{Bmatrix}
%  L & S & J \\
%  L & S & J \\
%  1 & 1 & 0 
% \end{Bmatrix}
%,
%\end{eqnarray}
%and 
\begin{eqnarray}
% \langle l^{4l+2-N} &&\alpha LSJM_J|\hat{H}_{\rm so}|l^{4l+2-N} \alpha' L'S'J'M_J'\rangle 
 \langle l^{4l+2-N} &&\alpha_{LS} LS JM_J|\hat{H}_{\rm so}|l^{4l+2-N} \alpha_{LS}' L'S' J'M_J'\rangle 
%&=&
\nonumber\\
 &&=
 -\delta_{LL'} \delta_{SS'} \delta_{JJ'} \delta_{M_J M_J'} 
 \zeta \frac{\Pi_{lJ}\sqrt{l(l+1)}}{\sqrt{2}} 
\nonumber\\
% &\times&
 &&\times
 \begin{Bmatrix}
  L & S & J \\
  L & S & J \\
  1 & 1 & 0 
 \end{Bmatrix}
% \langle l^N \alpha LS\Vert V^{11} \Vert l^N \alpha' LS\rangle
 \langle l^N \alpha_{LS} LS\Vert \{\hat{c}^\dagger \otimes \bar{c}\}_1^1 \Vert l^N \alpha_{LS}' LS\rangle
\nonumber\\
\end{eqnarray}
for $N \le 2l+1$.
Here, the curly bracket with $3 \times 3$ elements are the $9j$ symbol (\ref{Eq:9j}).
%and $V^{11}$ is the operator defined and tabulated in Ref. \onlinecite{Nielson1963}.
Therefore, the spin-orbit splitting is proportional to the reduced matrix element
of operator $\{\hat{c}^\dagger \otimes \bar{c}\}_{1-m}^{1m}$
\footnote{The reduced matrix elements of $V_{m-m}^{11} = -\sqrt{6}
\{\hat{c}^\dagger \otimes \bar{c}\}_{1-m}^{1m}$ is tabulated in Ref. \onlinecite{Nielson1963}.
See, for example, Sec. 6.2 in Ref. \onlinecite{Judd1967} 
for the relation between the spin-orbit coupling and $V^{11}$.}.
%of the operator $V^{11}$ % tabulated in Ref. \onlinecite{Nielson1963}
%(see, for example, Sec. 6.2 in Ref. \onlinecite{Judd1967} 
%for the relation between the spin-orbit coupling and $V^{11}$).

\subsection{Kinetic exchange through monoatomic bridge}
\label{Sec:example1}
As a simple example, consider an exchange-coupled Dy$^{3+}$ dimer with axial bridging geometry (Fig. \ref{Fig1}a).
The largest transfer parameter ($t$) is expected between $f_{5z^3-3r^2z}$ $(m=0)$ orbitals because of their sigma bonding to the $p_z$ orbital of the bridging ligand atom (Fig. \ref{Fig1}b). Then, according to the rule (\ref{Eq:q}), $q_{\rm max}=1$, while Eq. (\ref{Eq:tt}) gives $q=-q'$. The resulting form of the exchange 
Hamiltonian $\hat{H}_{\rm ex}$, %$\hat{H}_{\rm ex}$ 
after expanding the Stevens operators in Eq. (\ref{Eq:Hbar_tensor}), is 
\begin{eqnarray}
 \hat{H}_{\rm ex} &=& 
 \hat{K}^{(1)}
 + \hat{\mathbf{J}}_1 \cdot \hat{\mathbf{J}}_2 \hat{K}^{(2)}
 + \hat{K}^{(2)} \hat{\mathbf{J}}_1 \cdot \hat{\mathbf{J}}_2,
\label{Eq:Hlinear}
\\
 \hat{K}^{(1)} &=& 
 \sum_{k,k'=0}^7 \mathcal{K}^{(1)}_{kk'} \hat{J}_{1z}^k \hat{J}_{2z}^{k'},
\label{Eq:K1}
\\
 \hat{K}^{(2)} &=& \sum_{k,k'=1}^7\mathcal{K}^{(2)}_{kk'} \hat{J}_{1z}^{k-1} \hat{J}_{2z}^{k'-1},
\label{Eq:K2}
\end{eqnarray}
where $k+k'$=even. We can see that, even in this simplest case, $\hat{H}_{\rm ex}$ does not reduce to the isotropic form (1) because Ising ($\propto \hat{K}^{(1)}$) and mixed Ising-Heisenberg ($\propto \hat{K}^{(2)}$) terms, both involving high powers of momentum projection operators of two sites. 
The parameters $\mathcal{K}_{kk'}^{(1)}, \mathcal{K}_{kk'}^{(2)}$ %of the exchange Hamiltonian 
%for the linear Dy complex are tabulated in Table \ref{Table:K}. 
are tabulated in Table \ref{Table:K}. 
When the eigenvalue of $\hat{J}_{iz}$ is large, %close to the maximal value $15/2$, 
the higher order terms are significantly enhanced and 
contribute to the exchange interaction rather than the bilinear term.
As a result the exchange spectra calculated with the full Hamiltonian (\ref {Eq:Hlinear}) and with its Heisenberg-type part (\ref{Eq:JJ}) show large discrepancy between them (Fig. \ref{Fig1}(c)).
%Eq. (\ref{Eq:Hlinear}) contains Ising part (\ref{Eq:K1}) 
%as well as Heisenberg like part in contrast to Eq. (\ref{Eq:JJ}).
%Contrary to the spin system, both $\hat{K}^{(1)}$ and $\hat{K}^{(2)}$ depend on $\hat{J}_{iz}$ and
%include higher order terms up to 7, in general, 
%the interaction cannot be approximated by the Heisenberg form (\ref{Eq:JJ}).
%Moreover, the eigenstates of (\ref{Eq:Hlinear}) and (\ref{Eq:JJ}) also differ significantly. % \cite{SM}.
The discrepancy is also seen in their eigenstates. 
%Therefore, the form of the Heisenberg type interaction (\ref{Eq:JJ}) is not adequate.
%The exchange Hamiltonian between $J$-multiplet and isotropic spin has similar form.
%In the latter case, as we have seen above, Eq. (\ref{Eq:HJS}), the spin operator is up to first order. 
%The exchange levels of the Hamiltonian are shown in Fig. \ref{Fig:energy0}. 
%Both the Ising term (12) and the Heisenberg term (13) are important in the exchange interaction.
The difference between the exchange states of (\ref{Eq:Hlinear}) with those of $\hat{H}_\text{Heis}$
(\ref{Eq:JJ}) is compared by expanding the former by the latter.
The solution of $\hat{H}_\text{Heis}$ for two site system is given 
\begin{eqnarray}
|J_{12}M_{12}\rangle = \sum_{M_1,M_2}|J_1M_1,J_2M_2\rangle C_{J_1M_1J_2M_2}^{J_{12}M_{12}},
\label{Eq:JM_Heis}
\end{eqnarray}
where $J_{12}$ and $M_{12}$ are the total angular momentum for the dimer and its projection.
The low-energy exchange states of $\hat{H}_{\rm ex}$ (\ref{Eq:Hlinear}) 
are written in the basis of $\{|J_{12}M_{12}\rangle\}$ as follows: % (\ref{Eq:JM_Heis}),
\begin{eqnarray*}
 |\Psi_1, A_{1g}\rangle &\approx& 0.574|0,0\rangle + 0.773|2,0\rangle + 0.190|4,0\rangle 
\nonumber\\
 &-& 0.160|6,0\rangle,
\\
 |\Psi_2, A_{1u}\rangle &\approx& 0.846|1,0\rangle + 0.505|3,0\rangle -0.161|7,0\rangle,
\\
 |\Psi_{3,4}, E_{1u}\rangle &\approx& 0.352|1,\pm 1\rangle + 0.768|3,\pm 1\rangle 
 + 0.522|5,\pm 1\rangle 
\nonumber\\
 &+& 0.107|7, \pm 1\rangle,
\\
 |\Psi_{5,6}, E_{1g}\rangle &\approx& 0.636|2,\pm 1\rangle + 0.712|4,\pm 1\rangle + 0.294|6, \pm 1\rangle.
\end{eqnarray*}
Here, the irreducible representation $\Gamma$ of $D_{\infty h}$ is used, 
and the states $|\Psi_i,\Gamma\rangle$ belong to the eigenvalues 
$E_{1,2} = -0.278523$, $E_{3,4,5,6} = -0.243015$, respectively, in the units of $t^2/U_{12}$.
The low-energy exchange states are not necessarily mainly contributed by the ground state of the
antiferromagnetic $\hat{H}_\text{Heis}$, $|J_{12}M_{12}\rangle = |0,0\rangle$. 
Therefore, we conclude that the Heisenberg form of the interaction is not adequate 
to describe the exchange interaction between $J$-multiplets.

\begin{figure*}[tb]
\begin{tabular}{ll}
(a) & (b) \\
\includegraphics[bb=0 0 1040 704, width=8cm]{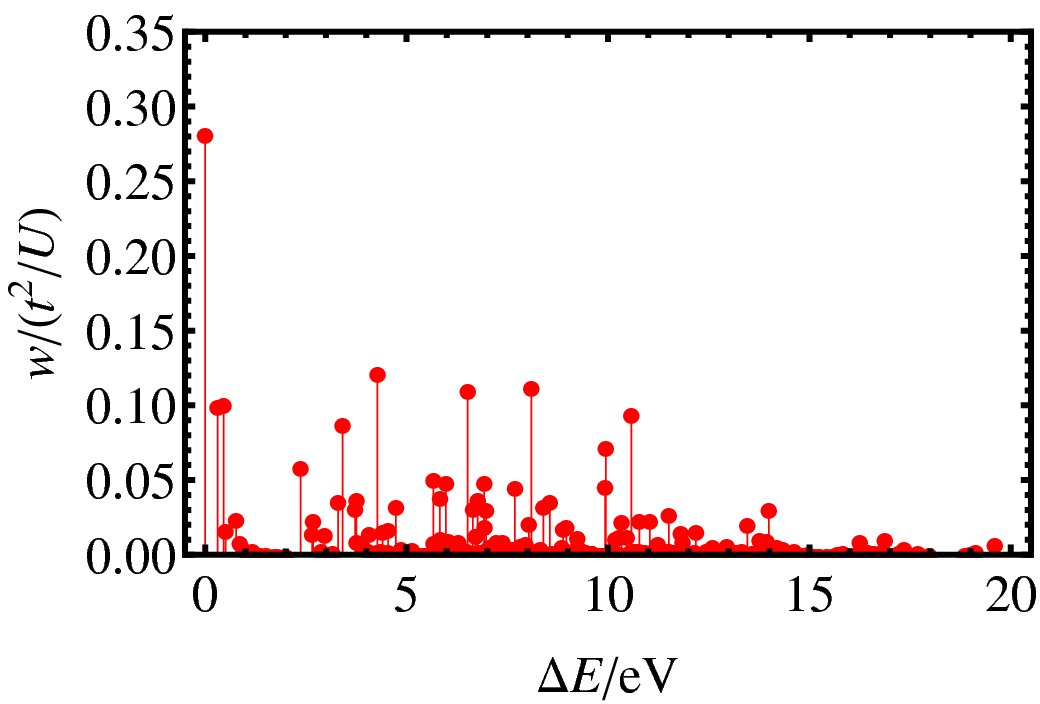}
&
\includegraphics[bb=0 0 1040 704, width=8cm]{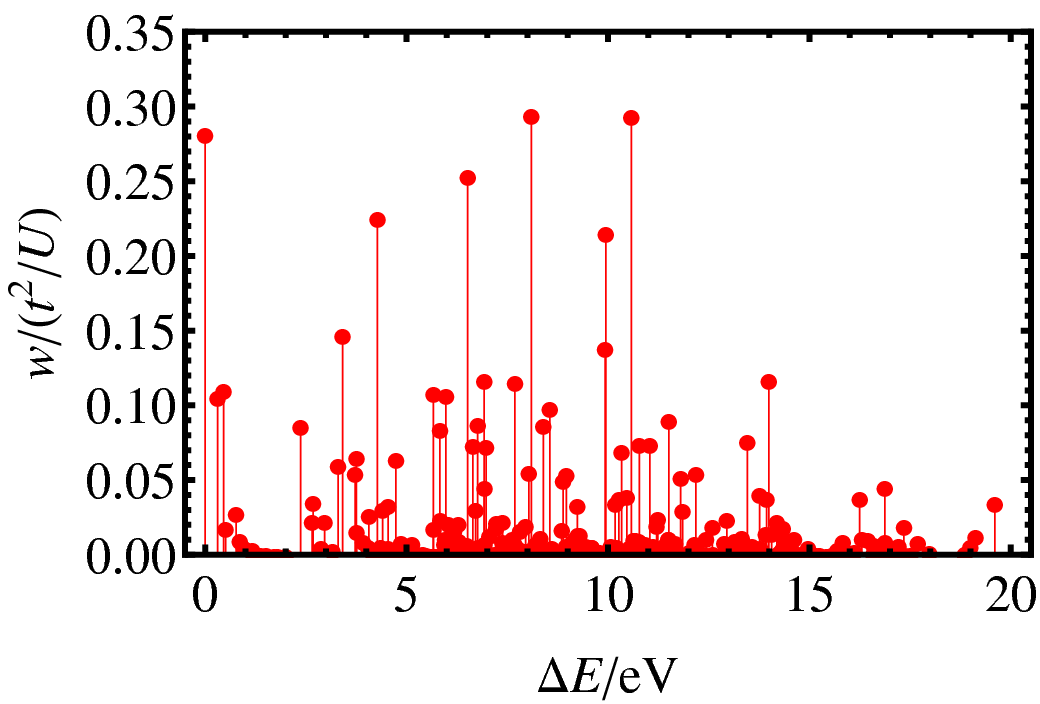}
\\
(c) & (d) \\
\includegraphics[bb=0 0 1040 704, width=8cm]{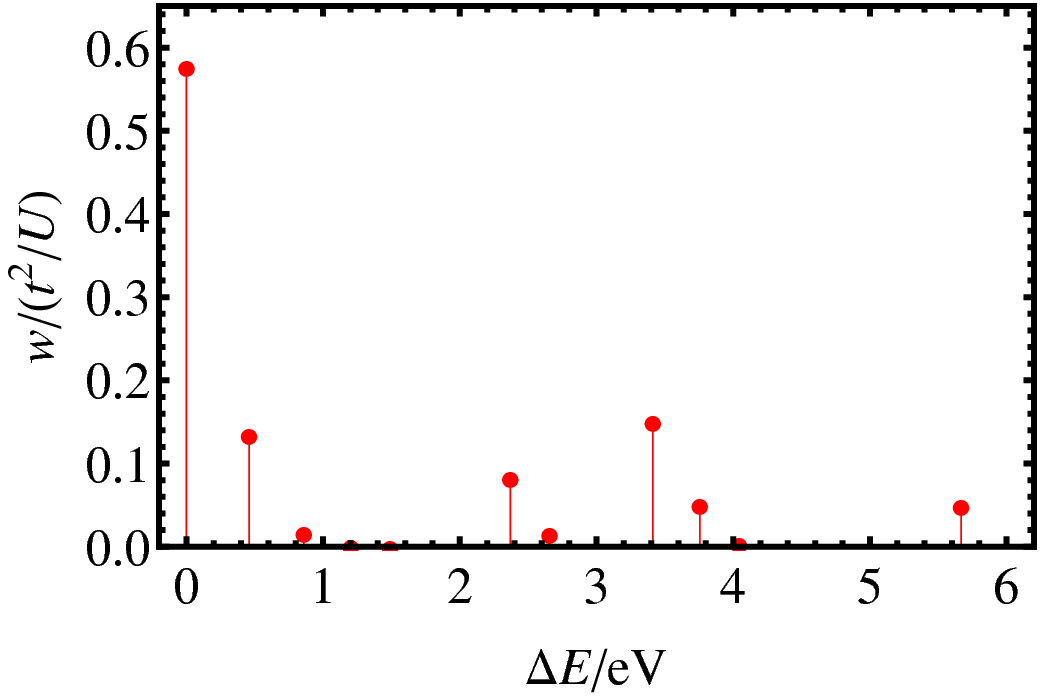}
&
\includegraphics[bb=0 0 1040 704, width=8cm]{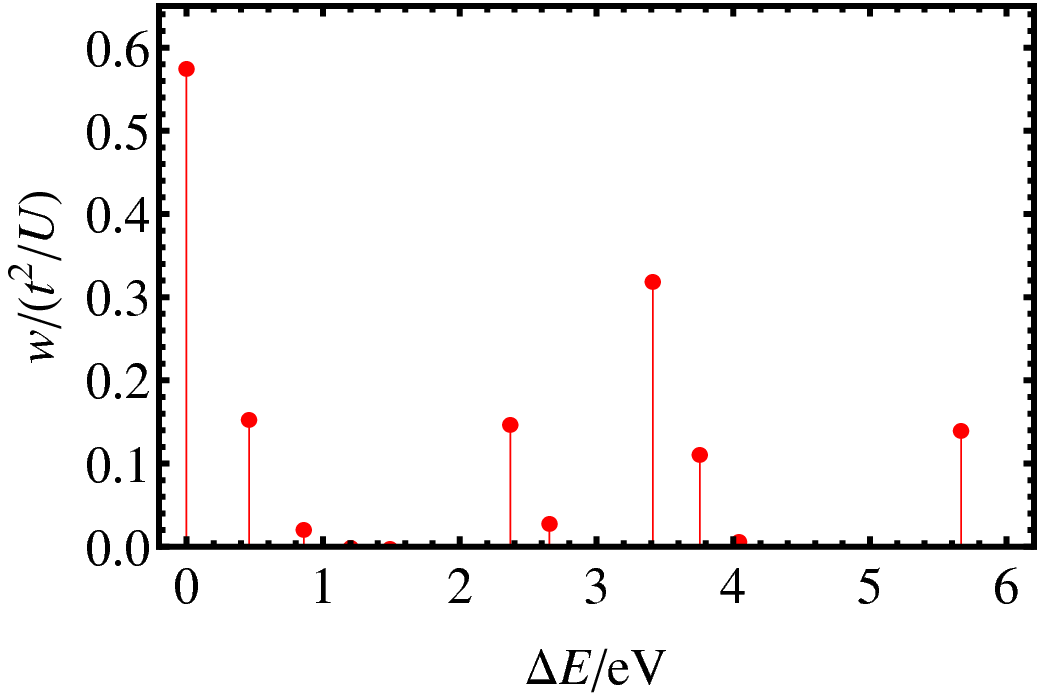}
\end{tabular}
\caption{The contributions from the intermediate states $w$ %(in $t^2/U$)
corresponding to the excitation energy $\Delta E$ 
($ = \Delta E_{\alpha_J J}^{N} + \Delta E_{\alpha_J' J'}^{N'}$)
for (a) the full exchange interaction $\hat{H}_{\rm ex}$ and 
(b) the exchange within the $1/\bar{U}$ approximation $\hat{H}_{\rm ex}^{1/\bar{U}}$
for Dy$^{3+}$ dimer bridged by N$_2^{2-}$
and 
%The contributions from the excited states to the net exchange interaction ($t^2/U$)
for (c) the full exchange interaction and (d) the exchange within the $1/\bar{U}$ approximation 
for the Dy-radical system ($n=3$).
%The contributions within the $1/U$ approximation are not scaled here.
The units of $w$ are $t^2/U_{21}$ for (a), (c) and $t^2/\bar{U}$ for (b), (d).
}
\label{Fig:weight}
\end{figure*}

\subsection{Kinetic exchange through biatomic bridge}
\label{Sec:example2}
Consider the exchange interaction in 
the Dy$^{3+}$ dimer bridged by the N$_2^{n-}$ $(n=2,3)$ anion (Fig. \ref{Fig2}(a)) \cite{Rinehart2011}.
In the case of $n=2$, $4f$ electrons of Dy$^{3+}$ ions would transfer 
between the metal sites via the highest occupied molecular orbital (HOMO) of N$_2^{2-}$ 
(Figs. \ref{Fig2}(b),(c)).
The HOMO overlaps with the $f_{(5z^2-r^2)x}$ ($|m|=1$) and the $f_{x^3-3xy^2}$ ($|m|=3$)
metal orbitals, the former interaction being dominant. 
Hence, we only consider the electron transfer between the orbitals with $|m|=1$
(Fig. \ref{Fig2}(b)).
%In the present system, the exchange interaction has not only the linear terms with respect to 
For them $\Delta_{\rm max}=2$ and we obtain according to Eq. (\ref{Eq:q}) $q_{\rm max}=3$. Then $\hat{H}_{\rm ex}$ will include
powers of $\hat{J}_{i\pm}$ $(= \hat{J}_{ix}\pm i \hat{J}_{iy})$ for each center
%appearing in Eq. (\ref{Eq:Hlinear}) but also the higher terms 
up to third order.

Figure \ref{Fig2}(e) shows the calculated exchange spectrum for full $\hat{H}_{\rm ex}$, and 
its first-rank contribution, and for one single promotion energy $\bar{U}$ ($1/\bar{U}$ approximation) 
\cite{Elliott1968, Santini2009}. %with $U_{ij}=5$ eV. 
Although the first-rank contribution is bilinear in $\hat{J}_{i\gamma}$,
it is not isotropic and 
the corresponding spectrum does not resemble the pattern of levels 
of Heisenberg-type Hamiltonian (\ref{Eq:JJ}).
%Nonetheless, this is not the main part in $\hat{H}_{\rm ex}$.
%The width and the distribution of the 1st rank levels
%are significantly different from those of the exact exchange level ($\hat{H}_{\rm ex}$). 
%The small splitting of the former indicates the importance of 
%the higher order rank parts in $\hat{H}_{\rm ex}$.
%The distribution is mainly explained by the levels for the ferromagnetic Heisenberg type Hamiltonian
%(\ref{Eq:JJ}), although we consider the superexchange mechanism.
%These discrepancies show that the exchange interaction between $J$-multiplets cannot be described by 
%the Heisenberg type interaction (\ref{Eq:JJ}).
%
Also the spectrum is quite different when the $1/\bar{U}$ approximation is applied.
This approximation 
%leads to the interchange of the exchange states.
%The discrepancy between the exact exchange states and the $1/U$ states is 
%due to the complicated electronic structure of the charge transferred intermediate state.
%Since the $1/U$ approximation 
neglects the splitting of the $LS$-terms 
which exceeds several times the minimal electron promotion energy.
As a result, the relative contributions to the exchange interaction 
from various intermediate states are significantly modified.
In order to see the variation of the contributions 
from the intermediate states to the kinetic exchange interaction,
we divide the kinetic exchange Hamiltonian as follows:
%The kinetic exchange Hamiltonian is a sum of the terms from the intermediate states,
\begin{eqnarray}
 \hat{H}_\text{ex} &=& \sum_{\alpha_J J} \sum_{\alpha_J' J'}
                       \hat{h}_\text{ex}(\alpha_J J, \alpha_J' J').
\end{eqnarray}
Here, $\hat{h}_\text{ex}(\alpha_J J, \alpha_J' J')$ indicates the term which only includes the contribution
from the set of the intermediate states $(\alpha_J J, \alpha_J' J')$.
The contribution from each such process can be measured by the width $w$ of the 
eigenvalues of $\hat{h}_\text{ex}(\alpha_J J, \alpha_J' J')$.
%The widths $w$ %for Dy dimer bridged by N$_2^{2-}$ 
%of the full Hamiltonian $\hat{H}_{\rm ex}$ and 
%that with $1/U$ approximation, $\hat{H}_\text{ex}^{1/U}$,
The widths $w$ for the full exchange Hamiltonian $\hat{H}_\text{ex}$ and those within $1/\bar{U}$ approximation
are shown in Figure \ref{Fig:weight}(a) and (b), respectively.
In comparison with the contributions to the full Hamiltonian,
those from the high energy states ($\Delta E \approx$ 5-10 eV) are exaggerated in $\hat{H}_\text{ex}^{1/\bar{U}}$.

In the case of N$_2^{3-}$ bridge, the main exchange coupling arises 
%The interchange of the exchange levels is clearly seen in the dimer with N$_2^{3-}$.
%the same complex (Fig. \ref{Fig2}(a)) 
%becomes an example containing $J$-multiplet and isotropic spin.
between the $f_{xyz}$ orbital of Dy and the unpaired electron occupying
the lowest unoccupied molecular orbital (LUMO) of N$_2^{2-}$
%The interaction is mainly due to the electron transfer between 
%the $f_{xyz}$ orbital and the lowest unoccupied molecular orbital (LUMO) of N$_2^{2-}$
(Fig. \ref{Fig2}(d)). %both of $b_{3u}$ symmetry.
The LUMO level in N$_2^{3-}$ has significantly higher energy compared 
to the orbital energy of $4f$ electrons in Dy$^{3+}$. 
On this reason and also due to a larger space distribution of the LUMO compared to the $4f$ orbitals, the minimal electron promotion energy from N$_2^{3-}$ to Dy$^{3+}$ ($U_{21}$) is expected to be much smaller than in the opposite direction ($U_{12}$). 
%Since the anti-bonding LUMO is unstable, 
%the energy gap between the LUMO and the $f$ orbital partially cancels the Coulomb repulsion
%due to the electron transfer from N$_2^{3-}$ to Dy$^{3+}$, leading to small promotion energy.
%Owing to the same reason, 
%the promotion energy for the electron transfer from the $f$ orbital to the LUMO becomes large.
%Hence the electron transfer process from N$_2^{3-}$ to Dy$^{3+}$ is dominant.
Hence, we neglect the latter process. 
Given that the Dy orbitals involved in the electron transfer have $|m|=2$, according to Eq. (\ref{Eq:q}) $q_{\rm max}=5$, %for large $k$, 
the same for the maximal power of $\hat{J}_{1\pm}$ in the exchange Hamiltonian. 

Figure \ref{Fig2}(f) shows the exchange levels obtained for full $\hat{H}_{\rm ex}$, 
its first-rank part, and for the $1/\bar{U}$ approximation.
In the present case, the first-rank part of $\hat{H}_{\rm ex}$ coincides with Eq. (\ref{Eq:JJ}), 
while the corresponding spectrum strongly differs from the full $\hat{H}_{\rm ex}$ indicating 
%However, the splitting of tme 1st rank levels is smaller than that of the exact ones
%and their distributions are different from each other, 
%which again indicates 
the importance of higher order terms.
As in the previous example, the $1/\bar{U}$ approximation modifies the relative contributions 
to the exchange interaction from intermediate states (Fig. \ref{Fig:weight} (c),(d))
and induces, in particular, the interchange 
%between the three-fold degenerate ground state 
%and the nondegenerate first excited state
of the three-fold degenerate ground and the nondegenerate first excited states 
(marked with arrow in Fig. \ref{Fig2}(f))
\footnote{Such interchange is also seen in the case for $n=2$
(see Appendix \ref{Appendix:example}).
}
\footnote{The three-fold degeneracy arises due to the accidental high symmetry of the 
exchange Hamiltonian: the latter is isomorphic to $O_h$ symmetry.}.
%In the present case, the ground state is three-fold degenerate while 
%the first excited state is nondegenerate, 
Because of the difference in the nature of the exchange states, 
the magnetic properties predicted by the exchange states 
of the full Hamiltonian and the $1/\bar{U}$ approximation 
%are different from the low-energy magnetic properties.
%gives rise to different predictions.
differ from each other.

%With the increase of $U_{21}$ in Eq. (\ref{Eq:JKE}), 
%the gap between the ground and the first excited levels diminish and they cross around 
%$U_{21} \approx 9.8$ eV. 
%The critical $U_{21}$ is more than three times larger than $U_{21}$ we took for the present example,
%while it is still smaller than the splitting of the $LS$ terms.
%This means that the value of $U_{21}$ adequate 
%%Then with the further evolution of $U_{21}$, the gap between the inverted exchange levels increases.
%%and the gap is further enhanced 
%%due to the limit $U\rightarrow \infty$
%%where the electronic structure in the virtual process is negligible.
%%The value of $U_{21}$ where the $1/U$ approximation applies is 
%This means that the assumption used in the $1/U$ approximation %does not seem to be physically adequate.
%is not physically adequate. 
%%Therefore, neither the Heisenberg form (\ref{Eq:JJ}) nor the exchange interaction 
%%with the $1/U$ approximation will adequately describe the magnetic properties.

There is another reason that the $1/\bar{U}$ approximation is not recommended: 
the Hund's rule coupling is completely neglected within this approximation, 
leading to the removal of the Goodenough's ferromagnetic exchange contribution
\cite{Goodenough1963} though the latter plays important role in many systems.

\section{Conclusion}
\label{Sec:Conclusion}
The main results of the present work can be summarized as follows:
\begin{enumerate}
 \item We derived the Hamiltonian of exchange interaction between $J$-multiplets ($J$-$J$)
       and between $J$-multiplet and isotropic spin ($J$-$S$) on the basis of a %complete electronic model. 
       complete electronic Hamiltonian, including the intrasite relativistic effects.
       The exchange parameters are expressed via microscopic quantities which can be extracted from 
       first principle calculations.
       Despite their microscopic character, the obtained expressions (\ref{Eq:JDE}) and (\ref{Eq:JKE}) 
       are general (i) for arbitrary choice of quantization axes on two magnetic sites 
       (which are not expected to coincide) and 
       (ii) for various magnetic ions, which can be lanthanides, actinides, transition metal ions under special conditions or any their combinations. 
       The only requirement is that the low-lying states on the sites are well approximated by crystal-field split eigenstates of a total angular momentum.
 \item The structure of the $J$-$J$ and $J$-$S$ exchange Hamiltonian is clarified 
       on the basis of derived exchange Hamiltonian.
       More specific, the maximal rank and the projections of the irreducible tensors appearing in the 
       exchange Hamiltonian are elucidated.
 \item The obtained form of the (kinetic) exchange Hamiltonian was %applied for 
       analyzed for different geometries of the bridge. 
%       magnetic complexes with different geometries.
       The relation between the geometry and the structure of the Hamiltonian %is discussed.
       was established.
 \item On the basis of considered examples, we found that the exchange spectrum in systems 
       with $J$-$J$ and $J$-$S$ interaction cannot be adequately described neither 
       by exchange Hamiltonian of isotropic form (\ref{Eq:JJ}) nor within the $1/\bar{U}$ approximation.
%       From the exchange spectra, neither the bilinear form (\ref{Eq:JJ}) nor 
%       the one within the $1/U$ approximation adequately expresses the exchange interaction.
 \item The contributions to the kinetic exchange Hamiltonian from the intermediate $J$-multiplets
       are analyzed. %This analysis shows 
       It is found that the $1/\bar{U}$ approximation exaggerates the terms from the excited states.
       Moreover, within the $1/\bar{U}$ approximation, 
       the term splitting which is larger than the average $U$ is neglected, 
       leading to %the variation of the exchange spectra including the interchange of the levels.
       the wrong order of exchange levels.
\end{enumerate}
%Given these reasons, 
In combination with {\it ab initio} and DFT extraction of microscopic electronic parameters,
the microscopic exchange Hamiltonians derived in this work can become a powerful tool for the investigation of strongly anisotropic materials containing metal ions with unquenched orbital momentum.
% 
%The obtained form of the Hamiltonian contains up to seventh order irreducible tensors 
%for $f$-electron systems, and hence the exchange Hamiltonian 
%cannot be merely described by a Heisenberg form.
%The exchange energy levels of the present Hamiltonian are compared with those of often assumed forms,
%i.e., a Heisenberg type Hamiltonian and the kinetic Hamiltonian within $1/U$ approximation.
%The low-energy states of the latter are qualitatively different from those of the exact exchange state,
%indicating that the simplified exchange interaction cannot appropriately describe the magnetic property.
%This work paves the way for the adequate description of the magnetism of the complexes and compounds with 
%unquenched orbital momentum.

\begin{acknowledgments}
N. I. would like to acknowledge the financial support from the 
the Fonds Wetenschappelijk Onderzoek - Vlaanderen (FWO) and the GOA grant from KU Leuven.
We thank Liviu Ungur for his help with {\it ab initio} calculations.
\end{acknowledgments}

\appendix
\section{Theoretical tools}
\label{Appendix:tool}
The transformation of the exchange Hamiltonian into the tensor form is done
using the theory of angular momentum \cite{Judd1967, Edmonds1974, Varshalovich1988}.
%More precisely, the reduction of the reducible operators,
%the method of equivalent operator (see for example Ref. \onlinecite{Abragam1970}).
%and the form of $J$-multiplet state. % (Eq. (4) in the main text). 
For the convenience of the readers, the tools necessary in the derivation are collected here.

\subsection{Coupling of angular momenta}
\label{Appendix:VC}
%Here, we fix the convention of phase of spherical harmonics, 
%and summarize the definitions of some symbols and the formula
%related to the coupling of the angular momenta.
For the phase of the spherical harmonics $Y_j^m(\theta, \phi)$,
we use the convention in Refs. \onlinecite{Edmonds1974, Varshalovich1988}.
With this phase convention, the complex conjugation of $Y_j^m$ is related to $Y_j^{-m}$ as 
\begin{eqnarray}
 \left[Y_j^m(\theta, \phi)\right]^* &=& (-1)^m Y_j^{-m}(\theta, \phi).
\label{Eq:Y}
\end{eqnarray}
Here, the subscript $j$ indicates the rank, the superscript $m$ is the component,
$\theta$ and $\phi$ are the spherical angular coordinates.
%$\mathbf{r} = r(\sin\theta \cos\phi, \sin\theta \sin\phi, \cos\theta)$.

%Consider the system consisting of two subsystems characterized by the angular momentum 
%(e.g. orbital and spin parts), $|Y_{j_i}^{m_i}\rangle$ $(i=1,2)$. 
%The system is also characterized by the angular momentum using 
%the Clebsch-Gordan coefficient $C_{j_1m_1j_2m_2}^{jm}$:
Consider two systems whose states are the eigenstates of the angular momentum, $|j_im_i\rangle$ $(i=1,2)$.
The coupled state characterized by the total angular momentum can be constructed using the 
Clebsch-Gordan coefficients, $C_{j_1m_1j_2m_2}^{jm}$: 
\begin{eqnarray}
 |jm\rangle &=& \sum_{m_1,m_2} |j_1m_1, j_2 m_2\rangle C_{j_1m_1j_2m_2}^{jm}.
\label{Eq:CG}
\end{eqnarray}
The Clebsch-Gordan coefficients have following symmetry properties
(Eqs. 8.4.3. (10), (11) in Ref. \onlinecite{Varshalovich1988}):
\begin{eqnarray}
 C_{j_1m_1j_2m_2}^{j_3m_3} &=& (-1)^{j_1+j_2-j_3} C_{j_2m_2j_1m_1}^{j_3m_3}
\nonumber\\
 &=& (-1)^{j_1-m_1} \frac{\Pi_{j_3}}{\Pi_{j_2}} C_{j_1m_1 j_3-m_3}^{j_2-m_2}
\nonumber\\
 &=& (-1)^{j_1-m_1} \frac{\Pi_{j_3}}{\Pi_{j_2}} C_{j_3m_3 j_1-m_1}^{j_2m_2}
\nonumber\\
 &=& (-1)^{j_2+m_2} \frac{\Pi_{j_3}}{\Pi_{j_1}} C_{j_3-m_3 j_2m_2}^{j_1-m_1}
\nonumber\\
 &=& (-1)^{j_2+m_2} \frac{\Pi_{j_3}}{\Pi_{j_1}} C_{j_2-m_2 j_3m_3}^{j_1m_1}
\nonumber\\
 &=& (-1)^{j_1+j_2-j_3}C_{j_1-m_1 j_2-m_2}^{j_3-m_3},
\label{Eq:CGsymm}
\end{eqnarray}
where $\Pi_j=\sqrt{2j+1}$.

Using the Clebsch-Gordan coefficients, 
the $6j$ and $9j$ symbols are defined as \cite{Varshalovich1988}
\begin{widetext}
\begin{eqnarray}
 \delta_{jj'}\delta_{mm'} 
 (-1)^{j_1+j_2+j_3+j} %\sqrt{(2j_{12}+1)(2j_{23}+1)}
 \Pi_{j_{12}j_{23}}
 \begin{Bmatrix}
  j_1 & j_2 & j_{12} \\
  j_3 & j & j_{23} 
 \end{Bmatrix}
 &=&
 \sum_{m_i m_{ij}} 
 C_{j_{12}m_{12}j_3m_3}^{jm} 
 C_{j_1m_1j_2m_2}^{j_{12}m_{12}} 
 C_{j_1m_1j_{23}m_{23}}^{j'm'} 
 C_{j_2m_2j_3m_3}^{j_{23}m_{23}},
\label{Eq:6j}
\end{eqnarray}
and 
\begin{eqnarray}
 \delta_{jj'}\delta_{mm'} 
 \Pi_{j_{12}j_{34}j_{13}j_{24}}
 \begin{Bmatrix}
  j_1 & j_2 & j_{12} \\
  j_3 & j_4 & j_{34} \\
  j_{13} & j_{24} & j
 \end{Bmatrix}
 &=&
%\sqrt{(2j_{12}+1)(2j_{34}+1)(2j_{13}+1)(2j_{24}+1)}
 \sum_{m_i m_{ij}} 
 C_{j_{12}m_{12}j_{34}m_{34}}^{jm} C_{j_1m_1 j_2m_2}^{j_{12}m_{12}} 
 C_{j_3m_3 j_4m_4}^{j_{34}m_{34}}
 C_{j_{13}m_{13}j_{24}m_{24}}^{j'm'} 
 C_{j_1m_1 j_3m_3}^{j_{13}m_{13}} C_{j_2m_2 j_4m_4}^{j_{24}m_{24}},
\nonumber\\
\label{Eq:9j}
\end{eqnarray}
respectively.
Here, 
$\sum_{m_i,m_{ij}}$ stands for the summation over all $m_i$ and $m_{ij}$ ($i,j=1,2,3,4$).
The $6j$ symbol (\ref{Eq:6j}) is symmetric with respect to the permutation of columns
and the interchange of the upper and lower components of two columns 
(Eq. 9.4.2. (2) in Ref. \onlinecite{Varshalovich1988}).

%The $6j$ symbol is invariant under any permutation of its columns
%or under interchange of the upper and lower arguments in each of any two columns
%(Eq. 9.4.2. (2) in Ref. \onlinecite{Varshalovich1988a}) and 
%%The $9j$ symbol (\ref{Eq:9j}) is symmetric under the permutations of two ....

From Eqs. (\ref{Eq:6j}), (\ref{Eq:9j}), we immediately obtain some formulae
involving $6j$ or $9j$ symbol.
%Multiplying $CC$ both sides of Eq. (\ref{Eq:6j}) and summing over $jm$ and $j'm'$, ...
Multiplying both sides of Eq. (\ref{Eq:6j}) by $C_{j_1m_1j_{23}m_{23}}^{j'm'}$ and summing over $j'm'$,
we obtain (Eq. 8.7.3. (12) in Ref. \onlinecite{Varshalovich1988})
\begin{eqnarray}
% (-1)^{j_1+j_2+j_3+j} %\sqrt{(2j_{12}+1)(2j_{23}+1)}
 (-1)^{j_1+j_2+j_3+j}
 \Pi_{j_{12}j_{23}}
 C_{j_1m_1j_{23}m_{23}}^{jm}
 \begin{Bmatrix}
  j_1 & j_2 & j_{12} \\
  j_3 & j & j_{23} 
 \end{Bmatrix}
 &=&
 \sum_{m_2m_3m_{12}} 
 C_{j_{12}m_{12}j_3m_3}^{jm} 
 C_{j_1m_1j_2m_2}^{j_{12}m_{12}} 
 C_{j_2m_2j_3m_3}^{j_{23}m_{23}}.
\label{Eq:sum3CG}
\end{eqnarray}
Similarly, multiplying both sides of Eq. (\ref{Eq:6j}) by 
$C_{j_{12}m_{12}j_3m_3}^{jm} C_{j_1m_1j_{23}m_{23}}^{j'm'}$ and summing over $jm,j'm'$,
we obtain (Eq. 8.7.3. (12) in Ref. \onlinecite{Varshalovich1988})
\begin{eqnarray}
 \sum_{jm}
 (-1)^{j_1+j_2+j_3+j}
 \Pi_{j_{12}j_{23}}
 %(-1)^{j_1+j_2+j_3+j} \sqrt{(2j_{12}+1)(2j_{23}+1)}
 C_{j_{12}m_{12}j_3m_3}^{jm} 
 C_{j_1m_1j_{23}m_{23}}^{jm}
 \begin{Bmatrix}
  j_1 & j_2 & j_{12} \\
  j_3 & j & j_{23} 
 \end{Bmatrix}
 &=&
 \sum_{m_2} 
 C_{j_1m_1j_2m_2}^{j_{12}m_{12}} 
 C_{j_2m_2j_3m_3}^{j_{23}m_{23}}.
\label{Eq:sum2CG}
\end{eqnarray}

%Here the unitary property (Eq. 8.1.1. (8) in Ref. \onlinecite{Varshalovich1988}) is used:
%\begin{eqnarray}
% \sum_{jm} C_{j_1m_1j_2m_2}^{jm}C_{j_1m_3j_2m_4}^{jm} &=& 
% \delta_{m_1m_3}\delta_{m_2m_4}.
%\end{eqnarray}

%Multiplying $CC$ both sides of Eq. (\ref{Eq:9j}) and summing over $jm$ and $j'm'$, ...
Multiplying both sides of Eq. (\ref{Eq:9j}) by $C_{j_{13}m_{13}j_{24}m_{24}}^{j'm'}$ 
and summing over $j'm'$, we obtain similar formula involving five Clebsch-Gordan coefficients:
\begin{eqnarray}
% \sqrt{(2j_{12}+1)(2j_{34}+1)(2j_{13}+1)(2j_{24}+1)}
 \Pi_{j_{12}j_{34}j_{13}j_{24}}
 C_{j_{13}m_{13}j_{24}m_{24}}^{jm}
 \begin{Bmatrix}
  j_1 & j_2 & j_{12} \\
  j_3 & j_4 & j_{34} \\
  j_{13} & j_{24} & j
 \end{Bmatrix}
&=&
 \sum_{m_i m_{ij}} 
 C_{j_{12}m_{12}j_{34}m_{34}}^{jm} C_{j_1m_1 j_2m_2}^{j_{12}m_{12}} 
 C_{j_3m_3 j_4m_4}^{j_{34}m_{34}}
 C_{j_1m_1 j_3m_3}^{j_{13}m_{13}} C_{j_2m_2 j_4m_4}^{j_{24}m_{24}}.
\label{Eq:sum5CG}
\end{eqnarray}
Multiplying both sides of Eq. (\ref{Eq:9j}) by 
$C_{j_{12}m_{12}j_{34}m_{34}}^{jm}  C_{j_{13}m_{13}j_{24}m_{24}}^{j'm'}$ 
and summing over $jm,j'm'$, we obtain a formula involving five Clebsch-Gordan coefficients
(Eq. 8.7.4. (26) in Ref. \onlinecite{Varshalovich1988}):
\begin{eqnarray}
% \sqrt{(2j_{12}+1)(2j_{34}+1)(2j_{13}+1)(2j_{24}+1)}
 \Pi_{j_{12}j_{34}j_{13}j_{24}}
 \sum_{jm} 
 C_{j_{12}m_{12}j_{34}m_{34}}^{jm} 
 C_{j_{13}m_{13}j_{24}m_{24}}^{jm}
 \begin{Bmatrix}
  j_1 & j_2 & j_{12} \\
  j_3 & j_4 & j_{34} \\
  j_{13} & j_{24} & j
 \end{Bmatrix}
&=&
 \sum_{m_i} 
 C_{j_1m_1 j_2m_2}^{j_{12}m_{12}} 
 C_{j_3m_3 j_4m_4}^{j_{34}m_{34}}
 C_{j_1m_1 j_3m_3}^{j_{13}m_{13}} 
 C_{j_2m_2 j_4m_4}^{j_{24}m_{24}}.
\label{Eq:sum4CG}
\end{eqnarray}
\end{widetext}

\subsection{Irreducible tensor operator}
\label{Appendix:ITO}
The irreducible tensor operator is defined as the operator $\hat{T}_{kq}$ 
which transforms as spherical harmonics $Y_k^q$ (\ref{Eq:Y}) under $SO(3)$ rotations:
\begin{eqnarray}
 \hat{R} \hat{T}_{kq} \hat{R}^\dagger &=& \sum_{q'=-k}^k \hat{T}_{kq'} D_{q'q}^k(R),
\label{Eq:RTR}
\end{eqnarray}
where $D_{q'q}^k(R) = \langle Y_k^{q'}|\hat{R}|Y_k^q\rangle$ 
is the Wigner $D$-function \cite{Varshalovich1988},
$R \in SO(3)$, and $\hat{R}$ is the rotational operator for $R$.
Since Eq. (\ref{Eq:RTR}) holds for any infinitesimal rotations, $\hat{T}_{kq}$ satisfies 
\begin{eqnarray}
 \left[\hat{J}_\mu, \hat{T}_{kq}\right] &=& \sqrt{k(k+1)} C_{kq1\mu}^{kq+\mu} \hat{T}_{kq+\mu}.
\label{Eq:JT}
\end{eqnarray}
The matrix element of $\hat{T}_{kq}$ with respect to the eigenstates of the angular momentum, 
$\{|JM\rangle\}$, is proportional to the Clebsch-Gordan coefficient (\ref{Eq:CG}).
The Wigner-Eckart theorem reads (Eq. 13.1.1. (2) in Ref. \onlinecite{Varshalovich1988}) 
\begin{eqnarray}
 \langle JM'|\hat{T}_{kq}|JM\rangle &=&
 \frac{(-1)^{2k}\langle J\Vert \hat{T}_k \Vert J\rangle}{\Pi_J} C_{JMkq}^{JM'}.
\label{Eq:WE}
\end{eqnarray}
%This fixes the phase convention of the reduced matrix elements.

One of the irreducible tensor operators
is the spherical tensor operator $Y_k^q(\hat{\mathbf{J}})$ 
which is constructed replacing the coordinates in the 
spherical harmonics $Y_k^q(\mathbf{r}/r)$ by the total angular momentum operator
($\mathbf{r}/r \rightarrow \hat{\mathbf{J}}$) and averaging it over all possible permutations
of $\hat{\mathbf{J}}$ operators \cite{Abragam1970}. 
For example, $\hat{J}_\mu \hat{J}_\nu$ is replaced by $(\hat{J}_\mu \hat{J}_\nu + \hat{J}_\nu \hat{J}_\mu)/2$.
%Here, $\mathbf{r}$ is the Cartesian coordinate.

When the system consists of two subsystems, double tensor \cite{Judd1967, Varshalovich1988} is used.
In this work, the subsystems are the orbital and the spin parts of the system.
The orbital and spin subsystems transform as irreducible tensor within 
$SO(3)$ and $SU(2)$ operations, respectively.
Thus, for the rotation of the system $R=R_1 R_2 \in SO(3) \otimes SU(2)$, 
where $R_1 \in SO(3)$ and $R_2 \in SU(2)$, 
the double tensor $\hat{T}^{k_1q_1}_{k_2q_2}$ of ranks $k_1$ and $k_2$ transform as 
\begin{eqnarray}
 \hat{R}\hat{T}^{k_1q_1}_{k_2q_2}\hat{R}^\dagger &=& 
 \sum_{q_1'=-k_1}^{k_1} \sum_{q_2'=-k_2}^{k_2} \hat{T}^{k_1q_1'}_{k_2q_2'}
 D_{q_1'q_1}^{k_1}(R_1) D_{q_2'q_2}^{k_2}(R_2),
\nonumber\\
\label{Eq:RTR_D}
\end{eqnarray}
and $\hat{T}^{k_1q_1}_{k_2q_2}$ fulfills
\begin{eqnarray}
 \left[\hat{L}_\mu, \hat{T}^{k_1q_1}_{k_2q_2}\right] &=& 
 \sqrt{k_1(k_1+1)} C_{k_1q_11\mu}^{k_1q_1+\mu} \hat{T}^{k_1q_1+\mu}_{k_2q_2},
\label{Eq:LT}
\\
 \left[\hat{S}_\mu, \hat{T}^{k_1q_1}_{k_2q_2}\right] &=& 
 \sqrt{k_2(k_2+1)} C_{k_2q_21\mu}^{k_2q_2+\mu} \hat{T}^{k_1q_1}_{k_2q_2+\mu}.
\label{Eq:ST}
\end{eqnarray}

One of the double tensor operator is the electron creation operator in atomic spin-orbital $(m,\sigma)$,
$\hat{c}_{im\sigma}^\dagger$.
This is clear since $\hat{c}_{im\sigma}^\dagger$ creates the one-electron state 
which transforms as the product of the spherical harmonics, 
$|lm,\frac{1}{2}\sigma\rangle$. %, where $l$ is the magnitude of the atomic orbital angular momentum. 
%It is irreducible in $SO(3) \otimes SU(2)$ group.
On the other hand, the annihilation operator $\hat{c}_{im\sigma}$ does not fulfill 
Eqs. (\ref{Eq:LT}) and (\ref{Eq:ST}),
%It has to be transformed as 
whereas $\bar{c}_{i-m-\sigma}$ defined below does \cite{Judd1967}:
\begin{eqnarray}
 \hat{c}_{im\sigma}
 &=& 
 (-1)^{l_i+m+\frac{1}{2}+\sigma}\bar{c}_{i-m-\sigma}.
\label{Eq:cbar}
\end{eqnarray}
Thus, $\bar{c}_{i-m-\sigma}$ instead of the annihilation operator is a double tensor.

\subsection{Method of equivalent operator}
\label{Sec:equiv}
Consider an irreducible tensor $\hat{T}_{kq}$ of rank $k$ and its argument $q$
acting on the spin degrees of freedom.
%Applying Wigner-Eckart theorem 
%\begin{eqnarray}
% \langle SM'|\hat{T}_{kq}|SM\rangle &=& \frac{(-1)^{2k} \langle S\Vert \hat{T}_k\Vert S\rangle}
% {\Pi_S} C_{SMkq}^{SM'},
%\end{eqnarray}
Replacing $J$ in Eq. (\ref{Eq:WE}) with $S$, we obtain an expression of the matrix element of $\hat{T}_{kq}$.
%where $\langle S\Vert \hat{T}_k\Vert S\rangle$ is a reduced matrix element of $\hat{T}_{kq}$,
%and $C_{SMkq}^{SM'}$ is the Clebsch-Gordan coefficient.
On the other hand, the matrix element of the spherical tensor operator $Y_k^q(\hat{\mathbf{S}})$ is 
written as 
\begin{eqnarray}
 \langle SM'|Y_k^q(\hat{\mathbf{S}})|SM\rangle &=& \frac{(-1)^{2k} 
 \langle S\Vert Y_k(\hat{\mathbf{S}})\Vert S\rangle}
 {\Pi_S} C_{SMkq}^{SM'},
\nonumber\\
\label{Eq:Ymat}
\end{eqnarray}
where $\hat{\mathbf{S}}$ is an abstract spin operator.
Comparing Eqs. (\ref{Eq:WE}) and (\ref{Eq:Ymat}), one finds the relation between 
tensor operators $\hat{T}_{kq}$ and $Y_k^q(\hat{\mathbf{S}})$:
\begin{eqnarray}
 \langle SM'|\hat{T}_{kq}|SM\rangle &=& \frac{\langle S\Vert \hat{T}_k\Vert S\rangle}
 {\langle S\Vert Y_k(\hat{\mathbf{S}})\Vert S\rangle}
 \langle SM'|Y_k^q(\hat{\mathbf{S}})|SM\rangle.
\nonumber\\
\end{eqnarray}
This equation holds for any matrix element, and hence, in the space of $\{|SM\rangle\}$, 
\begin{eqnarray}
 \hat{T}_{kq} &=& \frac{\langle S\Vert \hat{T}_k\Vert S\rangle}
 {\langle S\Vert Y_k(\hat{\mathbf{S}})\Vert S\rangle}
 Y_k^q(\hat{\mathbf{S}}).
\end{eqnarray}
The reduced matrix element in the denominator is simplified using Eq. (\ref{Eq:Ymat}) with 
$M=M'=S$ and $q=0$,
\begin{eqnarray}
 {\langle S\Vert Y_k(\hat{\mathbf{S}})\Vert S\rangle}
% &=& \frac{(-1)^{-2k}\Pi_S}{C_{SSk0}^{SS}}\langle SS|Y_k^0(\hat{\mathbf{S}})|SS\rangle
%\nonumber\\
 &=& \frac{(-1)^{-2k}\Pi_S}{C_{SSk0}^{SS}}Y_k^0(S).
\label{Eq:Y_red}
\end{eqnarray}
Consequently, $\hat{T}_{kq}$ is expressed as
\begin{eqnarray}
 \hat{T}_{kq} &=& \frac{(-1)^{2k}\langle S\Vert \hat{T}_k\Vert S\rangle}{\Pi_S}C_{SSk0}^{SS}
 \frac{Y_k^q(\hat{\mathbf{S}})}{Y_k^0(S)}.
\label{Eq:TY}
\end{eqnarray}
Eq. (\ref{Eq:TY}) holds for 
\begin{eqnarray}
 0 \le k \le 2S.
\end{eqnarray}
$Y_k^q(S)$ in the denominator of Eq. (\ref{Eq:TY}) is the scalar obtained by substituting 
$\hat{\mathbf{S}}^2=S(S+1)$ and $\hat{S}_z=S$ in the spherical harmonic tensor $Y_k^q(\hat{\mathbf{S}})$.

Now, we consider the case of double tensors $\hat{T}^{kq}_{k'q'}$ of rank $(k,k')$.
Assuming Eq. (\ref{Eq:JM}),
it is transformed into the tensor form within the space of the ground $J$-multiplet, $\{|JM\rangle\}$.
The matrix element of $\hat{T}^{kq}_{k'q'}$ is 
\begin{eqnarray}
 \langle JM'|\hat{T}^{kq}_{k'q'}|JM\rangle &=& 
 \sum_{M_L'M_S'} 
 \sum_{M_LM_S} 
 C_{LM_L'SM_S'}^{JM'} C_{LM_LSM_S}^{JM}
\nonumber\\
 &\times&
 \langle LM_L'SM_S'|\hat{T}^{kq}_{k'q'}|LM_LSM_S\rangle.
\nonumber\\
\end{eqnarray}
The Wigner-Eckart theorem (\ref{Eq:WE}) is applied to the orbital and the spin parts of the 
double tensor separately, 
\begin{eqnarray}
 \langle JM'|\hat{T}^{kq}_{k'q'}|JM\rangle &=& 
 \frac{(-1)^{2k+2k'} \langle LS\Vert \hat{T}^{k}_{k'}\Vert LS\rangle}
 {\Pi_{LS}}
\nonumber\\
 &\times&
 \sum_{M_L'M_S'} \sum_{M_LM_S} 
 C_{LM_L'SM_S'}^{JM'} 
 C_{LM_LSM_S}^{JM}
\nonumber\\
 &\times&
 C_{LM_Lkq}^{LM_L'} 
 C_{SM_Sk'q'}^{SM_S'}.
\end{eqnarray}
%Using Eq. 8.7.4. (26) in Ref. \onlinecite{Varshalovich1988},
Using Eq. (\ref{Eq:sum4CG}), the sum of the products of the Clebsch-Gordan coefficients 
reduces to the sum involving $9j$ symbol:
\begin{eqnarray}
 \langle JM'|\hat{T}^{kq}_{k'q'}|JM\rangle &=& 
 (-1)^{2k+2k'} \langle LS\Vert \hat{T}^{k}_{k'}\Vert LS\rangle 
 \sum_{nm}
 \Pi_{Jn} 
\nonumber\\
 &\times&
 C_{JMnm}^{JM'} C_{kqk'q'}^{nm} 
 \begin{Bmatrix}
  L & S & J \\
  L & S & J \\
  k & k' & n 
 \end{Bmatrix},
\end{eqnarray}
where $n$ is the rank, $m$ is its argument. 
The rest procedure is the same as the derivation of Eq. (\ref{Eq:TY}).
%Clebsch-Gordan coefficient 
$C_{JMnm}^{JM'}$ is replaced by the matrix element of 
the irreducible tensor operator $Y_n^m(\hat{\mathbf{J}})$, and 
$\hat{T}^{kq}_{k'q'}$ within $\{|JM\rangle\}$ is expressed as 
\begin{eqnarray}
 \hat{T}^{kq}_{k'q'}
 &=&
 (-1)^{2k+2k'}
 \langle LS\Vert 
 \hat{T}^{k}_{k'}
 \Vert LS\rangle
 \sum_{nm} \Pi_{Jn} 
\nonumber\\
 &\times&
 C_{JJn0}^{JJ} 
 C_{kq k'q'}^{nm}
 \begin{Bmatrix}
  L & S & J \\
  L & S & J \\
  k & k' & n 
 \end{Bmatrix}
 \frac{Y_n^m(\hat{\mathbf{J}})}{Y_n^0(J)}.
\label{Eq:T}
\end{eqnarray}
$k$, $k'$, and $n$ in Eq. (\ref{Eq:T}) obey
\begin{eqnarray}
 0 \le k \le 2L, \quad
 0 \le k' \le 2S, 
\nonumber\\
 \max[|k-k'|, 0] \le n \le \min[k+k', 2J].
\label{Eq:kkn}
\end{eqnarray}
In the derivation of Eq. (\ref{Eq:T}) we used Eq. (\ref{Eq:JM}), while it is not mandatory. 
However, without using Eq. (\ref{Eq:JM}), the result will have a more complicated form. %than Eq. (\ref{Eq:T}).
%Here, %the superscript and the subscript of $\hat{T}_{e\epsilon}^{c\gamma}$ express the 
%rank $(c,e)$ and the components $(\gamma,\epsilon)$ of the orbital and the spin parts, respectively,
%$\hat{\mathbf{J}}$ is the total angular momentum, 
%$L_i, S_i, J_i$ are the magnitude of the orbital, the spin, and the total angular momenta
%of the ground $J$-multiplet, respectively, 
%$O_k^q$ is the irreducible tensor operator 
%of rank $k$ and argument $q$ (Stevens operator),
%$C_{j_1m_1 j_2m_2}^{jm} = \langle j_1m_1j_2m_2|jm \rangle$ is Clebsch-Gordan coefficient,
%\cite{Varshalovich1988}
%$\langle L_iS_i\Vert \hat{T}_{e}^{c} \Vert L_iS_i\rangle$ is the reduced matrix element of 
%$\hat{T}_{e\epsilon}^{c\gamma}$,
%$\Pi_j=\sqrt{2j+1}$, and $\Pi_{jj'}=\Pi_j\Pi_{j'}$.
%In the derivation of Eq. (\ref{Eq:T}), formula 8.7.4. (26) in Ref. \onlinecite{Varshalovich1988} is used.

In the previous works, the so-called Stevens operator 
$O_k^q(\hat{\mathbf{J}}) = A_{kq} Y_k^q(\hat{\mathbf{J}})$ 
has been used instead of $Y_k^q$ where $A_{kq}$ is a coefficient which depends on both $k$ and $q$ 
\cite{Stevens1952}.
%Since the original form of the Stevens operator is not suitable to the Wigner-Eckart theorem (\ref{Eq:WE})
%because of the coefficient $A_{kq}$ which depends on $q$. 
However, the original Stevens operator does neither obey Eq. (\ref{Eq:JT}) nor
the Wigner-Eckart theorem (\ref{Eq:WE}).
%is not useful to make use of the Wigner-Eckart theorem (\ref{Eq:WE}).
%To avoid it, 
In order to use the Wigner-Eckart theorem, 
one can introduce such coefficient $A_{k}$ that only depends on $k$ and is independent from $q$
\cite{Chibotaru2012JCP, Chibotaru2013}.
%Another way to resolve the inconvenience is to 
Furthermore, we 
write the Stevens operator in the form of 
$O_k^q(\hat{\mathbf{J}})/O_k^0(J)$ which is equal to $Y_k^q(\hat{\mathbf{J}})/Y_k^0(J)$.
%However, each $Y_k^q(\hat{\mathbf{J}})$ is divided by $Y_k^q(J)$ in Eq. (\ref{Eq:T}), 
%and the choice of the constant is not important here.
%The form $O_k^q(\hat{\mathbf{J}})/O_k^0(J)$ is taken because it is useful to apply 
%the Wigner-Eckart theorem \cite{Varshalovich1988}:
Since the constant $A_{k}$ is canceled in this form, 
it is possible to apply the Wigner-Eckart theorem (\ref{Eq:WE}):
\begin{eqnarray}
 \langle JM'|\frac{O_k^q(\hat{\mathbf{J}})}{O_k^0(J)}|JM\rangle &=& \frac{C_{JMkq}^{JM'}}{C_{JJk0}^{JJ}}.
\end{eqnarray}
In this article, we use the latter (see for example Eq. (\ref{Eq:T})).

%\section{Interaction Hamiltonian in $J$-representation}
\section{Derivation of the exchange Hamiltonian in $J$-representation}
\label{Appendix:JJ}
\subsection{Direct exchange Hamiltonian}
\label{Appendix:DE}
Detailed calculation of $\mathcal{J}^\text{DE}$ is shown here.
Applying the method of equivalent operator (\ref{Eq:T}), 
$\{\hat{c}_i^\dagger \otimes \bar{c}_i\}^{a\alpha}_{b\beta}$ in Eq. (\ref{Eq:cc}) becomes
\begin{eqnarray}
 \left\{\hat{c}_i^\dagger \otimes \bar{c}_i\right\}^{a\alpha}_{b\beta}
 &=& 
 \sum_{kq} \Pi_{kJ_i} C_{J_iJ_ik0}^{J_iJ_i} C_{a\alpha b\beta}^{kq}
 \begin{Bmatrix}
  L_i & S_i & J_i \\
  L_i & S_i & J_i \\
  a & b & k \\
 \end{Bmatrix}
\nonumber\\
 &\times& 
 (-1)^{2a+2b} \langle L_iS_i\Vert \left\{\hat{c}_i^\dagger \otimes \bar{c}_i\right\}^a_b\Vert L_iS_i\rangle
\nonumber\\
 &\times&
 \frac{O_k^q(\hat{\mathbf{J}}_i)}{O_k^0(J_i)}.
\label{Eq:cxc_J}
\end{eqnarray}
Introducing $\mathcal{D}_{abk}^i$ defined by 
\begin{eqnarray}
 \mathcal{D}_{abk}^i &=& 
 \Pi_{kJ_i} C_{J_iJ_ik0}^{J_iJ_i} 
%C_{a\alpha b\beta}^{kq}
 \begin{Bmatrix}
  L_i & S_i & J_i \\
  L_i & S_i & J_i \\
  a & b & k \\
 \end{Bmatrix}
\nonumber\\
 &\times&
 (-1)^{2a+2b} 
 \langle L_iS_i\Vert \left\{\hat{c}_i^\dagger \otimes \bar{c}_i\right\}^a_b\Vert L_iS_i\rangle,
% {\Pi_{L_iS_i}}.
% \frac{(-1)^{2a+2b} 
% \langle L_iS_i\Vert \left\{\hat{c}_i^\dagger \otimes \bar{c}_i\right\}^a_b\Vert L_iS_i\rangle}
% {\Pi_{L_iS_i}}.
 \label{Eq:D}
\end{eqnarray}
Eq. (\ref{Eq:cxc_J}) reduces to Eq. (\ref{Eq:cxc}).

Replace $\{\hat{c}_i^\dagger \otimes \bar{c}_i\}^{a\alpha}_{b\beta}$ 
in Eq. (\ref{Eq:cc}) with Eq. (\ref{Eq:cxc}),
\begin{eqnarray}
 \hat{c}_{im\sigma}^\dagger \hat{c}_{in\sigma'}
 &=&
 (-1)^{l_i+n+\frac{1}{2}+\sigma'}
 \sum_{a\alpha b \beta} 
 \sum_{kq} C_{a\alpha b\beta}^{kq} \mathcal{D}^i_{abk} 
\nonumber\\
 &\times&
 C_{l_iml_i-n}^{a\alpha} C_{\frac{1}{2}\sigma \frac{1}{2}-\sigma'}^{b\beta}
 \frac{O_k^q(\hat{\mathbf{J}}_i)}{O_k^0(J_i)}.
\label{Eq:cc_J}
\end{eqnarray}
Substituting Eq. (\ref{Eq:cc_J}) in the direct exchange Hamiltonian (\ref{Eq:HDE}), 
\begin{eqnarray}
 \hat{H}_{\rm DE} &=& 
 -\sum_{kq} 
 \sum_{k'q'} 
 \sum_{a\alpha b \beta} 
 \sum_{a'\alpha' b' \beta'} 
\nonumber\\
 &\times&
 \sum_{mnm'n'}
  V_{mm'n'n} 
 (-1)^{l_1+n} (-1)^{l_2+n'}
\nonumber\\
 &\times&
 \sum_{\sigma \sigma'}
 (-1)^{\frac{1}{2}+\sigma'} (-1)^{\frac{1}{2}+\sigma}
 C_{\frac{1}{2}\sigma \frac{1}{2}-\sigma'}^{b\beta}
 C_{\frac{1}{2}\sigma' \frac{1}{2}-\sigma}^{b'\beta'}
\nonumber\\
 &\times&
 C_{a\alpha b\beta}^{kq} 
 C_{l_1ml_1-n}^{a\alpha} 
 C_{a'\alpha' b'\beta'}^{k'q'} 
 C_{l_2m'l_2-n'}^{a'\alpha'} 
\nonumber\\
 &\times&
\mathcal{D}^1_{abk} \mathcal{D}^2_{a'b'k'} 
 \frac{O_k^q(\hat{\mathbf{J}}_1) O_{k'}^{q'}(\hat{\mathbf{J}}_2)}{O_k^0(J_1) O_{k'}^0(J_2)}.
\end{eqnarray}
Since $(-1)^{\frac{1}{2}+\sigma}=(-1)^{-\frac{1}{2}-\sigma}$, $(-1)^{\sigma'-\sigma} =(-1)^{-\beta}$, and
%\begin{eqnarray}
$\sum_{\sigma \sigma'}
 C_{\frac{1}{2}\sigma \frac{1}{2}-\sigma'}^{b\beta}
 C_{\frac{1}{2}\sigma \frac{1}{2}-\sigma'}^{b'-\beta'}
%&=&
= \delta_{bb'} \delta_{\beta, -\beta'}$,
%\end{eqnarray}
using Eq. (\ref{Eq:V}),
\begin{eqnarray}
 \hat{H}_{\rm DE} &=& 
 -\sum_{kqk'q'} \sum_{aa'b} \mathcal{V}_{kqk'q'}^{aa'b}
%\nonumber\\
% &\times&
% \sum_{mnm'n'}
% \sum_{\alpha \alpha' \beta} 
%  V_{mm'n'n} 
% (-1)^{l_1+l_2+n+n'+\beta}
%\nonumber\\
% &\times&
% C_{a\alpha b\beta}^{kq} 
% C_{l_1ml_1-n}^{a\alpha} 
% C_{a'\alpha' b\beta}^{k'q'} 
% C_{l_2m'l_2-n'}^{a'\alpha'} 
%\nonumber\\
% &\times&
\mathcal{D}^1_{abk} \mathcal{D}^2_{a'bk'} 
 \frac{O_k^q(\hat{\mathbf{J}}_1) O_{k'}^{q'}(\hat{\mathbf{J}}_2)}
 {O_k^0(J_1) O_{k'}^0(J_2)}.
\nonumber\\
\end{eqnarray}
%From this form, the direct exchange coupling constant (\ref{Eq:JDE}) is obtained.
The coefficient of the operators is Eq. (\ref{Eq:JDE}).

%and the ranges of the ranks $a,b,k$ %in Eqs. (\ref{Eq:cc}), (\ref{Eq:cxc}) satisfy 
%$0 \le a \le 2\min[l_i, L_i], 0 \le b \le 2\min[1/2, S_i]$ and for a given set of $a,b$, 
%$k$ holds $|a-b| \le k \le \min[a+b, 2J_i]$.
%In the ground $LS$ term of the magnetic ion, 
%$L_i \ge l_i$ and $S_i \ge 1/2$ except for the case of the half-filling 
%($L_i=0$ and $J_i=S_i$).
%Thus, the ranks $a,b,k$ hold
%are given as 
%\begin{eqnarray}
% 0 \le a \le 2\min[l_i, L_i], \quad
% 0 \le b \le 1, 
%\nonumber\\
% 0 \le k \le \min[2l_i+1, 2L_i+1, 2J_i].
%\label{Eq:abk_DE}
%\end{eqnarray}
%From the definition of $a,b$ and Eq. (\ref{Eq:cc_J}), 
The ranks for the orbital ($a$) and the spin ($b$) parts of 
%the product of the creation and the annihilation operators 
$\{\hat{c}^\dagger_i \otimes \bar{c}_i\}^{a\alpha}_{b\beta}$
are bounded by $2l_i$ and $2 \times 1/2$, respectively.
Moreover, from the $9j$ symbol in Eq. (\ref{Eq:cxc_J}),
%$a,b$ are less than or equal to $2L_i$ and $2S_i$ 
$a \le 2L_i$ and $b \le 2S_i$ where $L_i$ and $S_i$ are the $LS$-term 
for the ground $J$-multiplet states, respectively. 
Thus, the ranges of ranks $a,b$ are given as 
\begin{eqnarray}
 0 \le a \le 2 \min[l_i, L_i],
\quad
 0 \le b \le 2 \min[1/2, S_i],
\label{Eq:ab_DE}
\end{eqnarray}
The maximum of $b$ is 1 because $S_i \ge 1/2$ for the magnetic ions.
For given $(a,b)$, $k$ is at most $a+b$.
Simultaneously $k$ is less than or equal to $2J_i$ (\ref{Eq:cxc_J}).
%On the other hand, $k$ is less than $2J_i$, and hence, 
Therefore, the range of $k$ is 
\begin{eqnarray}
 0 \le k \le \min[2l_i+1, 2L_i+1, 2J_i].
\label{Eq:k_DE}
\end{eqnarray}

\subsection{Kinetic exchange Hamiltonian}
\label{Appendix:KE}
In the kinetic exchange Hamiltonian $\hat{H}_{\rm KE}$, 
the operators appear as the form of 
$\hat{c}_{im\sigma}^\dagger \hat{P}^{N_i-1}_{i\alpha_J J} \hat{c}_{im'\sigma'}$ and
$\hat{c}_{im\sigma} \hat{P}^{N_i+1}_{i\alpha_J J} \hat{c}_{im'\sigma'}^\dagger$.
One should note that the projection operator $\hat{P}^N_{i\alpha_J J}$ 
is totally symmetric within $SO(3)$ group, whereas reducible within the $SO(3) \otimes SU(2)$ group.
Thus, $\hat{P}^N_{i\alpha_J J}$ is reduced within $SO(3) \otimes SU(2)$ group 
in order to simultaneously treat it with the other double tensors. 
With the use of Eq. (\ref{Eq:JM}),
the projection operator $\hat{P}^{N}_{i\alpha_J J} = \sum_M |iN\alpha_J JM\rangle \langle iN\alpha_J JM|$ is 
\begin{eqnarray}
\hat{P}^{N}_{i\alpha_J J} &=& 
                         %\sum_{M=-J}^J \sum_{M_L,M_L'=-L}^L \sum_{M_S,M_S'=-S}^S
                         \sum_M \sum_{M_L,M_S} \sum_{M_L',M_S'}
                         C_{LM_LSM_S}^{JM} 
                        C_{LM'_LSM'_S}^{JM} 
\nonumber\\
                       &\times& 
                        |iN \alpha_{LS} LM_LSM_S\rangle \langle iN \alpha_{LS} LM'_LSM'_S|.
%\nonumber\\
\end{eqnarray}
%transformed into the double tensor:
Introducing the irreducible double tensor $\hat{P}_{i\alpha_J a \alpha a'\alpha'}^N$ defined by 
\begin{eqnarray}
 \hat{P}_{i\alpha_J a \alpha a'\alpha'}^N
 &=&
 \sum_{m_Lm_S}\sum_{m_L'm_S'} (-1)^{L+m_L'+S+m_S'}
 C_{Lm_LL-m_L'}^{a\alpha} 
\nonumber\\
 &\times&
 C_{Sm_SS-m_S'}^{a'\alpha'} 
 |Lm_LSm_S\rangle \langle Lm_L'Sm_S'|,
\label{Eq:P2}
\end{eqnarray}
the projection operator $\hat{P}_{i\alpha_J J}^{N}$ is written as 
\begin{eqnarray}
\hat{P}^{N}_{i\alpha_J J} &=& %\sum_{M=-J}^J 
                         \sum_{M} \sum_{m_L m_S} \sum_{m_L' m_S'}
                         (-1)^{L+m_L'+S+m_S'} 
                         C_{Lm_LSm_S}^{JM} 
\nonumber\\
                    &\times&
                         C_{Lm'_LSm'_S}^{JM} 
                         \sum_{a\alpha, a'\alpha'}
%\nonumber\\
%                       &\times& |N\mu Lm_LSm_S\rangle \langle N\mu Lm'_LSm'_S|.
%                    &\times&
                         C_{Lm_LL-m_L'}^{a\alpha} C_{Sm_SS-m_S'}^{a'\alpha'} 
                        \hat{P}_{i\alpha_J  a \alpha a'\alpha'}^{N}.
\nonumber\\
\end{eqnarray}
Using the symmetry properties of the Clebsch-Gordan coefficients (\ref{Eq:CGsymm}) and 
the $6j$ symbol (\ref{Eq:6j}),
%(Eqs. 8.4.3. (10), (11) and 9.1 (5) in Ref. \onlinecite{Varshalovich1988}), 
\begin{eqnarray}
 \hat{P}^N_{i\alpha_J J} &=& \sum_{a \alpha} (-1)^{J-L-S+\alpha} \Pi_{JJ}
 \begin{Bmatrix}
  S & S & a \\
  L & L & J
 \end{Bmatrix}
 \hat{P}_{i\alpha_J a \alpha a-\alpha}^N,
\nonumber\\
\label{Eq:P1}
\end{eqnarray}
The range of the rank $a$ in Eq. (\ref{Eq:P1}) is 
\begin{eqnarray}
 0 \le a \le 2\min[S,L],
\label{Eq:a_KE}
\end{eqnarray}
and $-a \le \alpha \le a$.
%In the derivation of Eq. (\ref{Eq:P1}), the definition of the $6j$ symbol 
%(Eq. 9.1. (5) in Ref. \onlinecite{Varshalovich1988}) is used.
Substituting Eq. (\ref{Eq:P1}) into %the operator part, 
$\hat{c}_{im\sigma}^\dagger \hat{P}_{i\alpha_J J}^{N_i-1} \hat{c}_{in\sigma'}
 = (-1)^{l_i+n+\frac{1}{2}+\sigma'}\hat{c}_{im\sigma}^\dagger \hat{P}_{i\alpha_J J}^{N_i-1} \bar{c}_{i-n-\sigma'}$, 
\begin{widetext}
%in Eq. (\ref{Eq:HKE}),
\begin{eqnarray}
\hat{c}_{im\sigma}^\dagger \hat{P}_{i\alpha_J J}^{N_i-1} \hat{c}_{in\sigma'} 
 &=& 
 \sum_{a \alpha} (-1)^{J-L-S+\alpha} \Pi_{JJ}
 \begin{Bmatrix}
  S & S & a \\
  L & L & J
 \end{Bmatrix}
%\nonumber\\
% &\times&
 (-1)^{l_i+n+\frac{1}{2}+\sigma'}\hat{c}_{im\sigma}^\dagger 
 \hat{P}_{i\alpha_J a \alpha a-\alpha}^{N_i-1}
 \bar{c}_{i-n-\sigma'}.
\label{Eq:c+Pc1}
\end{eqnarray}
%Now, the operator part in Eq. (\ref{Eq:c+Pc1}) is a product of three double tensors,
The operator in Eq. (\ref{Eq:c+Pc1}),
$\hat{c}_{im\sigma}^\dagger 
 \hat{P}_{i\alpha_J a \alpha a-\alpha}^{N_i-1}
 \bar{c}_{i-n-\sigma'}$,
is reduced as follows:
%The product is reduced to irreducible double tensor as follows:
%\begin{widetext}
\begin{eqnarray}
 \hat{c}_{im\sigma}^\dagger 
 \hat{P}_{i\alpha_J a \alpha a-\alpha}^{N_i-1}
 \bar{c}_{i-n-\sigma'}
 &=&
 \sum_{b\beta d\delta}  
 C_{a\alpha l_i -n}^{b\beta}
 C_{a-\alpha \frac{1}{2} -\sigma'}^{d\delta}
 \hat{c}_{im\sigma}^\dagger 
 \left\{
 \hat{P}_{i\alpha_J a a}^{N_i-1}
 \otimes
 \bar{c}_{i}
 \right\}^{b\beta}_{d\delta}
\nonumber\\
 &=&
 \sum_{b\beta c\gamma d\delta e\epsilon}  
 C_{a\alpha l_i -n}^{b\beta}
 C_{a-\alpha \frac{1}{2} -\sigma'}^{d\delta}
 C_{l_i m b\beta}^{c\gamma} 
 C_{\frac{1}{2} \sigma d\delta}^{e\epsilon} 
 \left\{
 \hat{c}_{i}^\dagger \otimes
 \left\{
 \hat{P}_{i\alpha_J a a}^{N_i-1}
 \otimes
 \bar{c}_{i}
 \right\}^{b}_{d}
 \right\}^{c\gamma}_{e\epsilon}.
\end{eqnarray}
Using Eqs. (\ref{Eq:CGsymm}) and (\ref{Eq:sum2CG}), 
\begin{eqnarray}
 \hat{c}_{im\sigma}^\dagger 
 \hat{P}_{i\alpha_J a \alpha a-\alpha}^{N_i-1}
 \bar{c}_{i-n-\sigma'}
 &=&
 \sum_{b\beta c\gamma d e\epsilon}%  f\phi}
 C_{a\alpha l_i -n}^{b\beta}
 C_{l_i m b\beta}^{c\gamma} 
%\nonumber\\
 %&\times&
 \sum_{f\phi}
 (-1)^{\frac{1}{2}+2a+\alpha+d+f}\Pi_{de}
%\nonumber\\
% &&\times
 \begin{Bmatrix}
  \frac{1}{2} & \frac{1}{2} & f\\
  a & e & d\\
 \end{Bmatrix}
\nonumber\\
 &\times&
 C_{e\epsilon a\alpha}^{f\phi} 
 C_{\frac{1}{2}\sigma \frac{1}{2}-\sigma'}^{f\phi} 
 \left\{
 \hat{c}_{i}^\dagger \otimes
 \left\{
 \hat{P}_{i\alpha_J a a}^{N_i-1}
 \otimes
 \bar{c}_{i}
 \right\}^{b}_{d}
 \right\}^{c\gamma}_{e\epsilon}.
\label{Eq:c+Pc_irrep}
\end{eqnarray}
Here, the ranges of the ranks $b,c,d,e,f$ are 
\begin{eqnarray}
 |a-l_i| \le b \le a + l_i, \quad
 |a-1/2| \le d \le a + 1/2, \quad
 |b-l_i| \le c \le b + l_i, 
\nonumber\\
 |d-1/2| \le e \le d + 1/2, \quad
 \max[|a-e|, 0] \le f \le \min[a+e, 2 \times 1/2].
\label{Eq:bcdef_KE}
\end{eqnarray}
and their arguments satisfy 
$-b \le \beta \le b$,
$-d \le \delta \le d$,
$-c \le \gamma \le c$,
$-e \le \epsilon \le e$,
$-f \le \phi \le f$, respectively.
Note that $f$ is at the largest 1.
%In the last transformation, we used the symmetry property of the Clebsch-Gordan coefficient and 
%formula 8.7.3. (12) in Ref. \onlinecite{Varshalovich1988}.
Substituting Eq. (\ref{Eq:c+Pc_irrep}) into Eq. (\ref{Eq:c+Pc1}), 
%\begin{widetext}
\begin{eqnarray}
\hat{c}_{im\sigma}^\dagger \hat{P}_{i\alpha_J J}^{N_i-1} \hat{c}_{in\sigma'} 
 &=& 
 (-1)^{l_i+n+\frac{1}{2}+\sigma'} (-1)^{J-L-S}\Pi_{JJ}
%\nonumber\\
% &\times& 
 \sum_{a \alpha b \beta c\gamma d e \epsilon f \phi} 
 (-1)^{\frac{1}{2}+d+f}
 \Pi_{de}
\nonumber\\
 &\times&
 \begin{Bmatrix}
  S & S & a \\
  L & L & J
 \end{Bmatrix}
 \begin{Bmatrix}
  \frac{1}{2} & \frac{1}{2} & f \\
  a & e & d
 \end{Bmatrix}
 C_{a\alpha l_i-n}^{b\beta}C_{l_im b\beta}^{c\gamma} 
 C_{e\epsilon a\alpha}^{f\phi} 
 C_{\frac{1}{2}\sigma \frac{1}{2}-\sigma'}^{f\phi}
%\nonumber\\
% &\times&
 \left\{
 \hat{c}_{i}^\dagger 
 \otimes
 \left\{
 \hat{P}_{i\alpha_J a a}^{N_i-1}
 \otimes
 \bar{c}_{i}
 \right\}_d^b
 \right\}_{e\epsilon}^{c\gamma}.
\label{Eq:c+Pc}
\end{eqnarray}
Similarly, the operator for the other site in $\hat{H}_\text{KE}$ (\ref{Eq:HKE}) becomes
\begin{eqnarray}
\hat{c}_{jm'\sigma} \hat{P}_{j\alpha_J' J'}^{N_j+1} \hat{c}_{jn'\sigma'}^\dagger
 &=&
 (-1)^{l_j+m'+\frac{1}{2}+\sigma}(-1)^{J'-L'-S'}\Pi_{J'J'}
 \sum_{a\alpha b\beta c\gamma d e\epsilon f\phi} (-1)^{-\frac{1}{2}+d}\Pi_{de}
\nonumber\\
 &\times&
\begin{Bmatrix}
 S'& S'& a\\
 L'& L'& J'
\end{Bmatrix}
\begin{Bmatrix}
 \frac{1}{2} & \frac{1}{2} & f\\
 a & e & d\\
\end{Bmatrix}
 C_{a\alpha l_jn'}^{b\beta} C_{l_j-m' b\beta}^{c\gamma} C_{e\epsilon a\alpha}^{f\phi} 
 C_{\frac{1}{2}\sigma \frac{1}{2}-\sigma}^{f-\phi}
 \left\{
  \bar{c}_{j}
  \otimes
 \left\{
  \hat{P}_{j\alpha_J' a a}^{N_j+1}
  \otimes
  \hat{c}_{j}^\dagger
 \right\}^b_d
 \right\}^{c\gamma}_{e\epsilon}.
\label{Eq:cPc+}
\end{eqnarray}

The operators in Eq. (\ref{Eq:c+Pc}) is written %and (\ref{Eq:cPc+}) are written 
in terms of the total angular momentum using the method of equivalent operator.
Applying Eq. (\ref{Eq:T}) to the irreducible tensor,
$\{\hat{c}_{i}^\dagger \otimes \{\hat{P}_{i\alpha_J a a}^{N_i-1} \otimes \bar{c}_{i}\}_d^b\}_{e\epsilon}^{c\gamma}$
in Eq. (\ref{Eq:c+Pc}),
%and 
%$\{\bar{c}_{j} \otimes \{\hat{P}_{j\nu a a}^{N_j+1} \otimes \hat{c}_{j}^\dagger\}^b_d\}^{c\gamma}_{e\epsilon}$
%in Eq. (\ref{Eq:cPc+}), 
\begin{eqnarray}
\hat{c}_{im\sigma}^\dagger \hat{P}_{i\alpha_J J}^{N_i-1} \hat{c}_{in\sigma'} 
 &=& 
 (-1)^{l_i+n+\frac{1}{2}+\sigma'} (-1)^{J-L-S} %\Pi_{JJ}
 \sum_{a \alpha b \beta c\gamma d e \epsilon f \phi} 
 (-1)^{\frac{1}{2}+d+f}
 \Pi_{JJde}
\nonumber\\
 &\times&
 \begin{Bmatrix}
  S & S & a \\
  L & L & J
 \end{Bmatrix}
 \begin{Bmatrix}
  \frac{1}{2} & \frac{1}{2} & f \\
  a & e & d
 \end{Bmatrix}
 C_{a\alpha l_i-n}^{b\beta}C_{l_im b\beta}^{c\gamma} 
 C_{e\epsilon a\alpha}^{f\phi} 
 C_{\frac{1}{2}\sigma \frac{1}{2}-\sigma'}^{f\phi}
 \sum_{kq}
 \Pi_{J_ik} C_{J_iJ_ik0}^{J_iJ_i} C_{c\gamma e\epsilon}^{kq} 
 \begin{Bmatrix}
  L_i & S_i & J_i \\
  L_i & S_i & J_i \\
  c   & e   & k 
 \end{Bmatrix}
\nonumber\\
 &\times&
(-1)^{2c+2e}
\langle L_iS_i\Vert
 \left\{
 \hat{c}_{i}^\dagger 
 \otimes
 \left\{
 \hat{P}_{i\alpha_J a a}^{N_i-1}
 \otimes
 \bar{c}_{i}
 \right\}_d^b
 \right\}_{e}^{c}
\Vert L_iS_i\rangle
 \frac{O_k^q(\hat{\mathbf{J}}_i)}{O_k^0(J_i)}.
\label{Eq:c+Pc_J}
\end{eqnarray}
%using the symmetry of the Clebsch-Gordan coefficients (\ref{Eq:CGsymm}) 
%and Eq. (\ref{Eq:sum4CG}), we obtain
From Eq. (\ref{Eq:kkn}), $c$, $e$, and $k$ satisfy additional conditions:
\begin{eqnarray}
 0 \le c \le 2L_i, \quad
 0 \le e \le 2S_i, \quad
 \max[|c-e|, 0] \le k \le \min[c+e, 2J_i].
\label{Eq:cek}
\end{eqnarray}
The Clebsch-Gordan coefficients are replaced by the sum involving $9j$ symbol (\ref{Eq:sum4CG}):
\begin{eqnarray}
\hat{c}_{im\sigma}^\dagger \hat{P}_{i\alpha_J J}^{N_i-1} \hat{c}_{in\sigma'} 
 &=& 
 (-1)^{l_i+\frac{1}{2}+\sigma'} 
 \sum_{f\phi}\sum_{kq}\sum_{x\xi}
 (-1)^{2f} 
 C_{l_inkq}^{x\xi} C_{f\phi l_im}^{x\xi} 
 C_{\frac{1}{2}\sigma \frac{1}{2}-\sigma'}^{f\phi}
 \mathcal{F}_{\alpha_J Jfxk}^i
 \frac{O_k^q(\hat{\mathbf{J}}_i)}{O_k^0(J_i)}.
\label{Eq:c+Pc_OJ}
\end{eqnarray}
Similarly,
\begin{eqnarray}
\hat{c}_{jm'\sigma} \hat{P}_{j\alpha_J' J'}^{N_j+1} \hat{c}_{jn'\sigma'}^\dagger 
 &=& 
 (-1)^{l_j+\frac{1}{2}+\sigma+m'+n'} %(-1)^{J-L-S}
 \sum_{f\phi}\sum_{kq}\sum_{x\xi}
 (-1)^{-1+f}
 C_{l_j-n'kq}^{x\xi} C_{f\phi l_j-m'}^{x\xi} 
 C_{\frac{1}{2}\sigma \frac{1}{2}-\sigma'}^{f-\phi} %C_{J_jJ_jk0}^{J_jJ_j} 
 \mathcal{G}_{\alpha_J' Jfxk}^j
 \frac{O_k^q(\hat{\mathbf{J}}_j)}{O_k^0(J_j)}.
\label{Eq:cPc+_OJ}
\end{eqnarray}
Here, $\mathcal{F}^{i}_{\alpha_J Jfxk}$ in Eq. (\ref{Eq:c+Pc_OJ}) is defined by 
\begin{eqnarray}
 \mathcal{F}^{i}_{\alpha_J Jfxk}
 &=& 
 \sum_{a b c d e} 
 (-1)^{J-L-S+\frac{1}{2}+b+c+d-k}
 \Pi_{JJkkJ_ibcdef}
 C_{J_iJ_ik0}^{J_iJ_i}
 \begin{Bmatrix}
  S & S & a \\
  L & L & J
 \end{Bmatrix}
 \begin{Bmatrix}
  \frac{1}{2} & \frac{1}{2} & f \\
  a & e & d
 \end{Bmatrix}
 \begin{Bmatrix}
  a & b & l_i\\
  e & c & k \\
  f & l_i & x
 \end{Bmatrix}
 \begin{Bmatrix}
  L_i & S_i & J_i \\
  L_i & S_i & J_i \\
  c & e & k \\
 \end{Bmatrix}
\nonumber\\
 &\times&
 (-1)^{2c+2e}
 \langle L_iS_i\Vert
 \left\{\hat{c}_i^\dagger \otimes
 \left\{\hat{P}_{i\alpha_J aa}^{N_i-1}\otimes \bar{c}_i \right\}^b_d
 \right\}^{c}_{e}
 \Vert L_iS_i\rangle,
\label{Eq:F}
\end{eqnarray}
and $\mathcal{G}^{i}_{\alpha_J Jfxk}$ is obtained by replacing 
the reduced matrix element in Eq. (\ref{Eq:F}) by
$\langle L_iS_i\Vert
\{\bar{c}_i \otimes
 \{\hat{P}_{i\alpha_J aa}^{N_i+1}\otimes \hat{c}_i^\dagger \}^b_d
 \}^{c}_{e}
 \Vert L_iS_i\rangle$. 
The ranges of $x$ and its component $\xi$ are 
\begin{eqnarray}
 |f-l_i| \le x \le f+l_i, 
\label{Eq:x_KE}
\end{eqnarray}
and $-x \le \xi \le x$, respectively.

Substituting Eqs. (\ref{Eq:c+Pc_OJ}) and (\ref{Eq:cPc+_OJ}) into $\hat{H}_{\rm KE}$ (\ref{Eq:HKE}),
and first summing over $\sigma$ and $\sigma'$, the operator part of the numerator becomes 
\begin{eqnarray}
 \sum_{\sigma \sigma'}
 \left(\hat{c}_{im\sigma}^\dagger \hat{P}^{N_i-1}_{i\alpha_J J} \hat{c}_{in\sigma'}\right)
 \left(\hat{c}_{jm'\sigma} \hat{P}^{N_j+1}_{j\alpha_J' J'} \hat{c}_{jn'\sigma'}^\dagger \right)
 &=&
 -\sum_{f\phi}\sum_{kq}\sum_{x\xi}
 \sum_{k'q'}\sum_{x'\xi'}
 (-1)^{l_i-l_j-f+q'}
 \mathcal{F}_{\alpha_J Jfxk}^i
 \mathcal{G}_{\alpha_J' J'fx'k'}^j
\nonumber\\
 &\times&
 C_{l_inkq}^{x\xi} C_{f\phi l_im}^{x\xi} 
 C_{l_j-n'k'q'}^{x'\xi'} C_{f-\phi  l_j-m'}^{x'\xi'} 
 \frac{O_k^q(\hat{\mathbf{J}}_i) O_{k'}^{q'}(\hat{\mathbf{J}}_j)}{O_k^0(J_i)O_{k'}^0(J_j)}.
\end{eqnarray}
The kinetic exchange Hamiltonian (\ref{Eq:HKE}) is 
\begin{eqnarray}
 \hat{H}_\text{KE} &=& 
 \sum_{i\ne j} 
 \sum_{kq}
 \sum_{k'q'}
 \sum_{fxx'}
 \sum_{\alpha_J J} 
 \sum_{\alpha_J' J'} 
 \frac{\sum_{mn} \sum_{m'n'} \sum_{\xi \xi' \phi} (-1)^{l_i-l_j-f+q'} t_{mm'}^{ij} t_{n'n}^{ji}
 C_{l_inkq}^{x\xi} C_{f\phi l_im}^{x\xi} C_{l_j-n'k'q'}^{x'\xi'} C_{f-\phi  l_j-m'}^{x'\xi'}}
 {U_{ij} + \Delta E_{i\alpha_J J}^{N_i-1} + \Delta E_{j\alpha_J' J'}^{N_j+1}}
\nonumber\\
 &\times&
 \mathcal{F}_{\alpha_J Jfxk}^i
 \mathcal{G}_{\alpha_J' J'fx'k'}^j
 \frac{O_k^q(\hat{\mathbf{J}}_i) O_{k'}^{q'}(\hat{\mathbf{J}}_j)}{O_k^0(J_i)O_{k'}^0(J_j)}.
\end{eqnarray}
\end{widetext}
The numerator is replaced by $\{t \times t\}_{kqk'q'}^{fxx'}$ (\ref{Eq:tt}), and we obtain Eq. (\ref{Eq:JKE}).

The range of $k$ in Eqs. (\ref{Eq:c+Pc_OJ}) and (\ref{Eq:cPc+_OJ})
is $k \le \min[c+e, 2J_i, x + l_i]$ from Eqs. (\ref{Eq:a_KE}), (\ref{Eq:bcdef_KE}), %, (\ref{Eq:x_KE}).
and (\ref{Eq:F}).
Since $0 \le f \le 1$, $c+e \le 2a+2l_i+1$, and $x \le l_i + 1$, the range of $k$ becomes
\begin{eqnarray}
 0 \le k \le \min[2l_i+1, 2J_i].
\label{Eq:k_KE}
\end{eqnarray}
%From Eqs. (\ref{Eq:c+Pc_OJ}) and (\ref{Eq:cPc+_OJ}), we also obtain the relation of $c,e,k$:
%\begin{eqnarray}
% (-1)^{c+e+k}=1.
%\label{Eq:cek_KE}
%\end{eqnarray}

The range of $q$ is restricted by the transfer parameter as well as the maximal $k$ ($k_\text{max}$).
Considering the conservation law for the arguments of Clebsch-Gordan coefficients in Eq. (\ref{Eq:tt}),
$q=\phi+m+n$, and $|\phi| \le 1$,
\begin{eqnarray}
 |q| \le \min[k_\text{max}, 2m_\text{max}+1],
\end{eqnarray}
where $m_\text{max} (\ge 0)$ is the maximum projection of the
magnetic orbital that contributes to the electron transfer.
%In Eqs. (\ref{Eq:c+Pc_OJ}), (\ref{Eq:cPc+_OJ}), the nonzero terms satisfy simultaneously
%Eqs. (\ref{Eq:a_KE}), (\ref{Eq:bcdef_KE}), (\ref{Eq:x_KE}), (\ref{Eq:k_KE}), (\ref{Eq:cek_KE}).

\section{Property of $\mathcal{J}_{kqk'q'}$}
\label{Appendix:J}
By using the Hermiticity of $\bar{H}$ (\ref{Eq:Hbar_tensor}) and 
\begin{eqnarray}
 \left[{O}_k^q(\hat{\mathbf{J}})\right]^\dagger 
  &=& (-1)^{-q}
 {O}_k^{-q}(\hat{\mathbf{J}}),
\end{eqnarray}
we obtain 
\begin{eqnarray}
 (-1)^{-q-q'} \left(\mathcal{J}_{kqk'q'}\right)^*
 &=& \mathcal{J}_{k-qk'-q'}.
\label{Eq:Jsymm1}
\end{eqnarray}
On the other hand, using the time-reversal symmetry of $\bar{H}$ and 
\begin{eqnarray}
 \theta {O}_k^q(\hat{\mathbf{J}})\theta^{-1}
  &=& 
 \left[{O}_k^q(-\hat{\mathbf{J}})\right]^*
 = 
 (-1)^{k-q}
 {O}_k^{-q}(\hat{\mathbf{J}}),
\end{eqnarray}
where $\theta$ is time-reversal operator \cite{Abragam1970},
we obtain 
\begin{eqnarray}
 (-1)^{k+k'-q-q'} \left(\mathcal{J}_{kqk'q'}\right)^*
 &=& \mathcal{J}_{k-qk'-q'}.
\label{Eq:Jsymm2}
\end{eqnarray}
Comparing Eq. (\ref{Eq:Jsymm1}) and Eq. (\ref{Eq:Jsymm2}), 
\begin{eqnarray}
 (-1)^{k+k'} = 1.
\end{eqnarray}
Therefore, both of $k$ and $k'$ are even or odd.

\section{Exchange Hamiltonians for $J$-multiplet interacting with isotropic spin}
\label{Appendix:JS}
The exchange Hamiltonian between $J$-multiplet and isotropic spin is obtained 
replacing orbital angular momentum of the spin site ($i=2$) with zero.
When the spin state consists of some nondegenerate molecular orbitals, 
the orbital indices $r$ are introduced.
%Alternative way to obtain the exchange Hamiltonian is direct derivation.
%Here, we show the latter for spin part. 

\subsection{Direct exchange Hamiltonian}
Since $l_2=0$ and $L_2=0$, $J_2=S_2$, $a'=0$, $b=k'$, and 
the $9j$ symbol in $\mathcal{D}_{a'bk'}^2$ (\ref{Eq:D}) 
reduces to $1/\Pi_{S_2S_2k'}$. 
Therefore, $\mathcal{D}_{a'bk'}^2$ becomes
\begin{eqnarray}
 \tilde{\mathcal{D}}^2_{rr'k'} &=& C_{S_2S_2k'0}^{S_2S_2} 
 \frac{(-1)^{2k'}\langle S_2\Vert \left\{\hat{c}_{2r}^\dagger \otimes \bar{c}_{2r'}\right\}_{k'} 
 \Vert S_2\rangle}{\Pi_{S_2}}.
\nonumber\\
% &=&
% \langle S_2S_2|\left\{\hat{c}_{2r}^\dagger \otimes \bar{c}_{2r'}\right\}_{k'0} 
% |S_2S_2\rangle.
\label{Eq:Dtilde}
\end{eqnarray}
Here, $a'$ and $b$ are omitted for simplicity,
and the molecular orbital index $r$ is introduced.
%and Wigner-Eckart theorem (\ref{Eq:WE}) is used.
%Note that $\langle S_2S_2|\left\{\hat{c}_{2r}^\dagger \otimes \bar{c}_{2r'}\right\}_{k'0}|S_2S_2\rangle$  
%is zero when $r\ne r'$ because all the molecular orbitals taken into account are occupied 
%by up spin electrons, otherwise the low-energy states of site 2 are not described by isotropic spin.
On the other hand, $\mathcal{V}^{arr'}_{kqk'q'}$ is 
\begin{eqnarray}
 \mathcal{V}^{arr'}_{kqk'q'} &=&
 \sum_{mn}\sum_{\alpha} (-1)^{l_1+n-q'} V_{mrr'n} C_{l_1ml_1-n}^{a\alpha} C_{a\alpha k'-q'}^{kq},
\nonumber\\
\end{eqnarray}
where, $m'=n'=0$, $\alpha'=0$, and $q'=-\beta$ are used.
The direct exchange parameter between $J$-multiplet and isotropic spin is given by 
\begin{eqnarray}
 \mathcal{J}_{kqk'q'}^{\rm DE}
 &=&
 -\sum_{a}\sum_{r} \mathcal{V}^{arr'}_{kqk'q'} \mathcal{D}^1_{ak'k} \tilde{\mathcal{D}}^2_{rr'k'}.
\end{eqnarray}
From Eq. (\ref{Eq:k_DE}), the rank for the spin site $k'$ is 0 or 1.

\subsection{Kinetic exchange Hamiltonian}
Since $l_2=0$ and $L_2=0$, $J_2=S_2$, 
$a'=b'=c'=0$, $x'=f$, $e'=k'$ and $d'=1/2$ 
from the $6j$ and the $9j$ symbols in $\mathcal{F}$ (\ref{Eq:F}).
The values of the $6j$ and the $9j$ symbols are $(-1)^{2S'}/\Pi_{S'}$, $(-1)^{1+k'}/\Pi_{f\frac{1}{2}}$, 
$\delta_{fk'}/\Pi_{k'k'}$, and $1/S_2S_2k'$, respectively.
Thus, ${\mathcal{F}}^{2}_{\alpha_J Jfxk}$ becomes 
\begin{eqnarray}
 \tilde{\mathcal{F}}^{2}_{\alpha_S rr'k'}
 &=&
 (-1)^{2S} \frac{\Pi_S}{\Pi_{S_2}} C_{S_2S_2k'0}^{S_2S_2} 
 (-1)^{2k'}
\nonumber\\
&\times&
 \langle S_2\Vert \left\{\hat{c}_{2r}^\dagger \otimes 
 \hat{P}_{2\alpha_S 0}^{N_2-1} \bar{c}_{2r'} \right\}_{k'}\Vert S_2\rangle.
\label{Eq:F_tm}
\end{eqnarray}
%and %$\tilde{\mathcal{G}}^{2}_{\nu Srr'k'}$ is obtained replacing 
%$\hat{c}_{jr}^\dagger$, $\bar{c}_{jr'}$, $\hat{P}^{N_j-1}_{j\nu 0}$ in Eq. (\ref{Eq:F_tm}) by
%$\bar{c}_{jr}$, $\hat{c}_{jr'}^\dagger$, $\hat{P}^{N_j+1}_{j\nu 0}$, respectively.
%$\langle S_2\Vert \{\hat{c}_{2r}^\dagger \otimes 
%\hat{P}_{2\nu 0}^{N_2-1} \bar{c}_{2r'} \}_{k'}\Vert S_2\rangle$
%by 
%$\langle S_2\Vert \{\bar{c}_{2r} \otimes 
%\hat{P}_{2\nu 0}^{N_2+1} \hat{c}_{2r'}^\dagger \}_{k'}\Vert S_2\rangle$.
%$\mathcal{F}$ and $\mathcal{G}$ are the same as Eq. (9) in the main text.
Here, the rank for the orbital part in the double tensor projection operator is removed, 
$\hat{P}_{2\alpha_S 0}^{N_2-1}$ is irreducible tensor form which acts on spin state:
\begin{eqnarray}
 \hat{P}_{2\alpha_S a \alpha}^N
 &=&
 \sum_{m_Sm_S'} (-1)^{S+m_S'} C_{Sm_SS-m_S'}^{a\alpha} |Sm_S\rangle \langle Sm_S'|.
\nonumber\\
\end{eqnarray}
%With the use of the Wigner-Eckart theorem (\ref{Eq:WE}), 
%\begin{eqnarray}
% \tilde{\mathcal{F}}^{2}_{\nu rr'k'}
% &=&
% (-1)^{2S} \Pi_S
%\nonumber\\
% &\times&
% \langle S_2S_2| \left\{\hat{c}_{2r}^\dagger \otimes 
% \hat{P}_{2\nu 0}^{N_2-1} \bar{c}_{2r'} \right\}_{k'0}|S_2S_2\rangle.
%\end{eqnarray}
$\tilde{\mathcal{G}}^{2}_{\alpha_S rr'k'}$ is obtained by replacing 
$\langle S_2\Vert\{\hat{c}_{2r}^\dagger \otimes \hat{P}_{2\alpha_S 0}^{N_2-1} \bar{c}_{2r'}\}_{k'}\Vert S_2\rangle$
with 
$\langle S_2\Vert\{\bar{c}_{2r} \otimes \hat{P}_{2\alpha_S 0}^{N_2+1} \hat{c}_{2r'}^\dagger\}_{k'}\Vert S_2\rangle$.
On the other hand, $\{t \times t\}_{kqk'q'}^{fxx'}$ (\ref{Eq:tt}) reduces to 
\begin{eqnarray}
 \{t\times t\}^{xrr'}_{kq k'q'} &=& 
 (-1)^{l_1-k'+q'} \sum_{mm'} \sum_\xi t_{mr}^{12} t_{r'm'}^{21}
\nonumber\\
&\times&
 C_{l_1m'kq}^{x\xi} C_{k'-q' l_1m}^{x\xi}.
\label{Eq:txt_tm}
\end{eqnarray}
Therefore, the kinetic exchange coupling parameter is obtained as 
\begin{eqnarray}
 \mathcal{J}_{kqk'q'}^{\rm KE}
 &=&
 \sum_{xrr'} \sum_{\alpha_J J} \sum_{\alpha_S' S'} 
 \frac{\{t \times t\}^{xrr'}_{kq k'q'}
 \mathcal{F}^{1}_{\alpha_J Jk'xk}\tilde{\mathcal{G}}^{2}_{\alpha_S' rr'k'}}
  {U_{12} + \Delta E_{1\alpha_J J}^{N_1-1} + \Delta E_{2\alpha_S'S'}^{N_2+1}} 
%  {U_{AB} + \Delta E_{\mu J}^{N_A-1} + \Delta E_{\nu J'}^{N_B+1}} 
\nonumber\\
 &+&
% +
 \sum_{xrr'} \sum_{\alpha_J J} \sum_{\alpha_S' S'} 
 \frac{\{t \times t\}^{xrr'}_{kq k'q'}
 \mathcal{G}^{1}_{\alpha_J Jk'xk}
 \tilde{\mathcal{F}}^{2}_{\alpha_J' S'rr'k'}}
       {U_{21} + \Delta E_{1\alpha_J J}^{N_1+1} + \Delta E_{2\alpha_J' S'}^{N_2-1}}.
%       {U_{BA} + \Delta E_{\mu J}^{N_A+1} + \Delta E_{\nu J'}^{N_B-1}}
\nonumber\\
 \label{Eq:J_tm}
\end{eqnarray}
From Eq. (\ref{Eq:k_KE}), the rank for the spin site $k'$ is 0 or 1.

\begin{table*}[tb]
\begin{ruledtabular}
\caption{Reduced matrix elements of the creation operator for the ground ${}^6H$-term of Dy$^{3+}$.}
\label{Table:f9}
\begin{tabular}{cccccc}
%\multicolumn{4}{c}{$(-1)^{2(l+1/2)}\langle f^9, {}^6H|\hat{c}^\dagger|f^8, \Gamma \rangle/\Pi_{LS}$
% $(-1)^{2(l+1/2)}\langle f^9, {}^6H|\hat{c}^\dagger|f^8, \Gamma \rangle/\Pi_{LS}$}
%\multicolumn{4}{c}{$(-1)^{2(l+1/2)}\langle f^9, {}^6H|\hat{c}^\dagger|f^8, \Gamma \rangle/\Pi_{LS}$}
%& \multicolumn{2}{c}{$(-1)^{2(l+1/2)}\langle f^{10}, \Gamma|\hat{c}^\dagger|f^9, {}^6H \rangle/\Pi_{LS}$}
%\\
$\Gamma$ & $(-1)^{2(l+1/2)}\langle f^9, {}^6H|\hat{c}^\dagger|f^8, \Gamma \rangle/\Pi_{LS}$ & 
$\Gamma$ & $(-1)^{2(l+1/2)}\langle f^9, {}^6H|\hat{c}^\dagger|f^8, \Gamma \rangle/\Pi_{LS}$ & 
$\Gamma$ & $(-1)^{2(l+1/2)}\langle f^{10}, \Gamma|\hat{c}^\dagger|f^9, {}^6H \rangle/\Pi_{LS}$ \\
\hline
${}^7F$    & $\sqrt{\frac{7}{3}}$                & ${}^5G(2)$ & $-\frac{1}{7}\sqrt{\frac{65}{11}}$ 
& ${}^5D$   & $-2 \sqrt{\frac{11}{35}}$ \\
${}^5D(1)$ & $-\frac{1}{3}\sqrt{\frac{10}{7}}$   & ${}^5G(3)$ & $\frac{3}{7} \sqrt{\frac{3}{2}}$ 
& ${}^5F$   & $\sqrt{\frac{33}{35}}$ \\
${}^5D(2)$ & $-\frac{2 \sqrt{5}}{21}$            & ${}^5H(2)$ & $-1$
& ${}^5G$   & $-\sqrt{\frac{13}{7}}$ \\
${}^5D(3)$ & $\frac{3}{7} \sqrt{\frac{15}{11}}$  & ${}^5I(1)$ & $\frac{1}{3}\sqrt{\frac{91}{11}}$
& ${}^5I$   & $\sqrt{\frac{14}{5}}$ \\
${}^5F(1)$ & $-\frac{1}{\sqrt{6}}$ & ${}^5I(2)$ & $\frac{1}{3}\sqrt{\frac{26}{11}}$\\
${}^5F(2)$ & $-\sqrt{\frac{3}{22}}$ & ${}^5K$    & $-\sqrt{\frac{15}{11}}$ \\
${}^5G(1)$ & $-\sqrt{\frac{65}{154}}$ & ${}^5L$    & $\sqrt{\frac{17}{11}}$\\
%& $(-1)^{2(l+1/2)}\langle f^{10}, \Gamma|\hat{c}^\dagger|f^9, {}^6H \rangle/\Pi_{LS}$\\
\end{tabular}
\end{ruledtabular}
\end{table*}

%
% 1 column
%
%\begin{table}[tb]
%\begin{ruledtabular}
%\caption{Reduced matrix elements of the creation operator for the ground ${}^6H$-term of Dy$^{3+}$.}
%\label{Table:f9}
%\begin{tabular}{cc}
%  $\Gamma$ & Reduced matrix element\\
%\hline
%& $(-1)^{2(l+1/2)}\langle f^9, {}^6H|\hat{c}^\dagger|f^8, \Gamma \rangle/\Pi_{LS}$\\
%%\hline
%%           & $f^{9}({}^6H)\to f^{8}(\Gamma)$      & & $f^{9}({}^6H)\to f^{10}(\Gamma)$\\
%${}^7F$    & $\sqrt{\frac{7}{3}}$                \\ 
%${}^5D(1)$ & $-\frac{1}{3}\sqrt{\frac{10}{7}}$   \\
%${}^5D(2)$ & $-\frac{2 \sqrt{5}}{21}$            \\ 
%${}^5D(3)$ & $\frac{3}{7} \sqrt{\frac{15}{11}}$  \\ 
%${}^5F(1)$ & $-\frac{1}{\sqrt{6}}$ \\
%${}^5F(2)$ & $-\sqrt{\frac{3}{22}}$ \\
%${}^5G(1)$ & $-\sqrt{\frac{65}{154}}$ \\
%${}^5G(2)$ & $-\frac{1}{7}\sqrt{\frac{65}{11}}$ \\
%${}^5G(3)$ & $\frac{3}{7} \sqrt{\frac{3}{2}}$ \\
%${}^5H(2)$ & $-1$\\
%${}^5I(1)$ & $\frac{1}{3}\sqrt{\frac{91}{11}}$\\
%${}^5I(2)$ & $\frac{1}{3}\sqrt{\frac{26}{11}}$\\
%${}^5K$    & $-\sqrt{\frac{15}{11}}$ \\
%${}^5L$    & $\sqrt{\frac{17}{11}}$\\
%& $(-1)^{2(l+1/2)}\langle f^{10}, \Gamma|\hat{c}^\dagger|f^9, {}^6H \rangle/\Pi_{LS}$\\
%${}^5D$   & $-2 \sqrt{\frac{11}{35}}$ \\
%${}^5F$   & $\sqrt{\frac{33}{35}}$ \\
%${}^5G$   & $-\sqrt{\frac{13}{7}}$ \\
%${}^5I$   & $\sqrt{\frac{14}{5}}$ \\
%\end{tabular}
%\end{ruledtabular}
%\end{table}
%
\section{Reduced matrix elements of the creation operators}
\label{Appendix:c}
In order to calculate the exchange interaction parameters, 
the reduced matrix elements in Eqs. (\ref{Eq:D}), (\ref{Eq:F}) must be evaluated.
In the direct exchange interaction (\ref{Eq:JDE}), there appear
$(-1)^{2a+2b}\langle L_iS_i\Vert \{c_i^\dagger \otimes \bar{c}_i \}^a_b \Vert L_iS_i \rangle$ and 
$(-1)^{2a+2b}\langle L_jS_j\Vert \{\bar{c}_j \otimes \hat{c}_j^\dagger \}^a_b \Vert L_jS_j\rangle$.
However, note that the irreducible tensor operators are the same type as $V^{11}$ used 
for the calculations of the spin-orbit coupling \cite{Note5}.
On the other hand, for the calculations of the kinetic exchange interactions (\ref{Eq:JKE}), 
$(-1)^{2c+2e}\langle L_iS_i\Vert \{c_i^\dagger \otimes
 \{ \hat{P}^{N_i-1}_{i\alpha_J aa} \otimes \bar{c}_i \}^b_d \}^c_e
 \Vert L_iS_i \rangle$ and 
$(-1)^{2c+2e}\langle L_jS_j\Vert \{ \bar{c}_j \otimes \{ \hat{P}^{N_j+1}_{j\alpha_J aa}\otimes 
 \hat{c}_j^\dagger \}^b_d \}^c_e \Vert L_jS_j\rangle$
have to be evaluated.
By straightforward calculations, the former is 
\begin{widetext}
%\begin{eqnarray}
%(-1)^{2c+2e}
% &&
%\langle L_iS_i\Vert 
% \left\{
%  c_i^\dagger \otimes
% \left\{
%  \hat{P}^{N_i-1}_{i\mu aa} \otimes \bar{c}_i
% \right\}^b_d
% \right\}^c_e
% \Vert L_iS_i \rangle
%\nonumber\\
% &&=
% \frac{\Pi_{L_iS_i}}{C_{L_iM_Lc\gamma}^{L_iM_L'}C_{S_iM_Se\epsilon}^{S_iM_S'}} 
% \left|\frac{(-1)^{2(l_i+1/2)}\langle l_i^{N_i+1},LS\Vert \hat{c}^\dagger_i 
% \Vert l_i^{N_i},L_iS_i\rangle}{\Pi_{LS}}\right|^2
%\nonumber\\
% &&\times
% \sum_{mn} \sum_{\alpha \beta} \sum_{N_LN_L'} 
% (-1)^{l_i+m+L+N_L} 
% C_{l_i-mb\beta}^{c\gamma} 
% C_{a\alpha l_in}^{b\beta} 
% C_{LN_L'L-N_L}^{a\alpha} 
% C_{L_iM_L'l_im}^{LN_L'}
% C_{L_iM_Ll_in}^{LN_L}
%\nonumber\\
% &&\times
% \sum_{\sigma \sigma'} \sum_{\alpha' \delta} \sum_{N_SN_S'} 
% (-1)^{\frac{1}{2}+\sigma+S+N_S} 
% C_{\frac{1}{2}-\sigma d\delta}^{e\epsilon} 
% C_{a\alpha' \frac{1}{2}\sigma'}^{d\delta} 
% C_{SN_S'S-N_S}^{a\alpha'} 
% C_{S_iM_S'\frac{1}{2}m}^{SN_S'}
% C_{S_iM_S\frac{1}{2}n}^{SN_S},
%\end{eqnarray}
\begin{eqnarray}
%(-1)^{2c+2e}
%&&
% \langle L_iS_i\Vert \left\{ \bar{c}_i \otimes
% \left\{ \hat{P}^{N_i+1}_{i\mu aa}\otimes \hat{c}_i^\dagger \right\}^b_d
% \right\}^c_e \Vert L_iS_i\rangle
%\nonumber\\
% &&=
%&=&
 \frac{\Pi_{L_iS_i}}{C_{L_iM_Lc\gamma}^{L_iM_L'}C_{S_iM_Se\epsilon}^{S_iM_S'}} 
%\nonumber\\
%&&\times
 &&
 \left|\frac{(-1)^{2(l_i+1/2)}\langle l_i^{N_i},L_iS_i\Vert \hat{c}^\dagger_i 
 \Vert l_i^{N_i-1},\alpha LS\rangle}{\Pi_{L_iS_i}}\right|^2
\nonumber\\ 
 &&\times
 \sum_{mn} \sum_{\alpha \beta} \sum_{N_LN_L'} 
 (-1)^{l_i+n+L+N_L} 
 C_{l_imb\beta}^{c\gamma} 
%\nonumber\\ 
% &&\times 
 C_{a\alpha l_i-n}^{b\beta} 
 C_{LN_L'L-N_L}^{a\alpha} 
 C_{L_iM_L'l_im}^{LN_L'}
 C_{L_iM_Ll_in}^{LN_L}
\nonumber\\
 &&\times
% &\times&
 \sum_{\sigma \sigma'} \sum_{\alpha' \delta} \sum_{N_SN_S'} 
 (-1)^{\frac{1}{2}+\sigma'+S+N_S} 
 C_{\frac{1}{2}\sigma d\delta}^{e\epsilon} 
%\nonumber\\ 
% &&\times 
 C_{a\alpha' \frac{1}{2}-\sigma'}^{d\delta} 
 C_{SN_S'S-N_S}^{a\alpha'} 
 C_{S_iM_S'\frac{1}{2}m}^{SN_S'}
 C_{S_iM_S\frac{1}{2}n}^{SN_S},
\end{eqnarray}
and the latter is 
\begin{eqnarray}
% &&
 \frac{\Pi_{L_jS_j}}{C_{L_jM_Lc\gamma}^{L_jM_L'}C_{S_jM_Se\epsilon}^{S_jM_S'}} 
%\nonumber\\
% &&\times
 &&
 \left|\frac{(-1)^{2(l_j+1/2)}\langle l_j^{N_j+1},\alpha LS\Vert \hat{c}^\dagger_j 
 \Vert l_j^{N_j},L_jS_j\rangle}{\Pi_{LS}}\right|^2
\nonumber\\
 &&\times
 \sum_{mn} \sum_{\alpha \beta} \sum_{N_LN_L'} 
 (-1)^{l_j+m+L+N_L} 
 C_{l_j-mb\beta}^{c\gamma} 
%\nonumber\\
% &&\times
 C_{a\alpha l_jn}^{b\beta} 
 C_{LN_L'L-N_L}^{a\alpha} 
 C_{L_jM_L'l_jm}^{LN_L'}
 C_{L_jM_Ll_jn}^{LN_L}
\nonumber\\
 &&\times
 \sum_{\sigma \sigma'} \sum_{\alpha' \delta} \sum_{N_SN_S'} 
 (-1)^{\frac{1}{2}+\sigma+S+N_S} 
 C_{\frac{1}{2}-\sigma d\delta}^{e\epsilon} 
%\nonumber\\
% &&\times
 C_{a\alpha' \frac{1}{2}\sigma'}^{d\delta} 
 C_{SN_S'S-N_S}^{a\alpha'} 
 C_{S_jM_S'\frac{1}{2}m}^{SN_S'}
 C_{S_jM_S\frac{1}{2}n}^{SN_S}.
\end{eqnarray}
\end{widetext}
Here, the components $M_L,M_L',M_S,M_S',\gamma,\epsilon$ are chosen so that 
$C_{L_iM_Lc\gamma}^{L_iM_L'}C_{S_iM_Se\epsilon}^{S_iM_S'} \ne 0$ is satisfied.
For the calculations of the equations, 
the reduced matrix elements of the creation operators, $\hat{c}_{im\sigma}^\dagger$, are necessary.
The are calculated as \cite{Judd1967}
\begin{eqnarray}
 \frac{(-1)^{2(l+1/2)}\langle f^N \alpha L S\Vert \hat{c}^\dagger \Vert f^{N-1} \alpha' L'S'\rangle}{\Pi_{LS}}
\nonumber\\
 = 
 \sqrt{N} \langle f^N \alpha L S \{|f^{N-1}(\alpha'L'S')fLS\rangle,
\end{eqnarray}
where $\langle f^N \alpha L S \{|f^{N-1}(\alpha'L'S')fLS\rangle$ 
is the coefficient of fractional parentage,
which is tabulated in Ref. \onlinecite{Nielson1963}.
%The c.f.p.'s in the tables are those for $f^n$ $(1 \le n \le 7)$.
%The c.f.p.'s for $f^n$ with $n \ge 8$ are given as follows:
%\begin{eqnarray}
% \langle f^{4l+3-n} \alpha L S \{|f^{4l+2-n}(\alpha'L'S')fLS\rangle
% &=& (-1)^{S+S'+L+L'-l-\frac{1}{2}}
% \sqrt{\frac{n}{4l+3-n}}\frac{\Pi_{L'S'}}{\Pi_{LS}}
%\nonumber\\
% &\times& \langle f^n \alpha' L' S' \{|f^{n-1}(\alpha LS)fL'S'\rangle,
%\end{eqnarray}
%The $LS$-term $\Gamma$ in the table corresponds to that in Ref. \onlinecite{Nielson1963}.
The reduced matrix elements of $\hat{c}^\dagger$ 
necessary for the present examples are shown in Table \ref{Table:f9}.

\section{Exchange states of N$_2^{2-}$ bridged Dy$^{3+}$ dimer}
\label{Appendix:example}
The difference between the eigenstates of the full exchange Hamiltonian $\hat{H}_{\rm ex}$, 
$|\Psi_i,\Gamma\rangle$, and those of the Heisenberg type Hamiltonian (\ref{Eq:JJ}), $|J_{12}M_{12}\rangle$, 
are compared as in the case of the linear system.
The low-energy states are, in the basis of $\{|J_{12}M_{12}\rangle\}$ (\ref{Eq:JM_Heis}),
\begin{widetext}
\begin{eqnarray*}
 |\Psi_1, b_{1u}\rangle &\approx& 
 0.767 |1,0\rangle + 0.493 |3,0\rangle 
 + 0.108 (|3,-2\rangle + |3,2\rangle) 
 + 0.345|5,0\rangle 
 + 0.118|7,0\rangle,
\\
 |\Psi_2, a_g\rangle &\approx&
 0.552|0,0\rangle + 0.661|2,0\rangle + 0.400|4,0\rangle 
 + 0.250|6.0\rangle,
\\
 |\Psi_3, b_{2g}\rangle &\approx&
 0.586 (-|2,-1\rangle + |2,1\rangle) 
 - 0.260 (-|4,-1\rangle + |4,1\rangle) 
 + 0.256 (-|4,-3\rangle + |4,3\rangle) 
 + 
 0.108 (-|8,-1\rangle + |8,1\rangle),
\\
 |\Psi_4, b_{3u}\rangle &\approx& 
 -0.545 (-|1,-1\rangle + |1,1\rangle)
 - 0.158 (-|3,-1\rangle + |3,1\rangle) 
 - 0.161 (-|3,-3\rangle + |3,3\rangle) 
 + 0.293 (-|5,-1\rangle + |5,1\rangle) 
\nonumber\\
 &-&
  0.218 (-|5,-3\rangle + |5,3\rangle) 
 - 0.129 (-|7,-1\rangle + |7,1\rangle),
\\
 |\Psi_5, b_{1u}\rangle &\approx&
 -0.285 |1,0\rangle + 0.680 |3,0\rangle - 0.448 (|3,-2\rangle + |3,2\rangle),
\\
 |\Psi_6, a_g\rangle &\approx&
 0.446 |0,0\rangle - 0.331 |2,0\rangle + 0.347 (|2,-2\rangle + |2,2\rangle)
 -0.460 |4,0\rangle 
 + 0.255 (|4,-2\rangle + |4,2\rangle)
 + 0.181 |6,0\rangle 
\nonumber\\
&-& 0.155 (|6,-2\rangle + |6,2\rangle).
\end{eqnarray*}
Here, the irreducible representation $\Gamma$ of $D_{2h}$ symmetry is used.
The exchange states belong to the eigenvalues 
$E_1 = -0.274979$, $E_2 = -0.274977$, $E_3 = -0.262325$, $E_4 = -0.262299$, 
$E_5 = -0.260815$, $E_6 = -0.260805$, respectively.
As in the axial system, the low-energy states are not well described by $\hat{H}_{\rm Heis}$.
Within the $1/\bar{U}$ approximation, the low-energy states become
%\begin{widetext}
\begin{eqnarray*}
 |\Psi^{1/\bar{U}}_1, a_g \rangle &=& 0.509|0,0\rangle + 0.696|2,0\rangle 
 + 0.434|4,0\rangle + 0.222|6,0\rangle, 
\\
 |\Psi^{1/\bar{U}}_2, b_{1u} \rangle &=& 0.740|1,0\rangle + 0.559|3,0\rangle + 0.330|5,0\rangle + 0.115|7,0\rangle,
\\
 |\Psi^{1/\bar{U}}_3, b_{3u} \rangle &=& 0.554(-|1,-1\rangle + |1,1\rangle) + 0.205(-|3,-1\rangle + |3,1\rangle)
 + 0.193(-|3,-3\rangle + |3,3\rangle) 
 - 0.234(-|5,-1\rangle + |5,1\rangle)
\nonumber\\
 &+& 0.220 (-|5,-3\rangle + |5,3\rangle),
\\
 |\Psi^{1/\bar{U}}_4, b_{2g} \rangle &=& 0.608(-|2,-1\rangle + |2,1\rangle)
 -0.188 (-|4,-1\rangle + |4,1\rangle) 
%\nonumber\\
 + 0.279 (-|4,-3\rangle + |4,3\rangle), 
\\
 |\Psi^{1/\bar{U}}_5, a_g \rangle &=& 0.489|0,0\rangle - 0.263 |2,0\rangle - 0.374(-|2,-2\rangle + |2,2\rangle) 
 - 0.430 |4,0\rangle 
 - 0.265(-|4,-2\rangle + |4,2\rangle) 
 + 0.141|6,0\rangle 
\nonumber\\
 &+& 0.127(-|6,-2\rangle + |6,2\rangle) 
 - 0.103(-|6,-4\rangle + |6,4\rangle),
\\
 |\Psi^{1/\bar{U}}_6, b_{1u} \rangle &=& 
 0.351|1,0\rangle - 0.618 |3,0\rangle + 0.471(|3,-2\rangle + |3,2\rangle) + 0.109 (|5,-4\rangle + |5,4\rangle).
\end{eqnarray*}
\end{widetext}
The exchange states $|\Psi_i^{1/\bar{U}}\rangle$ are also ordered as the increase of the energy.
Some levels are interchanged due to the $1/\bar{U}$ approximation. 
$|\Psi_i\rangle$ and $|\Psi_i^{1/\bar{U}}\rangle$ with the same representation 
quantitatively differ from each other.
%As discussed main text, the order of the exchange states is changed by this approximation.

% Reference
\bibliographystyle{apsrev4-1}
\bibliography{ref}

\end{document}